\documentclass[aps,prx,twocolumn,showpacs,superscriptaddress,groupedaddress]{revtex4-2}

\usepackage{bm}
\usepackage{graphicx}
\usepackage{dcolumn}
\usepackage{lineno}  
\usepackage{siunitx}
\usepackage{amsmath}
\usepackage{graphicx,amssymb} 
\usepackage{tabularx}
\usepackage{booktabs}
\usepackage{setspace}

\usepackage[caption=false,font=footnotesize,captionskip=-0mm,farskip=0cm]{subfig}
\begin{document}

\title{Strong anomalous diffusion for free-ranging birds}

\author{Ohad Vilk$^{a,b,\ast}$, Motti Charter$^{c}$, Sivan Toledo$^{d}$, Eli Barkai$^{e}$, Ran Nathan$^{b}$}

\affiliation{a Racah Institute of Physics, The Hebrew University of Jerusalem, Jerusalem, Israel}
\affiliation{b Movement Ecology Lab, Department of Ecology, Evolution and Behavior, Alexander Silberman Institute of Life Sciences, Faculty of Science, The Hebrew University of Jerusalem, Jerusalem, Israel.}
\affiliation{c The Shamir Research Institute and School of Environmental Sciences, University of Haifa, 199 Aba Hushi Boulevard, Mount Carmel, Haifa, Israel.}
\affiliation{d Blavatnik School of Computer Science, Tel-Aviv University, Israel.}
\affiliation{e Department of Physics, Institute of Nanotechnology and Advanced Materials, Bar-Ilan University, Ramat-Gan 52900, Israel.}
\affiliation{$^{\ast}$To whom correspondence should be addressed: ohad.vilk@mail.huji.ac.il}

\begin{abstract}
  Diffusion and anomalous diffusion are widely observed and used to study movement across organisms, resulting in extensive use of the mean and mean-squared displacement (MSD). However, these measures -- corresponding to specific displacement moments -- do not capture the full complexity of movement behavior. Using high-resolution data from over 70 million localizations of young and adult free-ranging Barn Owls (\textit{Tyto alba}), we reveal strong anomalous diffusion as nonlinear growth of displacement moments. The moment spectrum function $\lambda_t(q)$ -- defined by $\left<|\bm{x}(t)|^q\right> \sim t^{\lambda_t(q)}$ -- displays piecewise linearity in $q$, with a critical moment marking the crossover between scaling regimes. This highlights the need of a broad spectrum of displacement moments to characterize movement, which we link to age-specific ecological drivers. Furthermore, a characteristic timescale of five minutes marks an unexpected transition from a convex to a concave $\lambda_t(q)$. Using two stochastic models -- a bounded Lévy walk and a multi-mode behavioral model -- we account for the observed phenomena, showing good agreement with data, relating age-specific behavioral states to environmentally confined movement, and demonstrating how Lévy walk-like patterns can arise from underlying behavioral structure. Finally, we discuss the ecological significance of our results, arguing that strong anomalous diffusion may be widespread in animal movement. 
\end{abstract}

\maketitle

\section{Introduction} \label{sec:introduction}

Understanding the movement of organisms is a central challenge across scientific disciplines, playing an essential role in a wide range of biological and ecological systems. For over a century, diffusion models have been instrumental in addressing this challenge. Karl Pearson first proposed the random walk model in a 1905 Nature letter to solve the challenge of mosquito dispersion in forests \cite{pearson1905problem}. Lord Rayleigh subsequently linked this model to the central limit theorem, noting that for a moderately large number of displacements the distribution of a random walker approaches a Gaussian distribution \cite{rayleigh1905problem}. These theories have since become central in a wide array of scientific analyses, from error quantification to the diffusion of diseases \cite{klafter2011first}. More recently, it has been recognized that the random walk model is applicable primarily to Markovian processes, whereas many natural processes do not exhibit Markovian properties \cite{sokolov2005diffusion}. Consequently, the study of so-called anomalous diffusion has become crucial in biological and ecological systems \cite{gonzalez2008understanding,sims2008scaling, metzler2000random, zaburdaev2015levy}. Nevertheless, most diffusion-based movement studies are still primarily focus on the mean and mean-squared displacement (MSD), often overlooking higher displacement moments~\cite{metzler2014anomalous, vilk2022unravelling, ran2022BigData, polev2022large}.   

Currently, diffusion and anomalous diffusion models, see definitions below, describe the movement of particles, molecules, and animals in a wide range of physical, chemical, biological, and ecological systems~\cite{okubo2001diffusion, ricciardi2013diffusion}. 
In the context of organism movement, diffusion models are used to explain how individuals disperse over space and time, affecting population dynamics, spatial distribution, and interaction with the environment \cite{holmes1994partial, hooten2017animal}. Diffusion is also relevant in ecological settings, where dispersal and other movement phenomena determine the structure and spread of the population, influence the dynamics of metapopulation, and underlie predator-prey interactions and the coexistence of competitors \cite{hanski1991metapopulation}. Understanding the diffusion processes that underlie the movement of organisms is also crucial for addressing applied ecological concerns related to habitat fragmentation, biodiversity conservation, species invasion, and the management of natural resources \cite{turchin1998quantitative, hooten2016hierarchical, trakhtenbrot2005importance}.
Furthermore, within movement ecology \cite{nathan2008movement, ran2022BigData}, models such as the random walk are essential to link movement metrics such as mean-displacement and travel-distance to underlying biological and environmental drivers. By employing a range of models, from minimal parameter models \cite{holmes1993diffusion, viswanathan2011physics, mendez2016stochastic} to more sophisticated, parameterized simulations that incorporate detailed ecological contexts and correlations \cite{mclane2011role, gurarie2016animal, gurarie2017framework}, empirical field data can be parsed into meaningful patterns and predictions. These models often offer insights into the mechanisms that govern movement and predict how organisms respond to environmental changes \cite{nathan2011spread,patterson2017statistical, mcgarigal2016multi}, thus bridging between empirical observation and theoretical understanding of movement.


Common observables in movement studies are the mean displacement and the MSD, both central to quantifying the diffusive movements of organisms \cite{nouvellet2009fundamental, ahmed2013measuring}. In ecological studies, MSD is used to characterize dispersal rates and movement strategies, offering insights into the underlying mechanisms of organism movement for both normal and anomalous diffusion processes \cite{bastille2016flexible, vilk2022unravelling, vilk2022classification, burte2023complex}. 
In normal diffusion, e.g., Pearson's random walk \cite{pearson1905problem}, the Gaussian central limit theorem results in an MSD increasing linearly with time according to 
$\langle \bm{x}^2(t) \rangle \sim 2 d Dt$, where $\bm{x}(t) = \vec{x}(t) - \vec{x}(0)$ is the position at time $t$, relative to the initial position, $d$ is the spatial dimension, and $D$ is the diffusion coefficient. Although this is a useful model for many natural processes, in many practical scenarios, the statistical properties of particle motion deviate from the classical Gaussian behavior \cite{metzler2014anomalous}. In these systems, with so-called \textit{anomalous} diffusion, the MSD exhibits a behavior inconsistent with linear scaling with time
according to
$
\langle \bm{x}^2(t) \rangle \sim t^{\alpha}
$,
with $\alpha$ being the diffusion exponent. Here, \(\alpha = 1\) corresponds to normal diffusion, \(\alpha < 1\) to subdiffusion with slow spread, and \(\alpha > 1\) to superdiffusion with faster-than-normal spread. 

In many studies, the MSD is used to characterize the motion, be it normal or anomalous. However, this may lead to a wrong conclusion, as the MSD is a meaningful characteristic for processes where the distribution scales like $(1/t^{\alpha/2})F(\bm{x}/t^{\alpha/2})$ (e.g., Gaussian processes). Such a behavior is called  mono-scaling, as $\bm{x}$ scales with $t^{\alpha/2}$. However, it has been shown by the work of Vulpiani and colleagues \cite{castiglione1999strong, andersen2000simple, vollmer2021displacement} that in a more general setting the MSD is only one point in the spectrum of exponents describing the growth of the moments of order $q$. Thus, a useful tool to analyze data is the moment spectrum function $\lambda_t(q)$ defined by
\begin{equation}
    \left<|\vec{x}(t) - \vec{x}(0)|^q\right> \equiv \left<|\bm{x}|^q\right>  \sim t^{\lambda_t(q)}, 
\end{equation}
where $q$ is a positive real number, and for the special case of $q=2$ we recover the MSD. For scale-invariant processes, the scaling exponent is time-independent and linear in $q$, e.g., for Brownian motion it follows $\lambda_t(q) = q/2$ and for fractional Brownian motion \cite{Man68fbm} $\lambda_t(q) = \alpha/2$ for all $q$. A nonlinear dependence of $\lambda_t(q)$ in $q$ is termed \textit{strong anomalous diffusion} (SAND) \cite{castiglione1999strong} and entails that the MSD is not indicative of displacement at varying spatial scales. 

SAND is evident in diverse theoretical processes such as transport in two-dimensional incompressible velocity fields \cite{castiglione1999strong}, dynamics in billiard systems \cite{sanders2006occurrence,  orchard2021diffusion}, avalanche behaviors in sandpile models \cite{carreras1999anomalous},  cold atoms in optical lattices \cite{dechant2012anomalous, afek2023colloquium}, and more \cite{cagnetta2015strong, wang2020strong, liu2022strong, samama2023statistics, bernardi2024anomalous}. 
Many of these systems share a piecewise linear scaling behavior, with two scaling exponents and a transition at a critical moment value, showcasing bi-linear scaling. 
Experimentally, SAND was found for tracers in cancerous cells~\cite{gal2010experimental}.
Recent experiments have also shown non-mono-scaling of moments for diffusion of light in a disordered medium \cite{pini2023breakdown}. 
To the best of our knowledge, empirical evidence of SAND is still limited and has not been shown in ecological systems. 

In this manuscript, we study and model the movement patterns of Barn Owls (\textit{Tyto alba}) based on the displacement moments of their movement. In Sec. \ref{sec:methods}, we describe the tracking, numerical, and analytical methods. Importantly, the same dataset was previously analyzed in Ref.~\cite{vilk2022ergodicity}, and is revisited here from a new theoretical and numerical perspective, applying the SAND framework and identifying movement scales not previously reported. In Sec. \ref{sec:results} we show empirical evidence for SAND with bi-linear scaling in the displacement moments, revealing that the MSD can be situated at the transition between scaling regimes, such that it does not represent either behavior accurately. 
Interestingly, SAND is found for both adult and young Barn Owls during the first few month after fledging (so-called fledglings) 
with a convex bi-linear $\lambda_s(q)$ at times shorter than 5 minutes. 
Contrary to the convex moment spectrum function at short times, we find a qualitatively different \textit{concave} function $\lambda_\ell(q)$ at longer times. In Sec. \ref{sec:simulations} we employ two stochastic models with varying degrees of ecological complexity to account for the bi-linear scaling and the transition from convex to concave moment spectrum functions. First, in Sec. \ref{subsec:BLW} a bounded L\'evy-walk (BLW) model is used as a zero-order model of the observed phenomena, coupling large displacements to bounded movement within a home range. 
Second, in Sec. \ref{sec:OU} we use an ecologically-oriented multi-mode correlated velocity model \cite{gurarie2016animal, breed2017predicting, gurarie2017framework, patterson2017statistical, jonsen2005robust} to show that the observed phenomena can be attributed to a mixture of two distinct ecological behaviors: local area-restricted searches and large-scale commutes~\cite{benhamou2014scales, vilk2022ergodicity}. We show that by parameterizing the multi-mode correlated velocity model from data we account for the observed phenomena, and for the similarities and differences between adults and fledglings.  
In Sec. \ref{sec:biology} we interpret the empirical SAND and scaling transitions in relation to age-specific foraging strategies and behavioral timescales \cite{wunderle1991age}, 
and discuss the generality of our results with respect to animal movement in light of the broader literature on scale-dependent movement~\cite{zaburdaev2015levy, benhamou2014scales}.
Finally, in Sec. \ref{sec:discussion} we summarize and establish the importance of accounting for a wide and continuous range of moments in movement studies.


\section{Methods} \label{sec:methods}
\subsection{Tracking} \label{sec:tracking}
Sixty Barn Owls were tracked in the Hula Valley, Israel (33.10N, 35.61E) between May and December 2018, using ATLAS (Advanced Tracking and Localization of Animals in Real-Life Systems), a reverse GPS system that localizes extremely light-weight, low-cost tags~\cite{weiser2016characterizing, toledo2016lessons, ran2022BigData}. Each ATLAS tag transmits a distinct radio signal which is detected by a network of base-stations distributed in the study area. Tag localization is computed using nanosecond-scale differences in signal time-of-arrival to each station, allowing for real-time tracking and alleviating the need to retrieve tags or have power-consuming remote-download capabilities. The individuals tracking frequency was 0.125 and 0.25 Hz, for the fledglings and adults, respectively. Localization errors are reported as a $2\times 2$  covariance matrix per localization. In this study we omit localizations with variance $> 50^2 \;m^2$, defined in terms of the trace over the covariance matrix. Furthermore, we filtered out nights in which many localizations are missing ($> 70\%$). Notably, in accordance with the typical error reported by the system ($\sigma \simeq5$ m)~\citep{weiser2016characterizing}, we assume $10$ m to be the noise limit in our measurements. While the noise is practically much smaller, this is treated as an upper limit for any significant results. 

Eighteen adults, breeding in nest boxes \cite{charter2022importance}, were tracked both during and post-breeding, and we used the hatching date to define the breeding season by the 90 days following hatching of nestlings, as for this period the nestlings still depend on their parents~\citep{taylor2004barn}. In addition, 42 fledglings were tracked up to five months after fledging. In total, our analyses incorporated high-quality data for 7,391 nights and over 70 million localizations. We limited the analyses to movement data collected during the nights, the activity hours of the Barn Owl.

In Figs.\ref{fig1:sub1} and \ref{fig1:sub5}, we present thirty-night movement tracks of an adult female and a fledgling, respectively, in the Hula Valley. The movement patterns exhibit both localized search behavior and long-distance commutes \cite{vilk2022ergodicity}. Additionally, the spatial concentration of activity suggests a distinct home range, while the fledgling appears to engage in more long-distance commutes than the adult female. We elaborate on these points below. Finally, we verified that the maximum nightly displacement -- $\max[|\bm{x}|]$ -- is $2.6 \pm 1.2$ km, consistent with previously reported values of $2.45 \pm 0.93$ km and $2.77 \pm 0.25$ km from other Barn Owl populations in Israel~\cite{cain2023movement, rozman2021movement}.

\subsection{Segmentation} \label{sec:segmentation}
To separate long-distance commutes from localized search behavior, daily tracks were segmented by detecting distinct switching points separating the two modes. We used the Penalized Contrast Method suggested by Barraquand and Benhamou \cite{barraquand2008animal}, a non-parametric method in which the initial number of segments is unknown and estimated by minimizing a penalized contrast function. First passage time (FPT) was used as the focal metric \cite{barraquand2008animal}. Each point was assigned an FPT outside a radius of $R_s$ and data was segmented such that points with similar FPT that were close in time were clustered together \cite{lavielle2005using}. Data are then split into commuting and searches according to a threshold on the mean FPT chosen in accordance with the animal's velocity. In our segmentation we choose $R_s = 100$ m and a threshold of $50$ s, 
see Ref. \cite{vilk2022ergodicity} for additional details.  

\subsection{Data analysis}

To calculate the displacement moments, we measure the absolute displacement \(|\bm{x}(t)| = |\vec{x}(t) - \vec{x}(0)| \) of an owl at time \( t \) from its initial position \(\vec{x}(0) \). We then raise this displacement to the power of the moment order \( q \), and take the average of these values for each \( t \) across multiple observations. The averaging is performed over all tracks of adults and similarly over all tracks of fledglings to construct the empirical moment curves for each group \footnote{Study of SAND for individual animals requires collecting additional data per individual over longer periods and employing time averages instead of ensemble averages, and is not discussed in this manuscript.} and each \(q\). Here,  $t \in [0, T]$ , where $T = 4$ hours is the chosen observational time-frame for this analysis.  We choose 4-hour windows to encompass multiple foraging events -- which are typically around 40 minutes long \cite{vilk2022ergodicity} -- while still being shorter than a full night -- which deterministically ends with the owls returning to their nests. Nevertheless, we have checked that the results shown below qualitatively hold for $T = 1$, $T = 4$ and $T =8$ hours; see Supplementary Information, Fig. S1 \cite{SIref}. 

Fits for all moments and for the moment spectrum functions below were performed using SciPy library's curve-fit (nonlinear least squares method) in python 3.12. In accordance with the observed bi-linear trends we have fitted a piecewise linear function to the observed curves, measuring the slopes at low and high moments and the transition point between the two slopes. As an error estimate we took one standard deviation of the fitted parameter. In general, across all panels shown below, the symbol sizes are larger than or comparable to the errors. Notably, in several empirical tracks we have missing data points; consequently, we only include in our analysis, time-series in which $>$90\% of the points are present. For any time-series that has $>$90\% of the points present, we fill each missing data point with a NaN (Not a Number), \textit{i.e.}, an empty placeholder which is naturally not included in the moment averaging. To verify that this choice does not affect the statistics, we checked that our results do not change when varying the 90\% threshold between 70\% and 95\%. Furthermore, we checked that in the simulations, randomly replacing 10\% of the data points in a simulated ensemble with NaNs does not significantly affect the results.

\section{Results} \label{sec:results}

\begin{figure*}[t]
\centering
\subfloat[]{\hspace{-4mm}
  \centering
 \includegraphics[width=.26\linewidth]{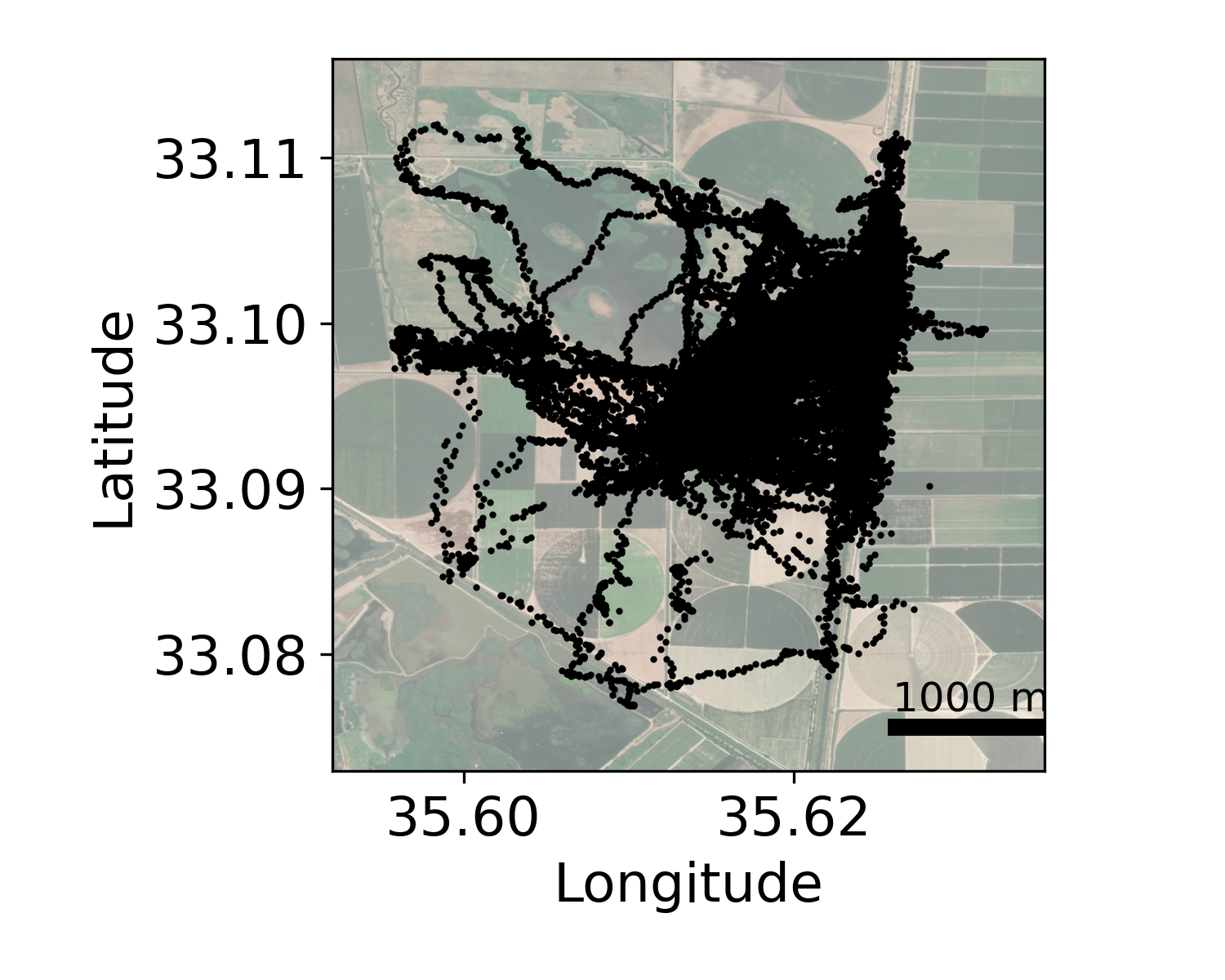}
  \label{fig1:sub1}
}
\subfloat[]{\hspace{-8mm}
  \centering
 \includegraphics[width=.26\textwidth]{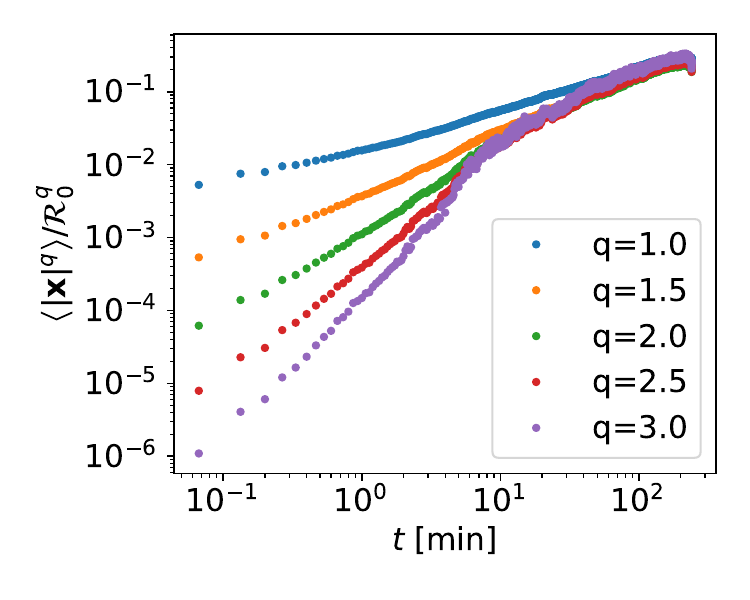}
  \label{fig1:sub2}
}
\subfloat[]{\hspace{-4mm}
  \centering
 \includegraphics[width=.26\textwidth]{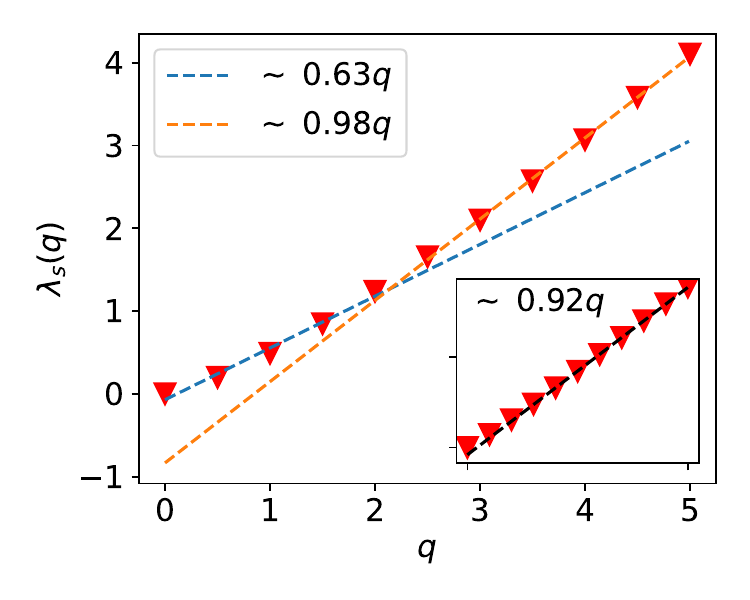}
  \label{fig1:sub3}
} 
\subfloat[]{\hspace{-4mm}
  \centering
 \includegraphics[width=.26\textwidth]{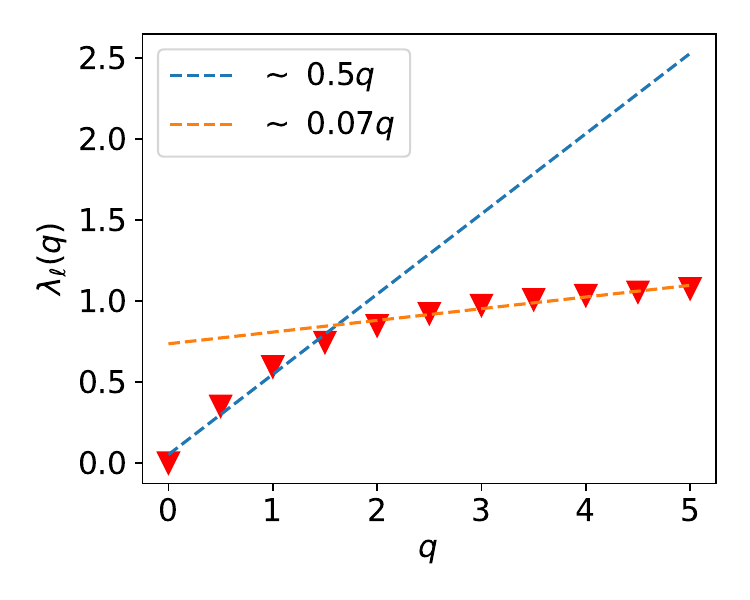}
  \label{fig1:sub4}
} \vspace{-1mm}
\centering
\subfloat[]{\hspace{-4mm}
  \centering
 \includegraphics[width=.26\linewidth]{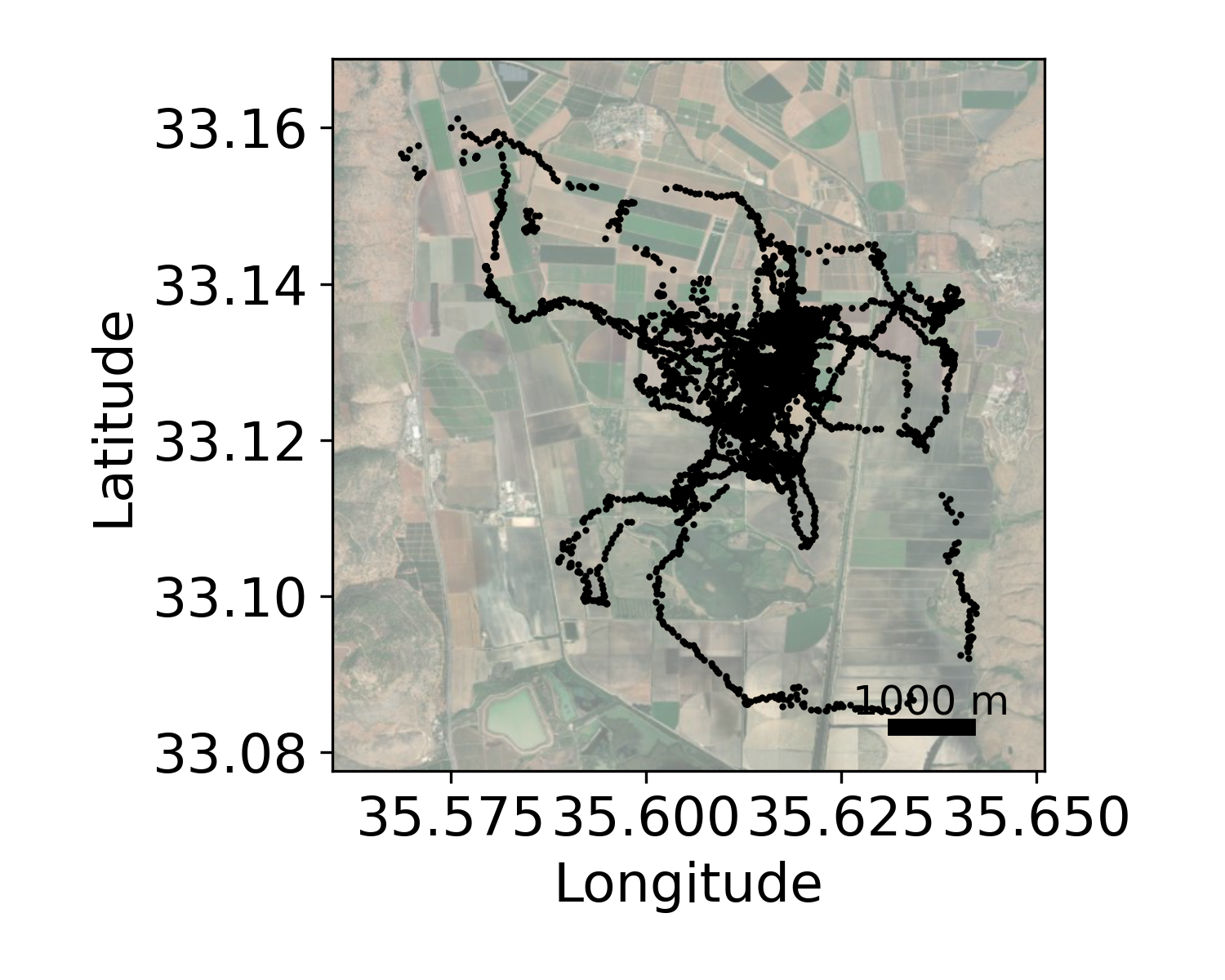}
  \label{fig1:sub5}
}
\subfloat[]{\hspace{-8mm}
  \centering
 \includegraphics[width=.26\textwidth]{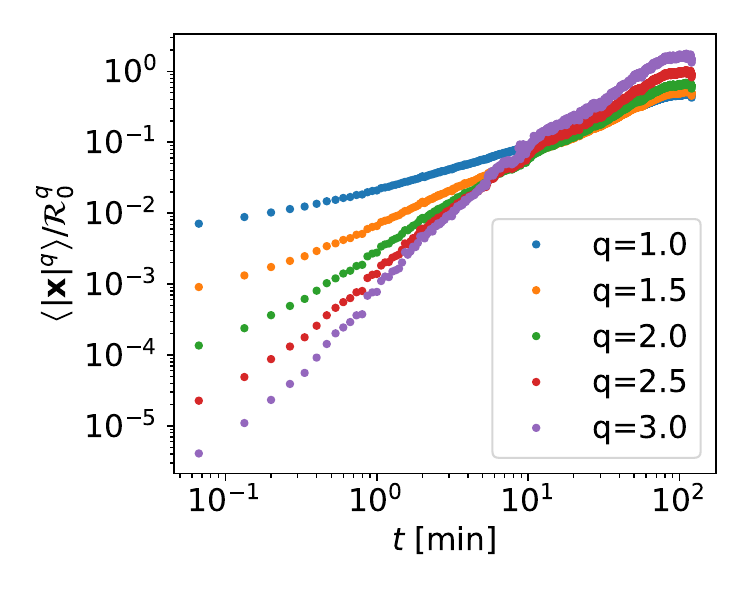}
  \label{fig1:sub6}
}
\subfloat[]{\hspace{-4mm}
  \centering
 \includegraphics[width=.26\textwidth]{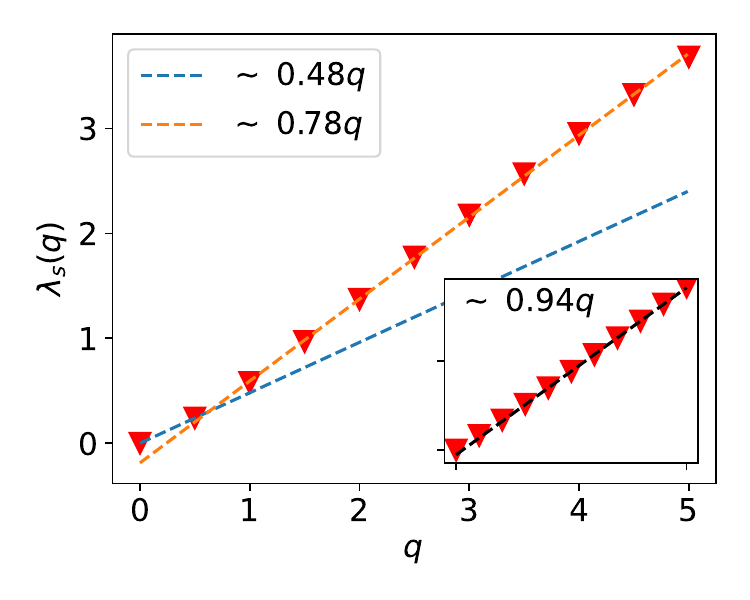}
  \label{fig1:sub7}
} 
\subfloat[]{\hspace{-4mm}
  \centering
 \includegraphics[width=.26\textwidth]{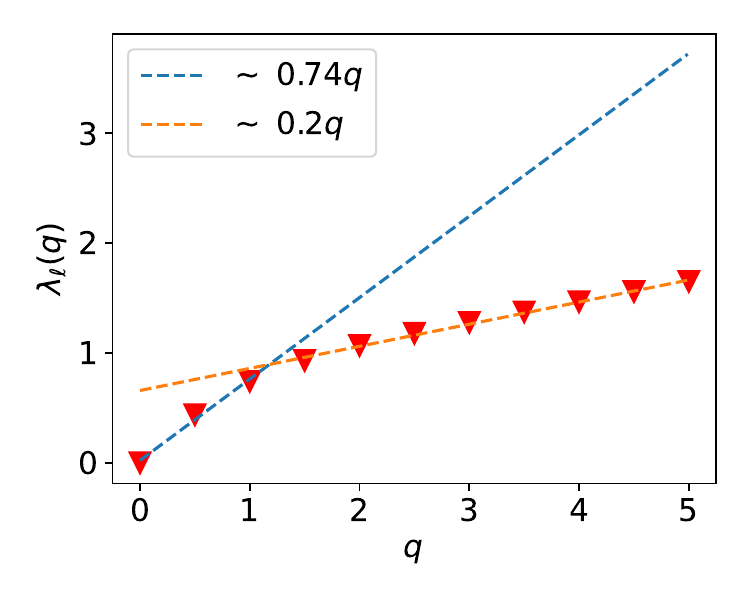}
  \label{fig1:sub8}
} \vspace{-1mm}
\centering
\caption{Results for the adult (a-d) and fledgling (e-h) Barn Owls. (a,e) Thirty nights of ATLAS relocations of a (a) Female and (e) fledgling Barn Owl in the Hula Valley at a mostly constant frequency of 0.25 Hz. Tracks are split into observational time-frames of $T = 4$ hours. (b,f) Log-log plot of the average absolute displacement moments \(\langle |\bm{x}|^q \rangle / \mathcal{R}_0^q\) with $\mathcal{R}_0 = 2.5$ km representing the average maximum displacement in a single night. Data points for moment orders $q=1, 1.5, 2, 2.5, 3$ are shown, see legend. We omit data for other moments in these panels, for clarity.  Moments show distinct scaling regimes, with a transition at $\sim 5$ minutes. (c,g) Moment spectrum function \(\lambda_s(q)\) at short times ($< 5$ minutes, red triangles) exhibiting convex behavior. The dashed blue and orange lines represent the slopes of the bi-linear fit for low and high moments, respectively, see values in legends. Insets: \(\lambda(q)\) for commuting flights alone (omitting area-restricted searches) as identified by segmenting the full tracks using behavioural change point analysis. The $q$ values along the horizontal axis in the inset are identical to those in the main panel and span the same range. In addition, the axis labels of the main panels apply also to the insets. (d,h) Moment spectrum function \(\lambda_\ell(q)\) for long time intervals ($> 5$ minutes, red triangles), exhibiting concave behavior. The dashed blue and orange lines represent the slopes of the bi-linear fit.  }
\label{fig1}
\end{figure*}

Results for the Barn Owls are shown in Fig. \ref{fig1}. In Figs. \ref{fig1:sub2} and \ref{fig1:sub6} we show the averaged displacement moments \( \langle |\bm{x}|^q \rangle / \mathcal{R}_0^q\) for $q$ values between 1 and 3. Here, \(\mathcal{R}_0 = 2.5\) km corresponds to the average maximum displacement that Barn Owls cover in a night (see Sec. \ref{sec:tracking}), and $\mathcal{R}_0^q$ is used to normalize the data, plotting the moments in a unitless format. For each moment $q$, we have performed linear fits for both short and long times, and for all moments we observe a crossover between two regimes at $\sim 5$ minutes. For example, the MSD ($q=2$) in Fig. \ref{fig1:sub2} follows $\langle  |\bm{x}|^2(t) \rangle \sim t^{1.25}$ for $t < 5$ minutes, entailing superdiffusive dynamics with faster-than-linear spread, and  $\langle \bm{x}^2(t) \rangle \sim t^{0.86}$ for $5< t < 240$ minutes, entailing subdiffusive dynamics with slower-than-linear spread. For the fledglings (Fig. \ref{fig1:sub6}) we have similar superdiffusive scaling $\langle  \bm{x}^2(t) \rangle \sim t^{1.37}$ for $t < 5$ minutes, but diffusive scaling $\langle  \bm{x}^2(t) \rangle \sim t^{1.04}$ at $5 < t < 240$ minutes. Importantly, the empirical 5-minute time scale separating the two regimes in Fig. \ref{fig1:sub2} is directly related to the birds' behavior and finite home range; see Sec. \ref{sec:biology} below. 

We repeated the calculations of the moments for $q \in [0,5]$ at 0.5 intervals to measure $\lambda_t(q)$ for both adults and fledglings. In the following we focus on $\lambda_t(q) = \lambda_s(q)$ at short times, $t < 5$ minutes, plotted in Fig. \ref{fig1:sub3} (returning below to $\lambda_t(q) = \lambda_\ell(q)$ at long times, $5 < t < 240$ minutes). The convex behavior of the moment spectrum function in Figs. \ref{fig1:sub3} and \ref{fig1:sub7} is one of the main results of this manuscript, entailing that free-ranging Barn Owls exhibit SAND. 
As we observe in these figures a bi-linear trend, we fit a piecewise linear function to the moment spectrum function \(\lambda_s(q)\), measuring the slopes at low and high moments and the transition point between the two slopes.
For both adults and fledglings the displacement moments can be described by 
\begin{equation} \label{eq:rebenshtok}
\left<|\bm{x}|^q\right>  \sim t^{\lambda_s(q)} \quad , \; \lambda_s(q)/q \sim
\begin{cases} 
s_1, & \text{if } q < q_s, \\
s_2, & \text{if } q \geq q_s , 
\end{cases}
\end{equation}
where $s_1 < s_2$ are the exponents, such that $\lambda_s(q)$ is convex, and $q_s$ is the critical moment of the piecewise scaling. For adults (Fig. \ref{fig1:sub3}) we find  
$s_1 = 0.63 \pm 0.06$ and $s_2=0.98 \pm 0.03$ for low and high moments, respectively, with $q_s = 2.14 \pm 0.18$. A qualitatively similar -- but quantitatively different -- convex moment spectrum function is observed for the fledglings, with $s_1 = 0.48 \pm 0.05$, $s_2 = 0.79 \pm 0.03$ and $q_s = 0.67 \pm 0.06$ (Fig. \ref{fig1:sub7}). Here, the fledglings have lower exponents for both low and high moments with a significantly lower value of $q_s$. The exponent values are summarized in Table \ref{tab:exponents}. 

\begin{table}\centering
\begin{tabular}{lcccccc}
\toprule
 & \multicolumn{3}{c}{\textbf{$t <$ 5 minutes}} & \multicolumn{3}{c}{\textbf{$t >$ 5 minutes}} \\
\cmidrule(lr){2-4} \cmidrule(lr){5-7}
 & \textbf{$s_1$} & \textbf{$s_2$} & \textbf{$q_s$} & \textbf{$\ell_1$} & \textbf{$\ell_2$} & \textbf{$q_\ell$} \\
\midrule
Adults        & 0.63 & 0.98 & 2.14 & 0.50 & 0.07 & 1.57 \\
Fledglings    & 0.48 & 0.79 & 0.67 & 0.74 & 0.2 & 1.19 \\
BLW           & 0.68 & 0.97 & 2.61 & 0.49 & 0.24 & 2.69 \\
multi-mode Adults       & 0.67 & 0.95 & 2.08 & 0.53 & 0.22 & 2.16 \\
multi-mode Fledglings  & 0.64 & 1.18 & 0.81 & 0.85 & 0.39 & 0.72 \\
\bottomrule
\end{tabular} 
\caption{Fit results for $\lambda_s(q)$ and $\lambda_\ell(q)$, for adults, fledglings, Bounded L\'evy Walk (BLW) model with $\alpha = 1.6$ and bounding radius $R = 2$ km, and the multi-mode model parameterized for adults and parameterized for fledglings. The errors for the adults and fledglings are given in the text, and we have checked that all other errors are $< 10\%$. }\label{tab:exponents}
\end{table}

A tempting explanation for the differences between adults and fledglings is that the former display ballistic-like flights while the latter do not fly in a straight path for prolonged periods. However, this interpretation is incorrect. To show this we analyze an ensemble of directed flights (commutes) between distant locations. To obtain these commutes, we used a behavioural change point analysis segmentation method \cite{gurarie2016animal} termed the Penalized Contrast Method~\citep{barraquand2008animal}, see Sec. \ref{sec:segmentation}. The segmentation yields two different ensembles: an ensemble of commutes and an ensemble of area-restricted searches. Following this behavior-based segmentation, we repeat the above analysis for only commutes, resulting in a single scaling of $0.92$ for the adults (Fig. \ref{fig1:sub3}, inset) and of $0.94$ for the fleglings (Fig. \ref{fig1:sub7}, inset), both suggesting near ballistic mono-scaling. Thus, ballistic commutes are similar between adults and fledglings and cannot account for the difference between their different exponents, see further discussion below.   

In contrast to the convex moment spectrum function for $t < 5$ minutes, at long times, $t > 5$ minutes, we find a \textit{concave} moment spectrum function for both adults and fledglings (Figs. \ref{fig1:sub4} and \ref{fig1:sub8}). This transition between a convex $\lambda_s(q)$ and concave $\lambda_\ell(q)$ is another key finding of this manuscript. Here, the moments also follow bi-linear behavior:
\begin{equation} \label{eq:rebenshtok_long}
\left<|\bm{x}|^q\right>  \sim t^{\lambda_\ell(q)} \quad , \; \lambda_\ell(q)/q \sim
\begin{cases} 
\ell_1, & \text{if } q < q_\ell, \\
\ell_2, & \text{if } q \geq q_\ell , 
\end{cases}
\end{equation}
with $\ell_1 > \ell_2$, resulting in a concave $\lambda_\ell(q)$. 
Fitting a piecewise linear function to $\lambda_\ell(q)$  for adults (Fig. \ref{fig1:sub4}) we find at lower moments $\ell_1 = 0.5 \pm 0.04$, transitioning at $q_\ell = 1.57 \pm 0.12$ to $\ell_2 = 0.07 \pm 0.02$. Notably, the low value of $\ell_2$ suggests that the long-time moment function nearly saturates at high moments. Similar transition between a convex and a concave function are observed for the fledglings (Fig. \ref{fig1:sub8}), with elevated values of $\ell_1 = 0.74 \pm 0.04$ and $\ell_2 = 0.2 \pm 0.02$, and a lower value of $q_\ell = 1.19 \pm 0.07$. The exponent values for adults and fledglings are summarized in Table \ref{tab:exponents}. 

Revisiting the short time-scale, $t < 5$ minutes, we note that the adult bi-linear moment spectrum function, $\lambda_s(q)$, is in agreement with a universal pattern previously identified for a broad class of L\'evy walk models \cite{rebenshtok2014non, andersen2000simple, burioni2013rare}. For these  models, previous studies have established that the bi-linear moment spectrum function follows 
\cite{andersen2000simple}: 
\begin{equation} \label{eq:rebenshtok_scaling}
\lambda_s(q)/q\sim 
\begin{cases} 
1/\xi, & \text{if } q < 2, \\
1, & \text{if } q \geq 2 . 
\end{cases}
\end{equation}
For the adults $s_1 = 1/\xi = 0.63 \pm 0.06$ resulting in \(\xi = 1.59 \pm 0.17\), where the physical meaning of the exponent $\xi$ is related to the broad tail of the flight time distribution, see details in Sec. \ref{sec:simulations} below. 
However \eqref{eq:rebenshtok_scaling} does not account for the results of the fledglings, see Table \ref{tab:exponents}, or for the concave $\lambda_\ell(q)$ at $t > 5$ minutes. These findings motivate the bounded  L\'evy walk model developed in Sec. \ref{subsec:BLW}. 

As discussed in Sec. \ref{sec:introduction}, the bi-linear dependence of the moment spectrum function \(\lambda_s(q)\) entails that the PDF does not adhere to mono-scaling in a self-similar fashion, which would imply a linear dependence of \(\lambda_s(q)\) with \(q\) \cite{andersen2000simple}. As the empirical data exhibit a bi-linear form of \(\lambda_s(q)\), we expect two distinct scaling regimes, a rescaled (self-similar) PDF $t^{1/\xi}P(|\bm{x}|)$ versus $|\bm{x}|/t^{1/\xi}$ for smaller displacements (bulk) and a rescaled PDF $t^{\xi}P(|\bm{x}|)$ versus $|\bm{x}|/t$  for larger displacements (the tails of the PDF) \cite{andersen2000simple}. Both rescaled regimes directly follow from \eqref{eq:rebenshtok_scaling}, whereas the latter indicates a ballisitic scaling. As such, this bi-scaling has important statistical consequences for the asymptotic behavior of the underlying probability densities. 

\begin{figure*}[t]
\centering
\hspace{-3.5mm}
 \includegraphics[width=.7\textwidth]{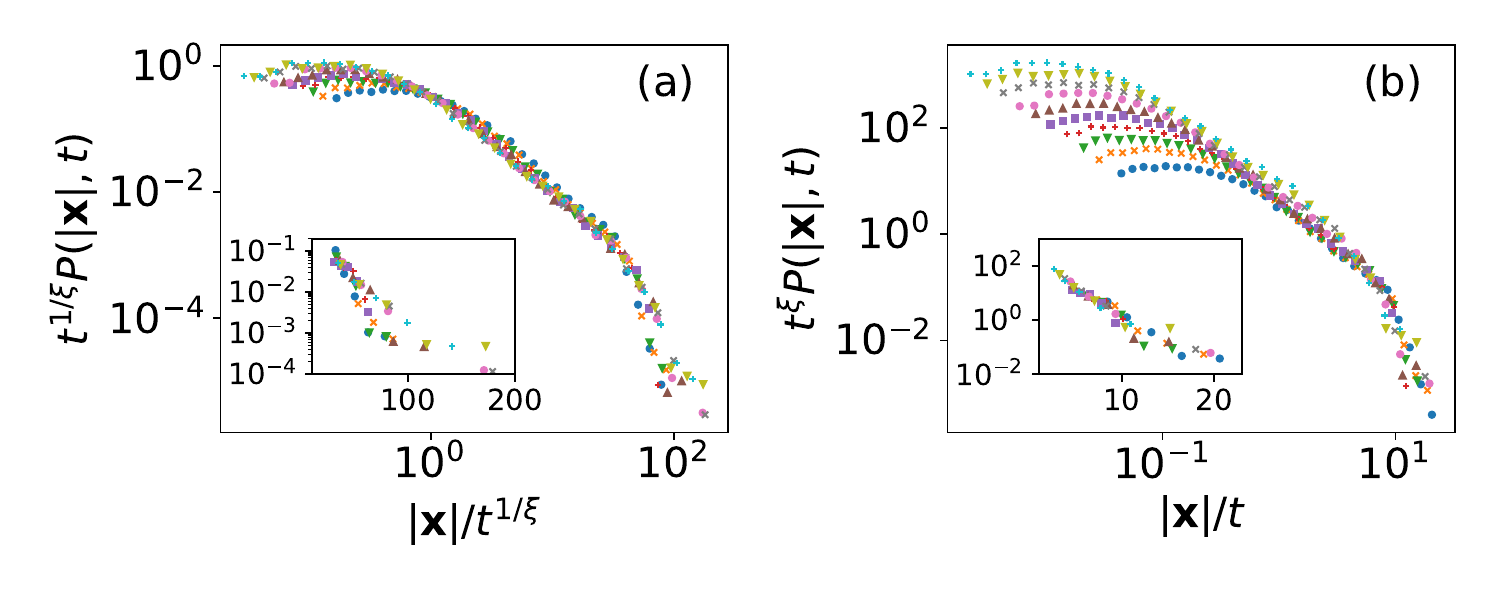}
  \label{fig2:sub1}
\vspace{-2mm}
\caption{Normalized PDFs for the displacement at different $t$, where different symbols correspond to different time windows -- ranging between 32 and 600 seconds (lower blue circles to upper green pluses, respectively). Panel (a) displays the PDF of the normalized variable \( |\bm{x}|/t^{1/\xi} \), while panel (b) shows the PDF of the ballistic scaling \( |\bm{x}|/t \). In the insets, we plot a normalized histogram of the top $2\%$ of the distances with x-axis in linear scale, where the axis labels of the main panels apply also to the insets. These two panels exhibit bi-scaling, the scaling in (a) for the bulk which corresponds to enhanced diffusion with $\xi = 1.59$, see Eq. \eqref{eq:rebenshtok_scaling}, and the scaling in (b) for the tail which displays ballistic scaling.}  
\label{fig1_2}
\end{figure*}

Analysis of the PDF of the displacements of the adult Barn Owls shows good agreement with these predictions. The normalized PDFs for different times $t$ are shown in Figs. \ref{fig1_2}a and \ref{fig1_2}b. In panel \ref{fig1_2}a we plot PDFs of the normalized variable \( |\bm{x}|/t^{1/\xi} \) for different times $t$, showing that the data at different times in the bulk of the PDF collapse onto universal curves. Here, $\xi=1.59$ is not directly fitted for the collapse, but rather obtained from analysis of the moments, see \eqref{eq:rebenshtok_scaling}. At the tails of the distribution, the scaling  \( |\bm{x}|/t^{1/\xi} \) is no longer observed. In panel \ref{fig1_2}b we plot PDFs of the normalized variable \( |\bm{x}|/t \), showing a collapse at the tails of the distribution. In the insets of both panels we plot a histogram of the upper $2\%$ of the displacement for each time $t$, highlighting that only the ballistic-like scaling \( |\bm{x}|/t \) in Fig. \ref{fig1_2}b, leads to a collapse of the tail at large displacements. Thus, we find that bi-scaling with ballistic motion for the tails of the propagator and  \( |\bm{x}|/t^{1/\xi} \) for the central part, are applicable to describe the dynamics.

As mentioned above, while SAND with a bi-linear convex $\lambda_s(q)$ has been observed in several systems, to the best of our knowledge, it has not been reported in ecological systems. Moreover, existing models of SAND do not accommodate the observed transition to a concave \(\lambda_\ell(q)\) at long timescales. Consequently, there is a need to both account for the underlying mechanisms responsible for such behavior and to develop an ecologically oriented model to understand the observed phenomena. Below, we employ two stochastic models that aim to meet these goals, enhancing our theoretical understanding of SAND and its application to ecological data.

\section{Simulations} \label{sec:simulations}

\subsection{Bounded Lévy walk} \label{subsec:BLW}
The first suggested model is \textit{bounded} Lévy walk (BLW), serving as a ``zero-order approach'', i.e., an effective model with a low-parameter phase space. The BLW is constructed to account for the finite home range of the animal and is based on the well-known Lévy walk model. Similarly to Lévy walk, the BLW is characterized by a power-law distribution of jump durations -- the times it takes the walker to move to a new location -- according to \(\psi(\tau) = 1/\tau^{1 + \alpha}\) for $\tau>1 $ and a constant velocity \(v\)~\cite{zaburdaev2015levy}. We define BLW by constraining movement within a fixed radius $R$ from the initial position; any sampled jump exceeding this boundary is resampled until a valid one is obtained (see Appendix \ref{appendix_LW}), such that $R$ serves to define a finite home range.  As it has been shown for unbounded Lévy walk that $\xi = \alpha$ \cite{rebenshtok2014non, andersen2000simple}, we use $\xi=1.59 \pm 0.17$ above (see below Eq. \eqref{eq:rebenshtok_scaling}) and set $\alpha = 1.6$. We have also checked that direct flights of the adult Barn Owls follow power law distributed $\psi(\tau)$ (Fig. S2 \cite{SIref}), and note that in the BLW model, due to the exclusion of steps beyond the predefined radius, the jump-time distribution is cutoff. In our simulations we set $v = 8$ m/s, in agreement with empirical data (Appendix \ref{appendix_LW}). In Fig. \ref{fig2:sub1} we plot an example of 4-hour long BLW for \(\alpha = 1.6\) and \(R = \mathcal{R}_0 = 2.5\) km, where the latter is approximately the birds' average maximum displacement, see Sec. \ref{sec:results}.
We stress that the BLW model has two parameters, $\alpha$ and $R$, which are chosen to match $\xi = 1.59 \pm 0.17$ and $\mathcal{R}_0 = 2.5$ km, respectively.

\begin{figure*}[t]
\centering
\subfloat[]{\hspace{-4mm}
  \centering
 \includegraphics[width=.26\linewidth]{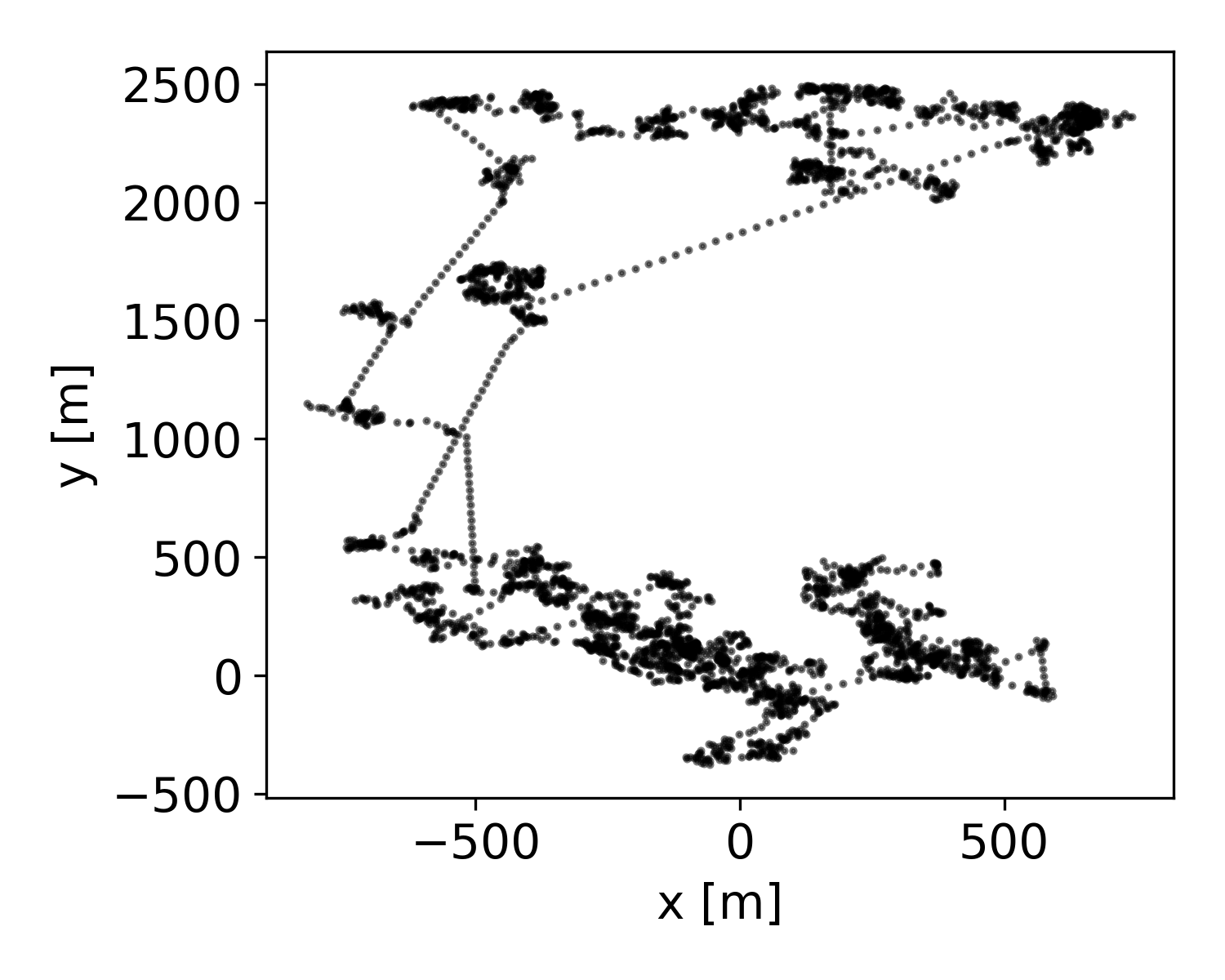}
  \label{fig2:sub1}
}
\subfloat[]{\hspace{-4mm}
  \centering
 \includegraphics[width=.26\textwidth]{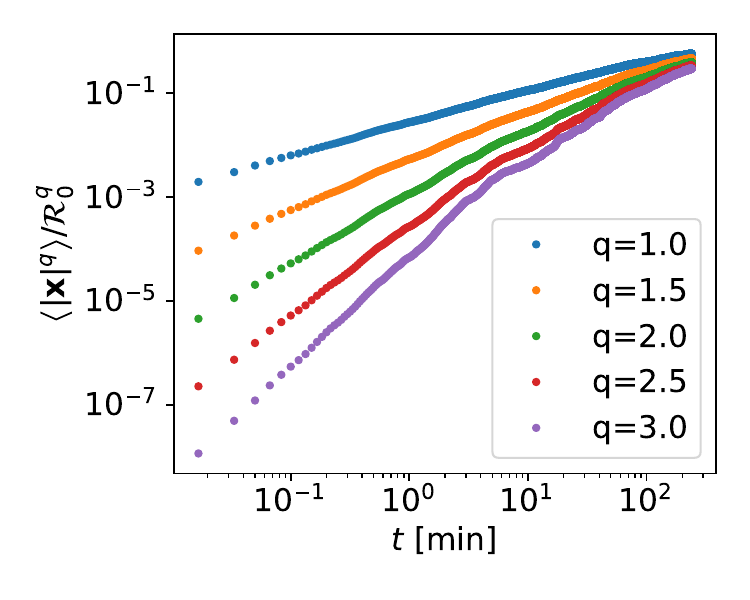}
  \label{fig2:sub2}
}
\subfloat[]{\hspace{-4mm}
  \centering
 \includegraphics[width=.26\textwidth]{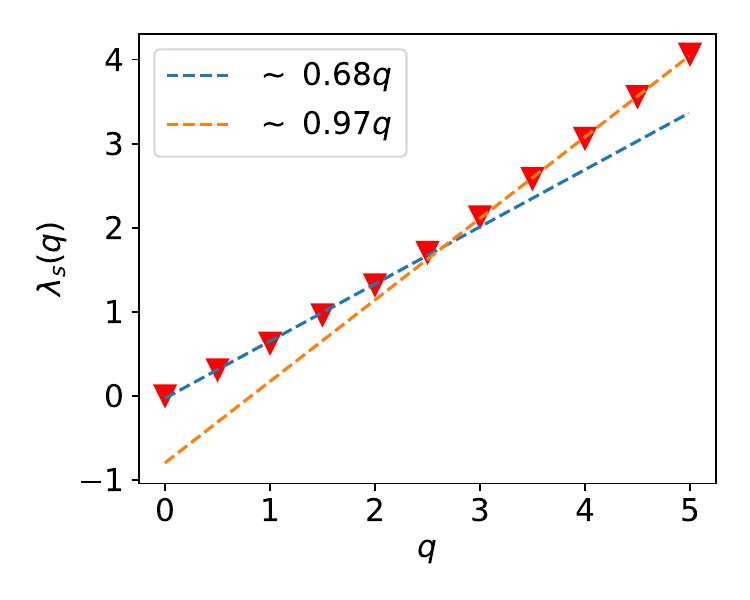}
  \label{fig2:sub3}
} 
\subfloat[]{\hspace{-4mm}
  \centering
 \includegraphics[width=.26\textwidth]{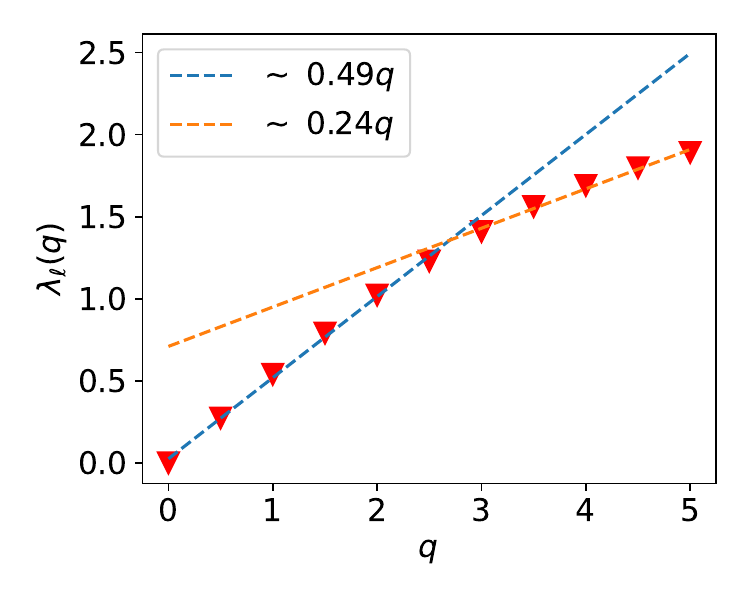}
  \label{fig2:sub6}
} \vspace{-1mm}
\centering
\caption{Analysis of BLW simulations with parameters \(\alpha = 1.6\) and \(R = 2.5\) gives results similar to empirical data. (a) 4-hour simulated tracks, sampled every 4 seconds. (b) Log-log plot of the average absolute displacement moments \(\langle |\bm{x}|^q \rangle / \mathcal{R}_0^q\) with $\mathcal{R}_0 = 2.5$ km representing the average maximum displacement in a single night. Data points for moment orders $q=1, 1.5, 2, 2.5, 3$ are shown, displaying two distinct scaling regimes for each moment $q$, with a transition at $\sim 5$ minutes. (c) \(\lambda_s(q)\) as a function of \(q\) for $t < 5$ minutes (red triangles). The dashed blue and orange lines represent the bi-linear fit slopes for low and high moments, respectively, where the slopes are denoted in the legends. 
(d) \(\lambda_\ell(q)\) as a function of \(q\) for $t > 5$ minutes (red triangles). The dashed blue and orange lines represent the bi-linear fit slopes for low and high moments, respectively.  }
\label{fig2}
\end{figure*}

We next repeat the analysis of the empirical data for the BLW simulations for \(\alpha = 1.6\) and \(R = 2.5\) km, and the results are shown in Fig. \ref{fig2} and summarized in Table \ref{tab:exponents}. The BLW shares several key similarities with the empirical data (Fig. \ref{fig1}), including: (a) a shift in the moments at a characteristic timescale of $\sim$5 minutes (Fig. \ref{fig2:sub2}); (b) convex moment spectrum function for $t < 5$ minutes quantitatively similar to the empirical results for the adult Barn Owls (Fig. \ref{fig2:sub3}, compare Fig. \ref{fig1:sub3}); 
and (c) a transition between a convex and a concave moment spectrum function (Fig. \ref{fig2:sub6}). 
Notably, for \(t < 5\) minutes, the BLW model closely aligns with adult Barn Owls in terms of \(s_1\), \(s_2\), and \(q_s\). For \(t > 5\) minutes, \(\ell_1\) is similar and both \(\ell_2\) and \(q_\ell\) are larger in the BLW model, compared to adults; however, the model still captures a concave \(\lambda_\ell(q)\).
When comparing fledglings to the BLW model, we find more discrepancies. For \(t < 5\) minutes, fledgling values for \(s_1\), \(s_2\), and \(q_s\) are lower, indicating an overestimation by the BLW model. For \(t > 5\) minutes, we find a small difference in \(\ell_1\) and a larger difference in \(q_\ell\).

In addition to results for specific values of $\alpha$ and $R$, we have checked that for $1<\alpha<2$, variations in \(R\) -- but not in $\alpha$ -- modify the form of the long-time $\lambda_\ell(q)$. Similarly, variations in $R$ affect the typical timescale for transitioning from convex to concave (Appendix \ref{appendix_LW}). 
Thus, the results of the BLW model with \(R = \mathcal{R}_0 = 2.5\) km, suggests that the observed phenomena are a mark of a finite home range, whose scale is given by the maximum displacement scale $\mathcal{R}_0 = 2.5$ km, see further discussion below. 
Finally, we note that for the BLW, $q_s> 2$ for all values of $R$ and $\alpha$ that we checked, in contrast to $q_s$ for the fledglings. This suggests that the BLW model does not as effectively represent fledgling movement compared to adults, although it still successfully models the transition from a convex \(\lambda_s(q)\) to a concave \(\lambda_\ell(q)\). In the following, we employ a more ecologically oriented model to better contextualize these phenomena in an ecological setting.  

\subsection{Multi-mode biased random-walk} \label{sec:OU}
\begin{figure*}[t]
\centering
\subfloat[]{\hspace{-4mm}
  \centering
 \includegraphics[width=.26\linewidth]{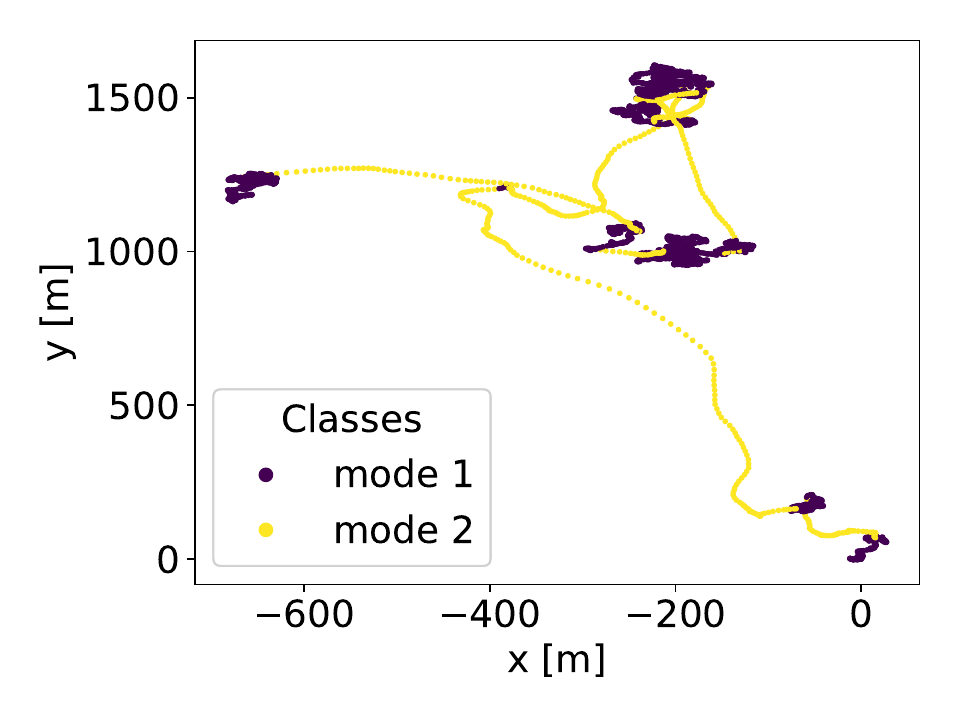}
  \label{fig3:sub1}
}
\subfloat[]{\hspace{-4mm}
  \centering
 \includegraphics[width=.26\textwidth]{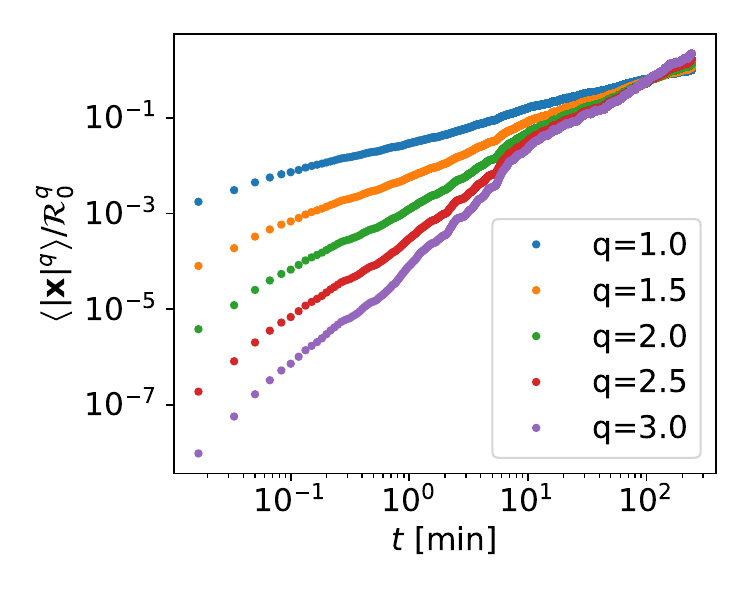}
  \label{fig3:sub2}
}
\subfloat[]{\hspace{-4mm}
  \centering
 \includegraphics[width=.26\textwidth]{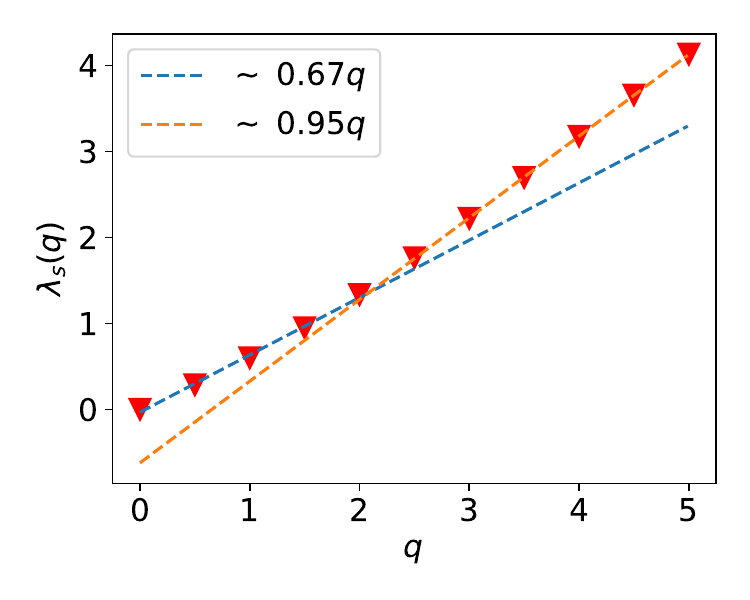}
  \label{fig3:sub3}
}
\subfloat[]{\hspace{-4mm}
  \centering
 \includegraphics[width=.26\textwidth]{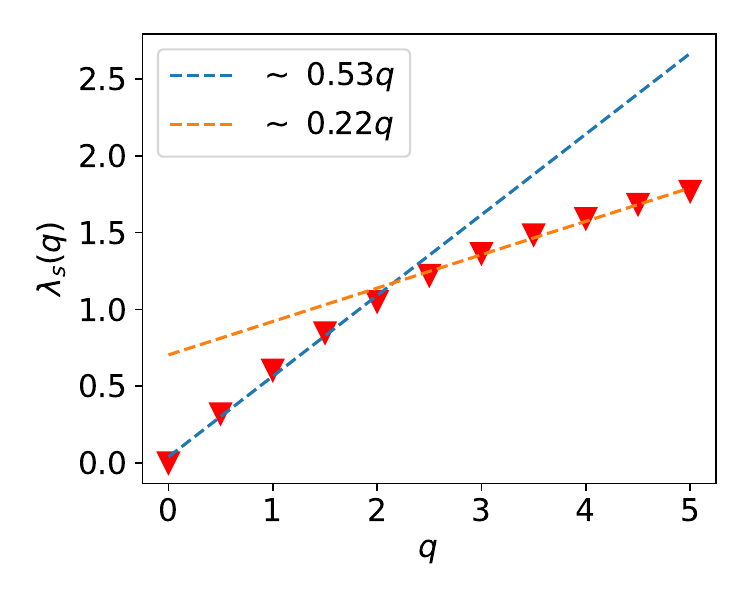}
  \label{fig3:sub6}
} \vspace{-1mm}
\centering
\caption{Analysis of Multi-mode simulations shows results similar to empirical data. (a) 4-hour simulated tracks. (b) Log-log plot of the average absolute displacement moments \(\langle |\bm{x}|^q \rangle / \mathcal{R}_0^q\) with $\mathcal{R}_0 = 2.5$ km representing the average maximum displacement of Barn Owls in a single night. Data points for moment orders $q=1, 1.5, 2, 2.5, 3$ are shown, displaying two distinct scaling regimes for each moment $q$, with a transition at $\sim 5$ minutes. (c) \(\lambda_s(q)\) as a function of \(q\). The red triangles represent the values over short time intervals of the simulations. The dashed blue and orange lines represent the bi-linear fit slopes for low and high moments, respectively, see legends. 
(f) \(\lambda_\ell(q)\) as a function of \(q\) for long time intervals ($> 5$ minutes, red triangles). The dashed blue and orange lines represent the bi-linear fit slopes for low and high moments, respectively.  }
\label{fig3}
\end{figure*}

We implement a multi-mode two-dimensional biased random-walk process for animal movement, similar to processes commonly used in the ecological literature \cite{blackwell1997random, hanks2011velocity, hooten2017animal, gurarie2017framework,  breed2017predicting, eisaguirre2021multistate}. Our model consists of a random walker that transitions between discrete behavioral states, where each state corresponds to a different mode of stochastic evolution \cite{jonsen2005robust, mcclintock2012general, hooten2016hierarchical}. Each behavioral state within the model is fully characterized by a correlation time, \(\tau_i\), and a mean velocity, \(\nu_i\), where the subscript \(i\) denotes different behavioral modes (e.g., feeding, resting, or evading predators). The transitions between these behavioral states are governed by a Markovian transition matrix (\(M\)), dictating the likelihood of an animal shifting from one behavior to another. Further details of the simulations are given in Appendix \ref{appendix_OU}, and in what follows we limit ourselves to two modes: commutes and local area-restricted searches. An example of a 4-hour multi-mode simulation, with two distinct modes is given in Fig. \ref{fig3:sub1}.
We note that here a bounding radius is not directly incorporated, although the transitions between modes has been reported to act as an inhibitor of large displacements \cite{eisaguirre2021multistate}.

Although the model is parameter-rich, estimation of the parameters can be guided by empirical data. 
Here we estimate the parameters by segmentation of the data into two modes: searches and commutes (see Sec. \ref{sec:segmentation} and Ref. \cite{gurarie2017framework}). Following the segmentation, we can derive a correlation time for each mode by examining the autocorrelation function of that mode \cite{dray2010exploratory}; the mean velocity is directly calculable from the displacements between adjacent relocation's; and the transition matrix \(M\) is inferred from the observed transitions between modes. 
This approach yields a direct estimation of the parameters, providing a basis to explore the model's behavior in relation to the empirical findings. The resulting parameters are summarized in Table \ref{table:OUsim}, and see Appendix \ref{appendix_OU} for more details. 

\begin{table}[t]
\centering
\begin{tabular}{lcccc}
\toprule
\textbf{Parameter} & \textbf{Adults} & \textbf{Fledglings} \\
\midrule
$\tau_1$ [s] & 2 & 1 \\
$\tau_2$ [s] & 20 & 45 \\
$\nu_1$ [m/s] & 5 & 2 \\
$\nu_2$ [m/s] & 9 & 8 \\
$M$ [\%] & $\begin{bmatrix} 99.85 & 0.15 \\ 0.62 & 99.38 \end{bmatrix}$ & $\begin{bmatrix} 99.93 & 0.07 \\ 0.29 & 99.71 \end{bmatrix}$ \\
\bottomrule
\end{tabular}
\centering
\caption{Simulation Parameters for multi-mode OU simulation based on segmentation of the datasets. $\tau_i$ are the correlation times, $v_i$ are the velocities, and $M$ is the probability matrix to transition between states.} \label{table:OUsim}
\end{table}

In Fig. \ref{fig3} and Table \ref{tab:exponents} we show the results for simulations with parameters inferred from the empirical tracks of the adults, applying the same analysis as performed for the empirical data above.  Similarly to the BLW model, the multi-mode model reproduces (a) the shift in the moments at a characteristic timescale of $\sim 5$ minutes (Fig. \ref{fig3:sub2}), (b) convex $\lambda_s(q)$, quantitatively similar to the empirical results for the adults (Fig. \ref{fig3:sub3}), and (c) 
the transition from a convex to a concave function (Fig. \ref{fig3:sub6}). We have also done a similar estimation for the fledglings (Table \ref{table:OUsim} and Fig. S3 \cite{SIref}) and the resulting exponents are given in Table \ref{tab:exponents}. As shown in Table \ref{tab:exponents}, for adults, the comparison between model and data shows strong alignment across most parameters, particularly for \(t < 5\) minutes, with \(s_1\), \(s_2\), and \(q_s\) closely matching. Over longer timescales (\(t > 5\) minutes), there are  variations in \(\ell_2\) and \(q_\ell\). For fledglings, the model also mostly aligns with the empirical data where the most notable variations are of elevated \(s_2\) and $\ell_2$ values. 

Importantly, some conclusions can be drawn from the multi-mode model parameterization of adult and fledgling data. In our parameterization (Table \ref{table:OUsim}) we have observed that fledglings exhibit shorter correlation times during searches (\(\tau_1 = 1\) vs \(\tau_1 = 2\) seconds), longer correlation times during commutes (\(\tau_2 = 45\) vs \(\tau_2 = 20\) seconds), and tend to remain in a specific mode --search or commute -- for slightly longer periods. In addition, fledglings demonstrate lower velocities in both modes (\(v_1 = 2\) and \(v_2 = 8\) vs \(v_1 = 5\) and \(v_2 = 9\) m/s). Taken together, these factors indicate a higher degree of separation between modes in fledglings compared to adults; that is, searches and commutes are more distinct from each other in fledglings than in adults. Based on this result we suggest that the lower crossover points for fledglings ($q_s$ and $q_\ell$), which are accurately modeled in the multi-mode model, can indicate a greater separation between modes.

\section{Ecological Significance } \label{sec:biology}

As shown above, observing SAND in adults and fledglings uncovered age-specific scaling regimes. At the same time, our analysis also highlighted commonalities: bi-linear scaling of the moment spectrum function and a characteristic temporal scale of approximately 5 minutes for both adults and fledglings. This shared timescale illustrates that while the structure and flexibility of movement modes differ ontogenetically, both age groups are constrained by a common temporal scale.

To understand the significance of this 5-minute timescale, we note that the commuting velocity for these birds is between $8$ and $9$ m/s (Table \ref{table:OUsim}) and the maximum displacement per night is $\sim 2.5$ km as detailed above. Factoring both the maximum displacement and the velocity, a bird flying at $8-9$ m/s will cover 2.5 km at approximately 5 minutes. Based on this observation and the models detailed above we offer two interpretation of this timescale. The first is a coupling between the individual and the environment as supported by the BLW model: due to the finite home range of the bird, straight long-range flights are limited to $\sim 5$ minutes, above which interactions with the edges of the home range become important and inhibit the increase of the displacement. This explanation is supported by direct analysis of the home range of the birds, 
where in Appendix \ref{appendix_HR} and Fig. S4 \cite{SIref} we show that indeed for the adult Barn Owls, the home range during breeding is $2-3$ km wide. 
On the other hand, we have found that -- on average -- the fledglings display significantly larger home ranges than the adults (Fig. S4 \cite{SIref}). 

Another closely related explanation for the 5-minute timescale is based on the multi-mode correlated-velocity model, and does not require incorporating explicit boundaries. In the multi-mode model, mechanical or preferential behavior of the bird limits its commuting behavior, resulting in SAND at short times and successfully producing a transition from convex $\lambda_s(q)$ to concave $\lambda_\ell(q)$ without an explicit home range. Notably, the multi-mode model can result in a relatively short (average) daily maximum displacement but would allow for larger excursions, as found for the fledglings. Here, an important finding is that the straightforward parameterization of the model -- without detailed fine-tuning -- accurately reflects our qualitative results as well as the quantitative differences between adults and fledglings, suggesting that our two-state multi-mode model captures the dynamics to a significant degree.

\subsection{Differences between age groups}

The differences in movement parameters between fledglings and adults -- namely, shorter correlation times during localized search ($\tau_1$), longer correlation times during commutes ($\tau_2$), and lower average velocities in fledglings compared to adults ($\nu_1$ and $\nu_2$) -- are consistent with documented patterns of age-specific foraging proficiency in birds. Juveniles of several species exhibit lower capture efficiency, less selective habitat use, and reduced foraging success due to inexperience, underdeveloped motor coordination, and limited spatial memory \citep{wunderle1991age,cristol2017age,martins2024age,grecian2018understanding,franks2018older}. The more sharply delineated behavioral modes observed in fledglings (stronger separation in $\tau_1$, $\tau_2$, $v_1$, and $v_2$) may reflect a compensatory strategy: discrete, simplified movement regimes may help manage energetic costs and reduce cognitive demands while foraging proficiency is still developing. This interpretation is supported by studies noting that juveniles often engage in more stereotyped search patterns, rely more heavily on trial-and-error or local enhancement, and are more vulnerable to suboptimal foraging decisions (e.g., overexploitation of poor patches or inefficient prey handling) \citep{wunderle1991age,martins2024age,franks2018older}. It is also consistent with the larger home ranges observed for fledglings in our data and in other species, where adults tend to maintain narrower home ranges, while juveniles continue to explore and sample broader areas for varying reasons~\citep{belthoff1993home, kruger2014differential,grecian2018understanding,franks2018older}.

From an adaptive perspective, the distinct convex $\lambda_s(q)$ and lower crossover point $q_s$ observed in fledglings could reflect limited flexibility in switching between exploratory and exploitative behaviors, possibly due to a need for extended learning or increased sensitivity to predation risk. Adults, by contrast, exhibit smoother transitions between movement modes, higher velocities, and more integrated search-commute patterns, likely facilitated by experience or by spatial knowledge of resource distributions. These differences may therefore reflect a maturation process that enhances both spatial and behavioral efficiency, ultimately conferring a fitness advantage through improved prey detection, reduced energy expenditure, and increased reproductive provisioning success \cite{wunderle1991age}. Our analysis and simulations provide quantitative support for these patterns; however, we note that the biological interpretations offered here are based primarily on parallels with findings from other species. Our results thus provide a biologically grounded framework for generating testable hypotheses about how developmental constraints shape foraging dynamics.

\subsection{Stochastic modeling} \label{sec:stochastic_model_discussion}

Further relating the two stochastic models to the data, we suggest the following explanation for the emergence of SAND: in a heterogeneous landscape, an animal may alternate between an extensive commuting mode -- search for resource-rich patches -- and an intensive, area-concentrated searching mode -- search for prey within a local patch \cite{benhamou2014scales}. Indeed, a major effort in the literature is to distinguish between local area-restricted searches and commutes \cite{paiva2010area, dorfman2022guide}. Thus, an analysis of unsegmented trajectories results in a mixture of directed flights and long stationary periods, which might resemble BLW dynamics \cite{zaburdaev2015levy, benhamou2014scales}. As discussed above, this mixture occurs at short times scales, up to $\sim 5$ minutes, either due to interactions with the environment -- at long time-scales commutes are limited by the animal's home-range -- or due to properties of the behavioral states such as energy-consumption or food-availability. The BLW model thus acts as a simple model for SAND, simplifying the more ecologically inclined multi-mode dynamics. We further stress that while the BLW does not aim to replicate specific behavioral processes, it captures the key statistical features of strong anomalous diffusion and serves as a useful reference point for interpreting the more complex, biologically grounded, OU simulations. In this sense, our findings offer another example of how Lévy walk-like movement patterns can emerge from underlying behavioral structure \cite{sims2019optimal, campeau2022evolutionary, alessandretti2020scales}, helping to explain the widespread reports of Lévy walks in ecological systems \cite{campeau2024intermittent}.

Indeed, Lévy walks have been widely discussed in the ecological literature~\cite{viswanathan2011physics, zaburdaev2015levy, viswanathan1996levy, campeau2022evolutionary}, in the human mobility literature \cite{pacheco2019nahua, song2010modelling, brockmann2006scaling, alessandretti2020scales}, and in the context of biological movement, including \textit{Dictyostelium discoideum} amoeba \cite{li2008persistent} and run-and-tumble bacteria \cite{matthaus2011origin}, where movement patterns have been approximated by alternating sequences of relatively straight trajectories and reorientations \citep{zaburdaev2015levy}. While these models are highly simplified representations of complex behavior, they are used to construct workable descriptions of motion in diverse systems \cite{okubo2001diffusion}. They have also contributed to theoretical discussions about the scales of movement, behavioral stationarity, and how to interpret trajectory data across spatial and temporal resolutions \cite{benhamou2014scales, ran2022BigData}. As shown above, one characteristic of Lévy walks -- and specifically our BLW model -- is the emergence of strong anomalous diffusion (SAND), suggesting that SAND may be a broadly relevant phenomenon.

Finally, our stochastic models highlight an important feature of movement analysis: the sensitivity of displacement statistics to temporal scale. The observed crossover around five minutes in $\lambda_t(q)$ reflects ecological timescales intrinsic to the system and is consistent across our BLW and multi-mode OU process. Crucially, the agreement between simulations and data arises from parameterizations that align these timescales, such as the bounding radius $R$ in the BLW model and the transition matrix $\mathbf{M}$ in the OU model. Modifying any of the model parameters may shift the crossover point between temporal regimes. Although a full exploration of this sensitivity is beyond the scope of the present study, future work should examine how high moments in multi-mode models, including the OU process and other run-and-tumble dynamics \cite{watkins2013evaluating, martens2012probability, cates2012diffusive}, respond to variation in temporal and spatial scales. 

\section{Discussion} \label{sec:discussion}

Analysis of a large ensembles of 4-hour movement tracks of adult and fledgling Barn Owls (over 10,000 4-hour tracks) revealed that the dynamics are described by SAND, indicating that mono-scaling theories and analysis based on the MSD and mean displacement are insufficient to characterize the dynamics. Indeed, our first main result is that the moment spectrum function at times $t < 5$ minutes follows SAND and is bi-linear, with a transition at $q_s \simeq 2.18$ and $q_s\simeq 0.67$ for adults and fledglings, respectively, and with only adults showing ballistic scaling at high moments. Here, the MSD only reflects the tail of the distribution for fledglings and sits at the transition between two scaling regimes for adults, thus failing to accurately represent either the diffusive or ballistic regimes. This finding underscores the necessity of extending beyond the MSD, a discrete value in a continuous moment spectrum, when studying organism movement.

Another main result is the unexpected transition from convex to concave behavior of the moment spectrum function, pointing to a typical time-scale of $\sim 5$ minutes and a closely related spatial home-range scale. Importantly, this transition was found for both adults and fledglings and was not previously observed in mathematical models. Here as well, our qualitative and quantitative results can be replicated using one of two stochastic models, both capturing key features of the empirical data.

This is the first time bi-linear scaling and a transition from convex to concave scaling, have been observed in an ecological system, where most previous works have found SAND only in mathematical models. We therefore developed two stochastic models that largely reproduce these features, although the different crossover points ($q_s$ and $q_\ell$) in adults and fledglings are reproduced only by the multi-mode model, suggesting that this model can be used to better account for the observation. In addition, based on the parameterization of data to the multi-mode model, we found a higher degree of separation between modes in fledglings, compared to adults, such that a lower crossover point indicates greater inter-mode separation. Notably, our models do not fully reproduce the fledglings' ballistic scaling (exponent $s_2=0.79$), and the near-saturation of moment spectrum function at long times (exponent $\ell_2=0.07$ for the adults). Nevertheless, we stress that the parameters of both the BLW and OU models were chosen independently -- based on short-time scaling (BLW) or behavioral segmentation (OU) -- and not optimized to fit the full empirical profile. The fact that these models reproduce key features of the data, including the crossover behavior and partial agreement with long-time scaling, underscores the interpretive value of our framework.


Importantly, many valuable approaches have been developed to analyze movement data, including step-length and turning-angle statistics, first-passage times, fractal dimensions, Hidden Markov Models, and composite or multi-scale movement models \cite{hooten2017animal, mendez2016stochastic, postlethwaite2013new, fleming2014fine, glennie2023hidden, langrock2012flexible,patterson2009classifying,fauchald2003using,johnson2008continuous,Pa08,Mo04}. While SAND highlights some limitations of MSD-based analyses, it does not replace these existing methods. Rather, it offers a generalization of the MSD framework that captures moment-dependent and scale-dependent behavior. In parallel, we do not aim to cover the full range of ecological models; instead, we focus on two representative simulation classes -- BLW and multi-mode correlated velocity models -- to illustrate how SAND can extract structural features across diverse regimes.

Finally, our findings highlight the significance of computing a continuous range of displacement moments when modeling animal movement. While MSD effectively captures behavior of mono-scaling processes, for behavior modeled by general processes it may only capture the bulk of the displacement distribution, its tail, or can reside at the crossover between the two regimes. Thus, by integrating the entire spectrum of moments, we can more accurately test hypotheses in alignment with movement data, and develop stochastic simulations to predict movement patterns and gain deeper ecological insights. Moreover, the analysis of higher moments revealed distinctions between adults and fledglings directly relatable to their multi-mode and home-range behavior. As the key features observed above can be effectively modeled using two different models, an important finding of this manuscript is the connection between empirical data, the BLW model, and an ecologically oriented multi-mode model. The fact that both these general models reproduced empirical SAND suggests the prevalence of SAND across animal movement behaviors, extending beyond the movement of Barn Owls discussed here, and offering a new perspective on understanding ecological dynamics with multi-mode behavioral strategies.


\section{Acknowledgments}
For fieldwork and technical assistance we thank Y. Bartan, A. Levi, S. Margalit, R. Shaish, G. Rozman and other members of the Movement Ecology Lab and the Minerva Center for Movement Ecology. ATLAS development, maintenance, and studies have been supported by the Minerva Center for Movement Ecology, the Minerva Foundation, and ISF grants 965/15 and 1919/19 to R.N and S.T.. R.N. also acknowledges support from Adelina and Massimo Della Pergola Chair of Life Sciences. E.B. acknowledges support of Israel Science Foundation's grant 1614/21.

\section{SUPPORTING INFORMATION}
Additional Supporting Information may be downloaded via the online version of this article.

\appendix

\section{Bounded Lévy Walk} \label{appendix_LW}
We implemented a two-dimensional bounded Lévy walk (BLW) simulation, constrained within a circular boundary of radius $R$. In our simulation, the time it takes the random walker to perform a step follows a power-law distribution, given by:
\[ \psi(\tau) \sim \frac{1}{\tau^{1 + \alpha}},  \]
for \(\tau \geq \tau_{\text{min}}\), where \(\psi(\tau)\) is the probability of walking for duration \(\tau\), and \(\tau_{\text{min}}\) is the minimum duration which we take to be $\tau_{\text{min}} = 1$ s in our simulations. During each walk, the walker has a constant velocity of \(v = 8\) m/s and the direction is determined randomly, ensuring an isotropic pattern of movement across the simulations. 
During each iteration of the simulation, a walk time is drawn and, together with the walker's velocity, used to calculate the distance of the next step. Here, if the new position exceeds the boundary defined by $R$, the step is recalculated until a valid position within the confines is found. 
We note that the value $\tau_{\text{min}} = 1$ s establishes the temporal scale for the simulations and is chosen arbitrarily, while the value of $\tau_{\text{min}} v = 8$ m sets the spatial scale and is chosen as a reasonable estimate for Barn Owls. We have checked that small variations in $v$ do not qualitatively change alter the results shown in the main text. Importantly, in the main text, we focus on variations in $R$, as it has direct physical significance related to the animal's home range, with constant $\tau_{\text{min}}$ and $v$. However, it is possible to define the model using the dimensionless variable $R/(\tau_{\text{min}} v)$, such that lower velocities $v$ would yield comparable results for smaller radii $R$.   

In Fig. S2 \cite{SIref} we present a comparison between the distribution of jump times in the BLW model and the observed data from adult and fledgling Barn Owls. To obtain the distributions for the two datasets, we utilized the segmentation method detailed in the main text and analyzed only the distribution of direct flights, selecting those with straightness index (defined as the ratio of net displacement to travel distance) exceeding 0.9. The distributions are normalized by $\tau_{min}$ as to allow for direct comparison between data and simulations. 

\section{Multi-mode correlated velocity model} \label{appendix_OU}

We implemented a multi-mode two-dimensional process as a more ecologically reflective model for animal movement, see main text. Each behavioral state within the model is characterized by a correlation time, \(\tau_i\), and a mean velocity, \(\nu_i\), where the subscript \(i\) denotes different behavioral modes such as feeding, resting, or evading predators. The transitions between these behavioral states are governed by a Markovian transition matrix (\(M\)), dictating the likelihood of an animal shifting from one behavior to another, thus capturing the stochastic yet structured decision-making processes inherent to animal behavior.

The velocity, \(v\), of the animal is updated in each time step according to its current state. The correlation time, \(\tau_i\), influences the velocity's relaxation time, effectively guiding the animal towards a baseline velocity appropriate for the current behavior, while \(\nu_i\) denotes the mean velocity for that state. Random fluctuations around this mean velocity are introduced through \(\xi\), a normally distributed random variable, reflecting the spontaneous variations in movement. The updating of the velocity at each time step, \(dt\), is governed by the equation:
\[ 
dv = -\frac{1}{\tau_i} \cdot v \cdot dt + \frac{2 \cdot \nu_i}{\sqrt{\pi \cdot \tau_i}} \cdot \xi \cdot dt  \;, \quad\quad v = v + dv ,
\]
and the position \(x\) is then updated using the new velocity:
\[
x = x + v \cdot dt 
\]
In the simulations shown in the main text, although explicit bounded movement is not directly incorporated, the transitions between the modes can act as a natural inhibitor of the animal's movement, preventing it from straying too far. 
However, for situations where an explicit boundary is necessary, one can introduce a bounding force that comes into effect once the animal crosses a certain radius, \( R \). This force acts to steer the animal back towards a point of origin, \( x_0 \), whenever it wanders outside the bounds of \( R \). The implementation of this bounding force in the simulation is as follows:
If the animal's position is outside radius \( R \), a restoring force is applied. This force is directed towards the origin and is proportional to the displacement of the animal from \( x_0 \), analogous to the restoring force in a harmonic potential given by \( (1/2) k (x - x_0)^2 \). The additional change in velocity due to this force, \( dv \), is calculated and added to the velocity as follows:
\[
dv = dv - \frac{k}{R} (x - x_0)  dt 
\]
Incorporating this force ensures that the simulated movements remain within the designated spatial constraints, effectively bounding the animal's movement and providing scenario where geographical limitations are present.


As detailed in the main text, we  parameterized the model for both adults and fledglings. The resulting SAND properties for the adults are shown in Fig 4 of the main text and are discussed therein. For the fledglings we plot the results of the parameterized multi-mode simulations in Fig. S3 \cite{SIref}, see discussion in the main text. 





\section{Area estimation} \label{appendix_HR}

To estimate home ranges for all individuals, we used two methods: Kernel Density Estimation (KDE) \cite{worton1989kernel} and Minimum Convex Polygon (MCP) \cite{bekoff1984simulation, fieberg2012could}. The KDE method applies a smoothing function to generate a continuous probability density surface from the recorded movement tracks, indicating regions of varying usage intensity \cite{worton1989kernel, seaman1996evaluation, fleming2015rigorous}. The MCP method calculates the smallest convex polygon that encompasses all the animal's recorded locations, offering a simple area-based estimate \cite{fieberg2012could}. Examples for KDE and MCP estimations are shown in Fig. S1(a-b) for an adult and a fledgling bird \cite{SIref}. 

The MCP was used to estimate the area differences between adult and fledgling birds, where we chose the MCP method over the Kernel Density Estimation (KDE) method because the MCP method accounts for relatively rare long-range excursions, which are significant in assessing the displacement moments for higher moments and are omitted in the KDE method. 
We assessed the normality of the area distributions for both groups using the Shapiro-Wilk test, which yielded p-values of 0.094 for fledglings and 0.098 for adults, both larger than $0.05$, indicating that the data distributions approximate normality. Further, Levene's test for equality of variances showed no significant differences in variance, with a p-value of 0.448. 
An independent samples t-test was subsequently performed, showing a statistically significant difference in MCP areas, with p-value of 0.002. This significant result points to distinct spatial usage patterns between fledglings and adults. The analysis is visualized in Fig. S1(c) \cite{SIref}, where we present a boxplot of the square root of MCP areas ($\sqrt{A}$) for the two groups. The square root values provide a scale for comparison: for adults, the average square root of the area was 4256 meters, while for fledglings it was significantly higher at 6263 meters. Notably, for the fledglings, some individuals' home ranges' square roots exceeded 10 kilometers.


\bibliography{bibliography}

\begin{thebibliography}{108}%
\makeatletter
\providecommand \@ifxundefined [1]{%
 \@ifx{#1\undefined}
}%
\providecommand \@ifnum [1]{%
 \ifnum #1\expandafter \@firstoftwo
 \else \expandafter \@secondoftwo
 \fi
}%
\providecommand \@ifx [1]{%
 \ifx #1\expandafter \@firstoftwo
 \else \expandafter \@secondoftwo
 \fi
}%
\providecommand \natexlab [1]{#1}%
\providecommand \enquote  [1]{``#1''}%
\providecommand \bibnamefont  [1]{#1}%
\providecommand \bibfnamefont [1]{#1}%
\providecommand \citenamefont [1]{#1}%
\providecommand \href@noop [0]{\@secondoftwo}%
\providecommand \href [0]{\begingroup \@sanitize@url \@href}%
\providecommand \@href[1]{\@@startlink{#1}\@@href}%
\providecommand \@@href[1]{\endgroup#1\@@endlink}%
\providecommand \@sanitize@url [0]{\catcode `\\12\catcode `\$12\catcode `\&12\catcode `\#12\catcode `\^12\catcode `\_12\catcode `\%12\relax}%
\providecommand \@@startlink[1]{}%
\providecommand \@@endlink[0]{}%
\providecommand \url  [0]{\begingroup\@sanitize@url \@url }%
\providecommand \@url [1]{\endgroup\@href {#1}{\urlprefix }}%
\providecommand \urlprefix  [0]{URL }%
\providecommand \Eprint [0]{\href }%
\providecommand \doibase [0]{https://doi.org/}%
\providecommand \selectlanguage [0]{\@gobble}%
\providecommand \bibinfo  [0]{\@secondoftwo}%
\providecommand \bibfield  [0]{\@secondoftwo}%
\providecommand \translation [1]{[#1]}%
\providecommand \BibitemOpen [0]{}%
\providecommand \bibitemStop [0]{}%
\providecommand \bibitemNoStop [0]{.\EOS\space}%
\providecommand \EOS [0]{\spacefactor3000\relax}%
\providecommand \BibitemShut  [1]{\csname bibitem#1\endcsname}%
\let\auto@bib@innerbib\@empty
\bibitem [{\citenamefont {Pearson}(1905)}]{pearson1905problem}%
  \BibitemOpen
  \bibfield  {author} {\bibinfo {author} {\bibfnamefont {K.}~\bibnamefont {Pearson}},\ }\bibfield  {title} {\bibinfo {title} {The problem of the random walk},\ }\href@noop {} {\bibfield  {journal} {\bibinfo  {journal} {Nature}\ }\textbf {\bibinfo {volume} {72}},\ \bibinfo {pages} {342} (\bibinfo {year} {1905})}\BibitemShut {NoStop}%
\bibitem [{\citenamefont {Rayleigh}(1905)}]{rayleigh1905problem}%
  \BibitemOpen
  \bibfield  {author} {\bibinfo {author} {\bibnamefont {Rayleigh}},\ }\bibfield  {title} {\bibinfo {title} {The problem of the random walk},\ }\href@noop {} {\bibfield  {journal} {\bibinfo  {journal} {Nature}\ }\textbf {\bibinfo {volume} {72}},\ \bibinfo {pages} {318} (\bibinfo {year} {1905})}\BibitemShut {NoStop}%
\bibitem [{\citenamefont {Klafter}\ and\ \citenamefont {Sokolov}(2011)}]{klafter2011first}%
  \BibitemOpen
  \bibfield  {author} {\bibinfo {author} {\bibfnamefont {J.}~\bibnamefont {Klafter}}\ and\ \bibinfo {author} {\bibfnamefont {I.~M.}\ \bibnamefont {Sokolov}},\ }\href@noop {} {\emph {\bibinfo {title} {First steps in random walks: from tools to applications}}}\ (\bibinfo  {publisher} {OUP Oxford},\ \bibinfo {year} {2011})\BibitemShut {NoStop}%
\bibitem [{\citenamefont {Sokolov}\ and\ \citenamefont {Klafter}(2005)}]{sokolov2005diffusion}%
  \BibitemOpen
  \bibfield  {author} {\bibinfo {author} {\bibfnamefont {I.~M.}\ \bibnamefont {Sokolov}}\ and\ \bibinfo {author} {\bibfnamefont {J.}~\bibnamefont {Klafter}},\ }\bibfield  {title} {\bibinfo {title} {From diffusion to anomalous diffusion: a century after einstein’s brownian motion},\ }\href@noop {} {\bibfield  {journal} {\bibinfo  {journal} {Chaos}\ }\textbf {\bibinfo {volume} {15}} (\bibinfo {year} {2005})}\BibitemShut {NoStop}%
\bibitem [{\citenamefont {Gonzalez}\ \emph {et~al.}(2008)\citenamefont {Gonzalez}, \citenamefont {Hidalgo},\ and\ \citenamefont {Barabasi}}]{gonzalez2008understanding}%
  \BibitemOpen
  \bibfield  {author} {\bibinfo {author} {\bibfnamefont {M.~C.}\ \bibnamefont {Gonzalez}}, \bibinfo {author} {\bibfnamefont {C.~A.}\ \bibnamefont {Hidalgo}},\ and\ \bibinfo {author} {\bibfnamefont {A.-L.}\ \bibnamefont {Barabasi}},\ }\bibfield  {title} {\bibinfo {title} {Understanding individual human mobility patterns},\ }\href@noop {} {\bibfield  {journal} {\bibinfo  {journal} {nature}\ }\textbf {\bibinfo {volume} {453}},\ \bibinfo {pages} {779} (\bibinfo {year} {2008})}\BibitemShut {NoStop}%
\bibitem [{\citenamefont {Sims}\ \emph {et~al.}(2008)\citenamefont {Sims}, \citenamefont {Southall}, \citenamefont {Humphries}, \citenamefont {Hays}, \citenamefont {Bradshaw}, \citenamefont {Pitchford}, \citenamefont {James}, \citenamefont {Ahmed}, \citenamefont {Brierley}, \citenamefont {Hindell} \emph {et~al.}}]{sims2008scaling}%
  \BibitemOpen
  \bibfield  {author} {\bibinfo {author} {\bibfnamefont {D.~W.}\ \bibnamefont {Sims}}, \bibinfo {author} {\bibfnamefont {E.~J.}\ \bibnamefont {Southall}}, \bibinfo {author} {\bibfnamefont {N.~E.}\ \bibnamefont {Humphries}}, \bibinfo {author} {\bibfnamefont {G.~C.}\ \bibnamefont {Hays}}, \bibinfo {author} {\bibfnamefont {C.~J.}\ \bibnamefont {Bradshaw}}, \bibinfo {author} {\bibfnamefont {J.~W.}\ \bibnamefont {Pitchford}}, \bibinfo {author} {\bibfnamefont {A.}~\bibnamefont {James}}, \bibinfo {author} {\bibfnamefont {M.~Z.}\ \bibnamefont {Ahmed}}, \bibinfo {author} {\bibfnamefont {A.~S.}\ \bibnamefont {Brierley}}, \bibinfo {author} {\bibfnamefont {M.~A.}\ \bibnamefont {Hindell}}, \emph {et~al.},\ }\bibfield  {title} {\bibinfo {title} {Scaling laws of marine predator search behaviour},\ }\href@noop {} {\bibfield  {journal} {\bibinfo  {journal} {Nature}\ }\textbf {\bibinfo {volume} {451}},\ \bibinfo {pages} {1098} (\bibinfo {year} {2008})}\BibitemShut {NoStop}%
\bibitem [{\citenamefont {Metzler}\ and\ \citenamefont {Klafter}(2000)}]{metzler2000random}%
  \BibitemOpen
  \bibfield  {author} {\bibinfo {author} {\bibfnamefont {R.}~\bibnamefont {Metzler}}\ and\ \bibinfo {author} {\bibfnamefont {J.}~\bibnamefont {Klafter}},\ }\bibfield  {title} {\bibinfo {title} {The random walk's guide to anomalous diffusion: a fractional dynamics approach},\ }\href@noop {} {\bibfield  {journal} {\bibinfo  {journal} {Phys. Rep.}\ }\textbf {\bibinfo {volume} {339}},\ \bibinfo {pages} {1} (\bibinfo {year} {2000})}\BibitemShut {NoStop}%
\bibitem [{\citenamefont {Zaburdaev}\ \emph {et~al.}(2015)\citenamefont {Zaburdaev}, \citenamefont {Denisov},\ and\ \citenamefont {Klafter}}]{zaburdaev2015levy}%
  \BibitemOpen
  \bibfield  {author} {\bibinfo {author} {\bibfnamefont {V.}~\bibnamefont {Zaburdaev}}, \bibinfo {author} {\bibfnamefont {S.}~\bibnamefont {Denisov}},\ and\ \bibinfo {author} {\bibfnamefont {J.}~\bibnamefont {Klafter}},\ }\bibfield  {title} {\bibinfo {title} {L{\'e}vy walks},\ }\href@noop {} {\bibfield  {journal} {\bibinfo  {journal} {Rev. Mod. Phys.}\ }\textbf {\bibinfo {volume} {87}},\ \bibinfo {pages} {483} (\bibinfo {year} {2015})}\BibitemShut {NoStop}%
\bibitem [{\citenamefont {Metzler}\ \emph {et~al.}(2014)\citenamefont {Metzler}, \citenamefont {Jeon}, \citenamefont {Cherstvy},\ and\ \citenamefont {Barkai}}]{metzler2014anomalous}%
  \BibitemOpen
  \bibfield  {author} {\bibinfo {author} {\bibfnamefont {R.}~\bibnamefont {Metzler}}, \bibinfo {author} {\bibfnamefont {J.-H.}\ \bibnamefont {Jeon}}, \bibinfo {author} {\bibfnamefont {A.~G.}\ \bibnamefont {Cherstvy}},\ and\ \bibinfo {author} {\bibfnamefont {E.}~\bibnamefont {Barkai}},\ }\bibfield  {title} {\bibinfo {title} {Anomalous diffusion models and their properties: non-stationarity, non-ergodicity, and ageing at the centenary of single particle tracking},\ }\href@noop {} {\bibfield  {journal} {\bibinfo  {journal} {Phys. Chem. Chem. Phys.}\ }\textbf {\bibinfo {volume} {16}},\ \bibinfo {pages} {24128} (\bibinfo {year} {2014})}\BibitemShut {NoStop}%
\bibitem [{\citenamefont {Vilk}\ \emph {et~al.}(2022{\natexlab{a}})\citenamefont {Vilk}, \citenamefont {Aghion}, \citenamefont {Avgar}, \citenamefont {Beta}, \citenamefont {Nagel}, \citenamefont {Sabri}, \citenamefont {Sarfati}, \citenamefont {Schwartz}, \citenamefont {Weiss}, \citenamefont {Krapf} \emph {et~al.}}]{vilk2022unravelling}%
  \BibitemOpen
  \bibfield  {author} {\bibinfo {author} {\bibfnamefont {O.}~\bibnamefont {Vilk}}, \bibinfo {author} {\bibfnamefont {E.}~\bibnamefont {Aghion}}, \bibinfo {author} {\bibfnamefont {T.}~\bibnamefont {Avgar}}, \bibinfo {author} {\bibfnamefont {C.}~\bibnamefont {Beta}}, \bibinfo {author} {\bibfnamefont {O.}~\bibnamefont {Nagel}}, \bibinfo {author} {\bibfnamefont {A.}~\bibnamefont {Sabri}}, \bibinfo {author} {\bibfnamefont {R.}~\bibnamefont {Sarfati}}, \bibinfo {author} {\bibfnamefont {D.~K.}\ \bibnamefont {Schwartz}}, \bibinfo {author} {\bibfnamefont {M.}~\bibnamefont {Weiss}}, \bibinfo {author} {\bibfnamefont {D.}~\bibnamefont {Krapf}}, \emph {et~al.},\ }\bibfield  {title} {\bibinfo {title} {Unravelling the origins of anomalous diffusion: from molecules to migrating storks},\ }\href@noop {} {\bibfield  {journal} {\bibinfo  {journal} {Phys. Rev. Res.}\ }\textbf {\bibinfo {volume} {4}},\ \bibinfo {pages} {033055} (\bibinfo {year} {2022}{\natexlab{a}})}\BibitemShut {NoStop}%
\bibitem [{\citenamefont {Nathan}\ \emph {et~al.}(2022)\citenamefont {Nathan}, \citenamefont {Monk}, \citenamefont {Arlinghaus}, \citenamefont {Adam}, \citenamefont {Al{\'o}s}, \citenamefont {Assaf}, \citenamefont {Baktoft}, \citenamefont {Beardsworth}, \citenamefont {Bertram}, \citenamefont {Bijleveld} \emph {et~al.}}]{ran2022BigData}%
  \BibitemOpen
  \bibfield  {author} {\bibinfo {author} {\bibfnamefont {R.}~\bibnamefont {Nathan}}, \bibinfo {author} {\bibfnamefont {C.~T.}\ \bibnamefont {Monk}}, \bibinfo {author} {\bibfnamefont {R.}~\bibnamefont {Arlinghaus}}, \bibinfo {author} {\bibfnamefont {T.}~\bibnamefont {Adam}}, \bibinfo {author} {\bibfnamefont {J.}~\bibnamefont {Al{\'o}s}}, \bibinfo {author} {\bibfnamefont {M.}~\bibnamefont {Assaf}}, \bibinfo {author} {\bibfnamefont {H.}~\bibnamefont {Baktoft}}, \bibinfo {author} {\bibfnamefont {C.~E.}\ \bibnamefont {Beardsworth}}, \bibinfo {author} {\bibfnamefont {M.~G.}\ \bibnamefont {Bertram}}, \bibinfo {author} {\bibfnamefont {A.~I.}\ \bibnamefont {Bijleveld}}, \emph {et~al.},\ }\bibfield  {title} {\bibinfo {title} {Big-data approaches lead to an increased understanding of the ecology of animal movement},\ }\href@noop {} {\bibfield  {journal} {\bibinfo  {journal} {Science}\ }\textbf {\bibinfo {volume} {375}},\ \bibinfo {pages} {eabg1780} (\bibinfo {year} {2022})}\BibitemShut {NoStop}%
\bibitem [{\citenamefont {Polev}\ \emph {et~al.}(2022)\citenamefont {Polev}, \citenamefont {Kolygina}, \citenamefont {Kandere-Grzybowska},\ and\ \citenamefont {Grzybowski}}]{polev2022large}%
  \BibitemOpen
  \bibfield  {author} {\bibinfo {author} {\bibfnamefont {K.}~\bibnamefont {Polev}}, \bibinfo {author} {\bibfnamefont {D.~V.}\ \bibnamefont {Kolygina}}, \bibinfo {author} {\bibfnamefont {K.}~\bibnamefont {Kandere-Grzybowska}},\ and\ \bibinfo {author} {\bibfnamefont {B.~A.}\ \bibnamefont {Grzybowski}},\ }\bibfield  {title} {\bibinfo {title} {Large-scale, wavelet-based analysis of lysosomal trajectories and co-movements of lysosomes with nanoparticle cargos},\ }\href@noop {} {\bibfield  {journal} {\bibinfo  {journal} {Cells}\ }\textbf {\bibinfo {volume} {11}},\ \bibinfo {pages} {270} (\bibinfo {year} {2022})}\BibitemShut {NoStop}%
\bibitem [{\citenamefont {Okubo}\ \emph {et~al.}(2001)\citenamefont {Okubo}, \citenamefont {Levin} \emph {et~al.}}]{okubo2001diffusion}%
  \BibitemOpen
  \bibfield  {author} {\bibinfo {author} {\bibfnamefont {A.}~\bibnamefont {Okubo}}, \bibinfo {author} {\bibfnamefont {S.~A.}\ \bibnamefont {Levin}}, \emph {et~al.},\ }\href@noop {} {\emph {\bibinfo {title} {Diffusion and ecological problems: modern perspectives}}},\ Vol.~\bibinfo {volume} {14}\ (\bibinfo  {publisher} {Springer},\ \bibinfo {year} {2001})\BibitemShut {NoStop}%
\bibitem [{\citenamefont {Ricciardi}(2013)}]{ricciardi2013diffusion}%
  \BibitemOpen
  \bibfield  {author} {\bibinfo {author} {\bibfnamefont {L.~M.}\ \bibnamefont {Ricciardi}},\ }\href@noop {} {\emph {\bibinfo {title} {Diffusion processes and related topics in biology}}},\ Vol.~\bibinfo {volume} {14}\ (\bibinfo  {publisher} {Springer Science \& Business Media},\ \bibinfo {year} {2013})\BibitemShut {NoStop}%
\bibitem [{\citenamefont {Holmes}\ \emph {et~al.}(1994)\citenamefont {Holmes}, \citenamefont {Lewis}, \citenamefont {Banks},\ and\ \citenamefont {Veit}}]{holmes1994partial}%
  \BibitemOpen
  \bibfield  {author} {\bibinfo {author} {\bibfnamefont {E.~E.}\ \bibnamefont {Holmes}}, \bibinfo {author} {\bibfnamefont {M.~A.}\ \bibnamefont {Lewis}}, \bibinfo {author} {\bibfnamefont {J.}~\bibnamefont {Banks}},\ and\ \bibinfo {author} {\bibfnamefont {R.}~\bibnamefont {Veit}},\ }\bibfield  {title} {\bibinfo {title} {Partial differential equations in ecology: spatial interactions and population dynamics},\ }\href@noop {} {\bibfield  {journal} {\bibinfo  {journal} {Ecology}\ }\textbf {\bibinfo {volume} {75}},\ \bibinfo {pages} {17} (\bibinfo {year} {1994})}\BibitemShut {NoStop}%
\bibitem [{\citenamefont {Hooten}\ \emph {et~al.}(2017)\citenamefont {Hooten}, \citenamefont {Johnson}, \citenamefont {McClintock},\ and\ \citenamefont {Morales}}]{hooten2017animal}%
  \BibitemOpen
  \bibfield  {author} {\bibinfo {author} {\bibfnamefont {M.~B.}\ \bibnamefont {Hooten}}, \bibinfo {author} {\bibfnamefont {D.~S.}\ \bibnamefont {Johnson}}, \bibinfo {author} {\bibfnamefont {B.~T.}\ \bibnamefont {McClintock}},\ and\ \bibinfo {author} {\bibfnamefont {J.~M.}\ \bibnamefont {Morales}},\ }\href@noop {} {\emph {\bibinfo {title} {Animal movement: statistical models for telemetry data}}}\ (\bibinfo  {publisher} {CRC press},\ \bibinfo {year} {2017})\BibitemShut {NoStop}%
\bibitem [{\citenamefont {Hanski}\ and\ \citenamefont {Gilpin}(1991)}]{hanski1991metapopulation}%
  \BibitemOpen
  \bibfield  {author} {\bibinfo {author} {\bibfnamefont {I.}~\bibnamefont {Hanski}}\ and\ \bibinfo {author} {\bibfnamefont {M.}~\bibnamefont {Gilpin}},\ }\bibfield  {title} {\bibinfo {title} {Metapopulation dynamics: brief history and conceptual domain},\ }\href@noop {} {\bibfield  {journal} {\bibinfo  {journal} {Biological journal of the Linnean Society}\ }\textbf {\bibinfo {volume} {42}},\ \bibinfo {pages} {3} (\bibinfo {year} {1991})}\BibitemShut {NoStop}%
\bibitem [{\citenamefont {Turchin}(1998)}]{turchin1998quantitative}%
  \BibitemOpen
  \bibfield  {author} {\bibinfo {author} {\bibfnamefont {P.}~\bibnamefont {Turchin}},\ }\href@noop {} {\emph {\bibinfo {title} {Quantitative analysis of movement: measuring and modeling population redistribution in animals and plants}}}\ (\bibinfo  {publisher} {Sinauer Associates},\ \bibinfo {year} {1998})\BibitemShut {NoStop}%
\bibitem [{\citenamefont {Hooten}\ \emph {et~al.}(2016)\citenamefont {Hooten}, \citenamefont {Buderman}, \citenamefont {Brost}, \citenamefont {Hanks},\ and\ \citenamefont {Ivan}}]{hooten2016hierarchical}%
  \BibitemOpen
  \bibfield  {author} {\bibinfo {author} {\bibfnamefont {M.~B.}\ \bibnamefont {Hooten}}, \bibinfo {author} {\bibfnamefont {F.~E.}\ \bibnamefont {Buderman}}, \bibinfo {author} {\bibfnamefont {B.~M.}\ \bibnamefont {Brost}}, \bibinfo {author} {\bibfnamefont {E.~M.}\ \bibnamefont {Hanks}},\ and\ \bibinfo {author} {\bibfnamefont {J.~S.}\ \bibnamefont {Ivan}},\ }\bibfield  {title} {\bibinfo {title} {Hierarchical animal movement models for population-level inference},\ }\href@noop {} {\bibfield  {journal} {\bibinfo  {journal} {Environmetrics}\ }\textbf {\bibinfo {volume} {27}},\ \bibinfo {pages} {322} (\bibinfo {year} {2016})}\BibitemShut {NoStop}%
\bibitem [{\citenamefont {Trakhtenbrot}\ \emph {et~al.}(2005)\citenamefont {Trakhtenbrot}, \citenamefont {Nathan}, \citenamefont {Perry},\ and\ \citenamefont {Richardson}}]{trakhtenbrot2005importance}%
  \BibitemOpen
  \bibfield  {author} {\bibinfo {author} {\bibfnamefont {A.}~\bibnamefont {Trakhtenbrot}}, \bibinfo {author} {\bibfnamefont {R.}~\bibnamefont {Nathan}}, \bibinfo {author} {\bibfnamefont {G.}~\bibnamefont {Perry}},\ and\ \bibinfo {author} {\bibfnamefont {D.~M.}\ \bibnamefont {Richardson}},\ }\bibfield  {title} {\bibinfo {title} {The importance of long-distance dispersal in biodiversity conservation},\ }\href@noop {} {\bibfield  {journal} {\bibinfo  {journal} {Diversity and Distributions}\ }\textbf {\bibinfo {volume} {11}},\ \bibinfo {pages} {173} (\bibinfo {year} {2005})}\BibitemShut {NoStop}%
\bibitem [{\citenamefont {Nathan}\ \emph {et~al.}(2008)\citenamefont {Nathan}, \citenamefont {Getz}, \citenamefont {Revilla}, \citenamefont {Holyoak}, \citenamefont {Kadmon}, \citenamefont {Saltz},\ and\ \citenamefont {Smouse}}]{nathan2008movement}%
  \BibitemOpen
  \bibfield  {author} {\bibinfo {author} {\bibfnamefont {R.}~\bibnamefont {Nathan}}, \bibinfo {author} {\bibfnamefont {W.~M.}\ \bibnamefont {Getz}}, \bibinfo {author} {\bibfnamefont {E.}~\bibnamefont {Revilla}}, \bibinfo {author} {\bibfnamefont {M.}~\bibnamefont {Holyoak}}, \bibinfo {author} {\bibfnamefont {R.}~\bibnamefont {Kadmon}}, \bibinfo {author} {\bibfnamefont {D.}~\bibnamefont {Saltz}},\ and\ \bibinfo {author} {\bibfnamefont {P.~E.}\ \bibnamefont {Smouse}},\ }\bibfield  {title} {\bibinfo {title} {A movement ecology paradigm for unifying organismal movement research},\ }\href@noop {} {\bibfield  {journal} {\bibinfo  {journal} {Proc. Nat. Acad. Sci.}\ }\textbf {\bibinfo {volume} {105}},\ \bibinfo {pages} {19052} (\bibinfo {year} {2008})}\BibitemShut {NoStop}%
\bibitem [{\citenamefont {Holmes}(1993)}]{holmes1993diffusion}%
  \BibitemOpen
  \bibfield  {author} {\bibinfo {author} {\bibfnamefont {E.~E.}\ \bibnamefont {Holmes}},\ }\bibfield  {title} {\bibinfo {title} {Are diffusion models too simple? a comparison with telegraph models of invasion},\ }\href@noop {} {\bibfield  {journal} {\bibinfo  {journal} {Am. Nat.}\ }\textbf {\bibinfo {volume} {142}},\ \bibinfo {pages} {779} (\bibinfo {year} {1993})}\BibitemShut {NoStop}%
\bibitem [{\citenamefont {Viswanathan}\ \emph {et~al.}(2011)\citenamefont {Viswanathan}, \citenamefont {Da~Luz}, \citenamefont {Raposo},\ and\ \citenamefont {Stanley}}]{viswanathan2011physics}%
  \BibitemOpen
  \bibfield  {author} {\bibinfo {author} {\bibfnamefont {G.~M.}\ \bibnamefont {Viswanathan}}, \bibinfo {author} {\bibfnamefont {M.~G.}\ \bibnamefont {Da~Luz}}, \bibinfo {author} {\bibfnamefont {E.~P.}\ \bibnamefont {Raposo}},\ and\ \bibinfo {author} {\bibfnamefont {H.~E.}\ \bibnamefont {Stanley}},\ }\href@noop {} {\emph {\bibinfo {title} {The physics of foraging: an introduction to random searches and biological encounters}}}\ (\bibinfo  {publisher} {Cambridge University Press},\ \bibinfo {year} {2011})\BibitemShut {NoStop}%
\bibitem [{\citenamefont {M{\'e}ndez}\ \emph {et~al.}(2016)\citenamefont {M{\'e}ndez}, \citenamefont {Campos},\ and\ \citenamefont {Bartumeus}}]{mendez2016stochastic}%
  \BibitemOpen
  \bibfield  {author} {\bibinfo {author} {\bibfnamefont {V.}~\bibnamefont {M{\'e}ndez}}, \bibinfo {author} {\bibfnamefont {D.}~\bibnamefont {Campos}},\ and\ \bibinfo {author} {\bibfnamefont {F.}~\bibnamefont {Bartumeus}},\ }\href@noop {} {\emph {\bibinfo {title} {Stochastic foundations in movement ecology}}}\ (\bibinfo  {publisher} {Springer},\ \bibinfo {year} {2016})\BibitemShut {NoStop}%
\bibitem [{\citenamefont {McLane}\ \emph {et~al.}(2011)\citenamefont {McLane}, \citenamefont {Semeniuk}, \citenamefont {McDermid},\ and\ \citenamefont {Marceau}}]{mclane2011role}%
  \BibitemOpen
  \bibfield  {author} {\bibinfo {author} {\bibfnamefont {A.~J.}\ \bibnamefont {McLane}}, \bibinfo {author} {\bibfnamefont {C.}~\bibnamefont {Semeniuk}}, \bibinfo {author} {\bibfnamefont {G.~J.}\ \bibnamefont {McDermid}},\ and\ \bibinfo {author} {\bibfnamefont {D.~J.}\ \bibnamefont {Marceau}},\ }\bibfield  {title} {\bibinfo {title} {The role of agent-based models in wildlife ecology and management},\ }\href@noop {} {\bibfield  {journal} {\bibinfo  {journal} {Ecol. Modell.}\ }\textbf {\bibinfo {volume} {222}},\ \bibinfo {pages} {1544} (\bibinfo {year} {2011})}\BibitemShut {NoStop}%
\bibitem [{\citenamefont {Gurarie}\ \emph {et~al.}(2016)\citenamefont {Gurarie}, \citenamefont {Bracis}, \citenamefont {Delgado}, \citenamefont {Meckley}, \citenamefont {Kojola},\ and\ \citenamefont {Wagner}}]{gurarie2016animal}%
  \BibitemOpen
  \bibfield  {author} {\bibinfo {author} {\bibfnamefont {E.}~\bibnamefont {Gurarie}}, \bibinfo {author} {\bibfnamefont {C.}~\bibnamefont {Bracis}}, \bibinfo {author} {\bibfnamefont {M.}~\bibnamefont {Delgado}}, \bibinfo {author} {\bibfnamefont {T.~D.}\ \bibnamefont {Meckley}}, \bibinfo {author} {\bibfnamefont {I.}~\bibnamefont {Kojola}},\ and\ \bibinfo {author} {\bibfnamefont {C.~M.}\ \bibnamefont {Wagner}},\ }\bibfield  {title} {\bibinfo {title} {What is the animal doing? tools for exploring behavioural structure in animal movements},\ }\href@noop {} {\bibfield  {journal} {\bibinfo  {journal} {J. Anim. Ecol.}\ }\textbf {\bibinfo {volume} {85}},\ \bibinfo {pages} {69} (\bibinfo {year} {2016})}\BibitemShut {NoStop}%
\bibitem [{\citenamefont {Gurarie}\ \emph {et~al.}(2017)\citenamefont {Gurarie}, \citenamefont {Cagnacci}, \citenamefont {Peters}, \citenamefont {Fleming}, \citenamefont {Calabrese}, \citenamefont {Mueller},\ and\ \citenamefont {Fagan}}]{gurarie2017framework}%
  \BibitemOpen
  \bibfield  {author} {\bibinfo {author} {\bibfnamefont {E.}~\bibnamefont {Gurarie}}, \bibinfo {author} {\bibfnamefont {F.}~\bibnamefont {Cagnacci}}, \bibinfo {author} {\bibfnamefont {W.}~\bibnamefont {Peters}}, \bibinfo {author} {\bibfnamefont {C.~H.}\ \bibnamefont {Fleming}}, \bibinfo {author} {\bibfnamefont {J.~M.}\ \bibnamefont {Calabrese}}, \bibinfo {author} {\bibfnamefont {T.}~\bibnamefont {Mueller}},\ and\ \bibinfo {author} {\bibfnamefont {W.~F.}\ \bibnamefont {Fagan}},\ }\bibfield  {title} {\bibinfo {title} {A framework for modelling range shifts and migrations: asking when, whither, whether and will it return},\ }\href@noop {} {\bibfield  {journal} {\bibinfo  {journal} {J. Anim. Ecol.}\ }\textbf {\bibinfo {volume} {86}},\ \bibinfo {pages} {943} (\bibinfo {year} {2017})}\BibitemShut {NoStop}%
\bibitem [{\citenamefont {Nathan}\ \emph {et~al.}(2011)\citenamefont {Nathan}, \citenamefont {Horvitz}, \citenamefont {He}, \citenamefont {Kuparinen}, \citenamefont {Schurr},\ and\ \citenamefont {Katul}}]{nathan2011spread}%
  \BibitemOpen
  \bibfield  {author} {\bibinfo {author} {\bibfnamefont {R.}~\bibnamefont {Nathan}}, \bibinfo {author} {\bibfnamefont {N.}~\bibnamefont {Horvitz}}, \bibinfo {author} {\bibfnamefont {Y.}~\bibnamefont {He}}, \bibinfo {author} {\bibfnamefont {A.}~\bibnamefont {Kuparinen}}, \bibinfo {author} {\bibfnamefont {F.~M.}\ \bibnamefont {Schurr}},\ and\ \bibinfo {author} {\bibfnamefont {G.~G.}\ \bibnamefont {Katul}},\ }\bibfield  {title} {\bibinfo {title} {Spread of north american wind-dispersed trees in future environments},\ }\href@noop {} {\bibfield  {journal} {\bibinfo  {journal} {Ecology letters}\ }\textbf {\bibinfo {volume} {14}},\ \bibinfo {pages} {211} (\bibinfo {year} {2011})}\BibitemShut {NoStop}%
\bibitem [{\citenamefont {Patterson}\ \emph {et~al.}(2017)\citenamefont {Patterson}, \citenamefont {Parton}, \citenamefont {Langrock}, \citenamefont {Blackwell}, \citenamefont {Thomas},\ and\ \citenamefont {King}}]{patterson2017statistical}%
  \BibitemOpen
  \bibfield  {author} {\bibinfo {author} {\bibfnamefont {T.~A.}\ \bibnamefont {Patterson}}, \bibinfo {author} {\bibfnamefont {A.}~\bibnamefont {Parton}}, \bibinfo {author} {\bibfnamefont {R.}~\bibnamefont {Langrock}}, \bibinfo {author} {\bibfnamefont {P.~G.}\ \bibnamefont {Blackwell}}, \bibinfo {author} {\bibfnamefont {L.}~\bibnamefont {Thomas}},\ and\ \bibinfo {author} {\bibfnamefont {R.}~\bibnamefont {King}},\ }\bibfield  {title} {\bibinfo {title} {Statistical modelling of individual animal movement: an overview of key methods and a discussion of practical challenges},\ }\href@noop {} {\bibfield  {journal} {\bibinfo  {journal} {AStA Adv. Stat.}\ }\textbf {\bibinfo {volume} {101}},\ \bibinfo {pages} {399} (\bibinfo {year} {2017})}\BibitemShut {NoStop}%
\bibitem [{\citenamefont {McGarigal}\ \emph {et~al.}(2016)\citenamefont {McGarigal}, \citenamefont {Wan}, \citenamefont {Zeller}, \citenamefont {Timm},\ and\ \citenamefont {Cushman}}]{mcgarigal2016multi}%
  \BibitemOpen
  \bibfield  {author} {\bibinfo {author} {\bibfnamefont {K.}~\bibnamefont {McGarigal}}, \bibinfo {author} {\bibfnamefont {H.~Y.}\ \bibnamefont {Wan}}, \bibinfo {author} {\bibfnamefont {K.~A.}\ \bibnamefont {Zeller}}, \bibinfo {author} {\bibfnamefont {B.~C.}\ \bibnamefont {Timm}},\ and\ \bibinfo {author} {\bibfnamefont {S.~A.}\ \bibnamefont {Cushman}},\ }\bibfield  {title} {\bibinfo {title} {Multi-scale habitat selection modeling: a review and outlook},\ }\href@noop {} {\bibfield  {journal} {\bibinfo  {journal} {Landsc. Ecol.}\ }\textbf {\bibinfo {volume} {31}},\ \bibinfo {pages} {1161} (\bibinfo {year} {2016})}\BibitemShut {NoStop}%
\bibitem [{\citenamefont {Nouvellet}\ \emph {et~al.}(2009)\citenamefont {Nouvellet}, \citenamefont {Bacon},\ and\ \citenamefont {Waxman}}]{nouvellet2009fundamental}%
  \BibitemOpen
  \bibfield  {author} {\bibinfo {author} {\bibfnamefont {P.}~\bibnamefont {Nouvellet}}, \bibinfo {author} {\bibfnamefont {J.}~\bibnamefont {Bacon}},\ and\ \bibinfo {author} {\bibfnamefont {D.}~\bibnamefont {Waxman}},\ }\bibfield  {title} {\bibinfo {title} {Fundamental insights into the random movement of animals from a single distance-related statistic},\ }\href@noop {} {\bibfield  {journal} {\bibinfo  {journal} {Am. Nat.}\ }\textbf {\bibinfo {volume} {174}},\ \bibinfo {pages} {506} (\bibinfo {year} {2009})}\BibitemShut {NoStop}%
\bibitem [{\citenamefont {Ahmed}\ \emph {et~al.}(2013)\citenamefont {Ahmed}, \citenamefont {Williams}, \citenamefont {Silver},\ and\ \citenamefont {Saif}}]{ahmed2013measuring}%
  \BibitemOpen
  \bibfield  {author} {\bibinfo {author} {\bibfnamefont {W.~W.}\ \bibnamefont {Ahmed}}, \bibinfo {author} {\bibfnamefont {B.~J.}\ \bibnamefont {Williams}}, \bibinfo {author} {\bibfnamefont {A.~M.}\ \bibnamefont {Silver}},\ and\ \bibinfo {author} {\bibfnamefont {T.~A.}\ \bibnamefont {Saif}},\ }\bibfield  {title} {\bibinfo {title} {Measuring nonequilibrium vesicle dynamics in neurons under tension},\ }\href@noop {} {\bibfield  {journal} {\bibinfo  {journal} {Lab on a Chip}\ }\textbf {\bibinfo {volume} {13}},\ \bibinfo {pages} {570} (\bibinfo {year} {2013})}\BibitemShut {NoStop}%
\bibitem [{\citenamefont {Bastille-Rousseau}\ \emph {et~al.}(2016)\citenamefont {Bastille-Rousseau}, \citenamefont {Potts}, \citenamefont {Yackulic}, \citenamefont {Frair}, \citenamefont {Ellington},\ and\ \citenamefont {Blake}}]{bastille2016flexible}%
  \BibitemOpen
  \bibfield  {author} {\bibinfo {author} {\bibfnamefont {G.}~\bibnamefont {Bastille-Rousseau}}, \bibinfo {author} {\bibfnamefont {J.~R.}\ \bibnamefont {Potts}}, \bibinfo {author} {\bibfnamefont {C.~B.}\ \bibnamefont {Yackulic}}, \bibinfo {author} {\bibfnamefont {J.~L.}\ \bibnamefont {Frair}}, \bibinfo {author} {\bibfnamefont {E.~H.}\ \bibnamefont {Ellington}},\ and\ \bibinfo {author} {\bibfnamefont {S.}~\bibnamefont {Blake}},\ }\bibfield  {title} {\bibinfo {title} {Flexible characterization of animal movement pattern using net squared displacement and a latent state model},\ }\href@noop {} {\bibfield  {journal} {\bibinfo  {journal} {Mov. Ecol.}\ }\textbf {\bibinfo {volume} {4}},\ \bibinfo {pages} {1} (\bibinfo {year} {2016})}\BibitemShut {NoStop}%
\bibitem [{\citenamefont {Vilk}\ \emph {et~al.}(2022{\natexlab{b}})\citenamefont {Vilk}, \citenamefont {Aghion}, \citenamefont {Nathan}, \citenamefont {Toledo}, \citenamefont {Metzler},\ and\ \citenamefont {Assaf}}]{vilk2022classification}%
  \BibitemOpen
  \bibfield  {author} {\bibinfo {author} {\bibfnamefont {O.}~\bibnamefont {Vilk}}, \bibinfo {author} {\bibfnamefont {E.}~\bibnamefont {Aghion}}, \bibinfo {author} {\bibfnamefont {R.}~\bibnamefont {Nathan}}, \bibinfo {author} {\bibfnamefont {S.}~\bibnamefont {Toledo}}, \bibinfo {author} {\bibfnamefont {R.}~\bibnamefont {Metzler}},\ and\ \bibinfo {author} {\bibfnamefont {M.}~\bibnamefont {Assaf}},\ }\bibfield  {title} {\bibinfo {title} {Classification of anomalous diffusion in animal movement data using power spectral analysis},\ }\href@noop {} {\bibfield  {journal} {\bibinfo  {journal} {J. Phys. A: Math. Theor.}\ }\textbf {\bibinfo {volume} {55}},\ \bibinfo {pages} {334004} (\bibinfo {year} {2022}{\natexlab{b}})}\BibitemShut {NoStop}%
\bibitem [{\citenamefont {Burte}\ \emph {et~al.}(2023)\citenamefont {Burte}, \citenamefont {Cointe}, \citenamefont {Perez}, \citenamefont {Mailleret},\ and\ \citenamefont {Calcagno}}]{burte2023complex}%
  \BibitemOpen
  \bibfield  {author} {\bibinfo {author} {\bibfnamefont {V.}~\bibnamefont {Burte}}, \bibinfo {author} {\bibfnamefont {M.}~\bibnamefont {Cointe}}, \bibinfo {author} {\bibfnamefont {G.}~\bibnamefont {Perez}}, \bibinfo {author} {\bibfnamefont {L.}~\bibnamefont {Mailleret}},\ and\ \bibinfo {author} {\bibfnamefont {V.}~\bibnamefont {Calcagno}},\ }\bibfield  {title} {\bibinfo {title} {When complex movement yields simple dispersal: behavioural heterogeneity, spatial spread and parasitism in groups of micro-wasps},\ }\href@noop {} {\bibfield  {journal} {\bibinfo  {journal} {Mov. Ecol.}\ }\textbf {\bibinfo {volume} {11}},\ \bibinfo {pages} {13} (\bibinfo {year} {2023})}\BibitemShut {NoStop}%
\bibitem [{\citenamefont {Castiglione}\ \emph {et~al.}(1999)\citenamefont {Castiglione}, \citenamefont {Mazzino}, \citenamefont {Muratore-Ginanneschi},\ and\ \citenamefont {Vulpiani}}]{castiglione1999strong}%
  \BibitemOpen
  \bibfield  {author} {\bibinfo {author} {\bibfnamefont {P.}~\bibnamefont {Castiglione}}, \bibinfo {author} {\bibfnamefont {A.}~\bibnamefont {Mazzino}}, \bibinfo {author} {\bibfnamefont {P.}~\bibnamefont {Muratore-Ginanneschi}},\ and\ \bibinfo {author} {\bibfnamefont {A.}~\bibnamefont {Vulpiani}},\ }\bibfield  {title} {\bibinfo {title} {On strong anomalous diffusion},\ }\href@noop {} {\bibfield  {journal} {\bibinfo  {journal} {Physica D: Nonlinear Phenomena}\ }\textbf {\bibinfo {volume} {134}},\ \bibinfo {pages} {75} (\bibinfo {year} {1999})}\BibitemShut {NoStop}%
\bibitem [{\citenamefont {Andersen}\ \emph {et~al.}(2000)\citenamefont {Andersen}, \citenamefont {Castiglione}, \citenamefont {Mazzino},\ and\ \citenamefont {Vulpiani}}]{andersen2000simple}%
  \BibitemOpen
  \bibfield  {author} {\bibinfo {author} {\bibfnamefont {K.}~\bibnamefont {Andersen}}, \bibinfo {author} {\bibfnamefont {P.}~\bibnamefont {Castiglione}}, \bibinfo {author} {\bibfnamefont {A.}~\bibnamefont {Mazzino}},\ and\ \bibinfo {author} {\bibfnamefont {A.}~\bibnamefont {Vulpiani}},\ }\bibfield  {title} {\bibinfo {title} {Simple stochastic models showing strong anomalous diffusion},\ }\href@noop {} {\bibfield  {journal} {\bibinfo  {journal} {The European Physical Journal B-Condensed Matter and Complex Systems}\ }\textbf {\bibinfo {volume} {18}},\ \bibinfo {pages} {447} (\bibinfo {year} {2000})}\BibitemShut {NoStop}%
\bibitem [{\citenamefont {Vollmer}\ \emph {et~al.}(2021)\citenamefont {Vollmer}, \citenamefont {Rondoni}, \citenamefont {Tayyab}, \citenamefont {Giberti},\ and\ \citenamefont {Mej{\'\i}a-Monasterio}}]{vollmer2021displacement}%
  \BibitemOpen
  \bibfield  {author} {\bibinfo {author} {\bibfnamefont {J.}~\bibnamefont {Vollmer}}, \bibinfo {author} {\bibfnamefont {L.}~\bibnamefont {Rondoni}}, \bibinfo {author} {\bibfnamefont {M.}~\bibnamefont {Tayyab}}, \bibinfo {author} {\bibfnamefont {C.}~\bibnamefont {Giberti}},\ and\ \bibinfo {author} {\bibfnamefont {C.}~\bibnamefont {Mej{\'\i}a-Monasterio}},\ }\bibfield  {title} {\bibinfo {title} {Displacement autocorrelation functions for strong anomalous diffusion: A scaling form, universal behavior, and corrections to scaling},\ }\href@noop {} {\bibfield  {journal} {\bibinfo  {journal} {Phys. Rev. Res.}\ }\textbf {\bibinfo {volume} {3}},\ \bibinfo {pages} {013067} (\bibinfo {year} {2021})}\BibitemShut {NoStop}%
\bibitem [{\citenamefont {Mandelbrot}\ and\ \citenamefont {Van~Ness}(1968)}]{Man68fbm}%
  \BibitemOpen
  \bibfield  {author} {\bibinfo {author} {\bibfnamefont {B.~B.}\ \bibnamefont {Mandelbrot}}\ and\ \bibinfo {author} {\bibfnamefont {J.~W.}\ \bibnamefont {Van~Ness}},\ }\bibfield  {title} {\bibinfo {title} {Fractional brownian motions, fractional noises and applications},\ }\href@noop {} {\bibfield  {journal} {\bibinfo  {journal} {SIAM review}\ }\textbf {\bibinfo {volume} {10}},\ \bibinfo {pages} {422} (\bibinfo {year} {1968})}\BibitemShut {NoStop}%
\bibitem [{\citenamefont {Sanders}\ and\ \citenamefont {Larralde}(2006)}]{sanders2006occurrence}%
  \BibitemOpen
  \bibfield  {author} {\bibinfo {author} {\bibfnamefont {D.~P.}\ \bibnamefont {Sanders}}\ and\ \bibinfo {author} {\bibfnamefont {H.}~\bibnamefont {Larralde}},\ }\bibfield  {title} {\bibinfo {title} {Occurrence of normal and anomalous diffusion in polygonal billiard channels},\ }\href@noop {} {\bibfield  {journal} {\bibinfo  {journal} {Phys. Rev. E}\ }\textbf {\bibinfo {volume} {73}},\ \bibinfo {pages} {026205} (\bibinfo {year} {2006})}\BibitemShut {NoStop}%
\bibitem [{\citenamefont {Orchard}\ \emph {et~al.}(2021)\citenamefont {Orchard}, \citenamefont {Rondoni}, \citenamefont {Mejia-Monasterio},\ and\ \citenamefont {Frascoli}}]{orchard2021diffusion}%
  \BibitemOpen
  \bibfield  {author} {\bibinfo {author} {\bibfnamefont {J.}~\bibnamefont {Orchard}}, \bibinfo {author} {\bibfnamefont {L.}~\bibnamefont {Rondoni}}, \bibinfo {author} {\bibfnamefont {C.}~\bibnamefont {Mejia-Monasterio}},\ and\ \bibinfo {author} {\bibfnamefont {F.}~\bibnamefont {Frascoli}},\ }\bibfield  {title} {\bibinfo {title} {Diffusion and escape from polygonal channels: Extreme values and geometric effects},\ }\href@noop {} {\bibfield  {journal} {\bibinfo  {journal} {J. Stat. Mech. Theo. Exp.}\ }\textbf {\bibinfo {volume} {2021}},\ \bibinfo {pages} {073208} (\bibinfo {year} {2021})}\BibitemShut {NoStop}%
\bibitem [{\citenamefont {Carreras}\ \emph {et~al.}(1999)\citenamefont {Carreras}, \citenamefont {Lynch}, \citenamefont {Newman},\ and\ \citenamefont {Zaslavsky}}]{carreras1999anomalous}%
  \BibitemOpen
  \bibfield  {author} {\bibinfo {author} {\bibfnamefont {B.}~\bibnamefont {Carreras}}, \bibinfo {author} {\bibfnamefont {V.}~\bibnamefont {Lynch}}, \bibinfo {author} {\bibfnamefont {D.}~\bibnamefont {Newman}},\ and\ \bibinfo {author} {\bibfnamefont {G.}~\bibnamefont {Zaslavsky}},\ }\bibfield  {title} {\bibinfo {title} {Anomalous diffusion in a running sandpile model},\ }\href@noop {} {\bibfield  {journal} {\bibinfo  {journal} {Phys. Rev. E}\ }\textbf {\bibinfo {volume} {60}},\ \bibinfo {pages} {4770} (\bibinfo {year} {1999})}\BibitemShut {NoStop}%
\bibitem [{\citenamefont {Dechant}\ and\ \citenamefont {Lutz}(2012)}]{dechant2012anomalous}%
  \BibitemOpen
  \bibfield  {author} {\bibinfo {author} {\bibfnamefont {A.}~\bibnamefont {Dechant}}\ and\ \bibinfo {author} {\bibfnamefont {E.}~\bibnamefont {Lutz}},\ }\bibfield  {title} {\bibinfo {title} {Anomalous spatial diffusion and multifractality in optical lattices},\ }\href@noop {} {\bibfield  {journal} {\bibinfo  {journal} {Phys. Rev. Lett.}\ }\textbf {\bibinfo {volume} {108}},\ \bibinfo {pages} {230601} (\bibinfo {year} {2012})}\BibitemShut {NoStop}%
\bibitem [{\citenamefont {Afek}\ \emph {et~al.}(2023)\citenamefont {Afek}, \citenamefont {Davidson}, \citenamefont {Kessler},\ and\ \citenamefont {Barkai}}]{afek2023colloquium}%
  \BibitemOpen
  \bibfield  {author} {\bibinfo {author} {\bibfnamefont {G.}~\bibnamefont {Afek}}, \bibinfo {author} {\bibfnamefont {N.}~\bibnamefont {Davidson}}, \bibinfo {author} {\bibfnamefont {D.~A.}\ \bibnamefont {Kessler}},\ and\ \bibinfo {author} {\bibfnamefont {E.}~\bibnamefont {Barkai}},\ }\bibfield  {title} {\bibinfo {title} {Colloquium: Anomalous statistics of laser-cooled atoms in dissipative optical lattices},\ }\href@noop {} {\bibfield  {journal} {\bibinfo  {journal} {Rev. Mod. Phys.}\ }\textbf {\bibinfo {volume} {95}},\ \bibinfo {pages} {031003} (\bibinfo {year} {2023})}\BibitemShut {NoStop}%
\bibitem [{\citenamefont {Cagnetta}\ \emph {et~al.}(2015)\citenamefont {Cagnetta}, \citenamefont {Gonnella}, \citenamefont {Mossa},\ and\ \citenamefont {Ruffo}}]{cagnetta2015strong}%
  \BibitemOpen
  \bibfield  {author} {\bibinfo {author} {\bibfnamefont {F.}~\bibnamefont {Cagnetta}}, \bibinfo {author} {\bibfnamefont {G.}~\bibnamefont {Gonnella}}, \bibinfo {author} {\bibfnamefont {A.}~\bibnamefont {Mossa}},\ and\ \bibinfo {author} {\bibfnamefont {S.}~\bibnamefont {Ruffo}},\ }\bibfield  {title} {\bibinfo {title} {Strong anomalous diffusion of the phase of a chaotic pendulum},\ }\href@noop {} {\bibfield  {journal} {\bibinfo  {journal} {Europhys. Lett.}\ }\textbf {\bibinfo {volume} {111}},\ \bibinfo {pages} {10002} (\bibinfo {year} {2015})}\BibitemShut {NoStop}%
\bibitem [{\citenamefont {Wang}\ \emph {et~al.}(2020)\citenamefont {Wang}, \citenamefont {Chen},\ and\ \citenamefont {Deng}}]{wang2020strong}%
  \BibitemOpen
  \bibfield  {author} {\bibinfo {author} {\bibfnamefont {X.}~\bibnamefont {Wang}}, \bibinfo {author} {\bibfnamefont {Y.}~\bibnamefont {Chen}},\ and\ \bibinfo {author} {\bibfnamefont {W.}~\bibnamefont {Deng}},\ }\bibfield  {title} {\bibinfo {title} {Strong anomalous diffusion in two-state process with l{\'e}vy walk and brownian motion},\ }\href@noop {} {\bibfield  {journal} {\bibinfo  {journal} {Phys. Rev. Res.}\ }\textbf {\bibinfo {volume} {2}},\ \bibinfo {pages} {013102} (\bibinfo {year} {2020})}\BibitemShut {NoStop}%
\bibitem [{\citenamefont {Liu}\ \emph {et~al.}(2022)\citenamefont {Liu}, \citenamefont {Zhu}, \citenamefont {Bao},\ and\ \citenamefont {Chen}}]{liu2022strong}%
  \BibitemOpen
  \bibfield  {author} {\bibinfo {author} {\bibfnamefont {J.}~\bibnamefont {Liu}}, \bibinfo {author} {\bibfnamefont {P.}~\bibnamefont {Zhu}}, \bibinfo {author} {\bibfnamefont {J.-D.}\ \bibnamefont {Bao}},\ and\ \bibinfo {author} {\bibfnamefont {X.}~\bibnamefont {Chen}},\ }\bibfield  {title} {\bibinfo {title} {Strong anomalous diffusive behaviors of the two-state random walk process},\ }\href@noop {} {\bibfield  {journal} {\bibinfo  {journal} {Phys. Rev. E}\ }\textbf {\bibinfo {volume} {105}},\ \bibinfo {pages} {014122} (\bibinfo {year} {2022})}\BibitemShut {NoStop}%
\bibitem [{\citenamefont {Samama}\ and\ \citenamefont {Barkai}(2023)}]{samama2023statistics}%
  \BibitemOpen
  \bibfield  {author} {\bibinfo {author} {\bibfnamefont {A.}~\bibnamefont {Samama}}\ and\ \bibinfo {author} {\bibfnamefont {E.}~\bibnamefont {Barkai}},\ }\bibfield  {title} {\bibinfo {title} {Statistics of long-range force fields in random environments: Beyond holtsmark},\ }\href@noop {} {\bibfield  {journal} {\bibinfo  {journal} {Phys. Rev. E}\ }\textbf {\bibinfo {volume} {108}},\ \bibinfo {pages} {044116} (\bibinfo {year} {2023})}\BibitemShut {NoStop}%
\bibitem [{\citenamefont {Bernardi}\ \emph {et~al.}(2024)\citenamefont {Bernardi}, \citenamefont {Pizzi},\ and\ \citenamefont {Rondoni}}]{bernardi2024anomalous}%
  \BibitemOpen
  \bibfield  {author} {\bibinfo {author} {\bibfnamefont {S.}~\bibnamefont {Bernardi}}, \bibinfo {author} {\bibfnamefont {M.}~\bibnamefont {Pizzi}},\ and\ \bibinfo {author} {\bibfnamefont {L.}~\bibnamefont {Rondoni}},\ }\bibfield  {title} {\bibinfo {title} {Anomalous heat transport and universality in macroscopic diffusion models},\ }\href@noop {} {\bibfield  {journal} {\bibinfo  {journal} {J. Therm. Anal. Calorim.}\ ,\ \bibinfo {pages} {1}} (\bibinfo {year} {2024})}\BibitemShut {NoStop}%
\bibitem [{\citenamefont {Gal}\ and\ \citenamefont {Weihs}(2010)}]{gal2010experimental}%
  \BibitemOpen
  \bibfield  {author} {\bibinfo {author} {\bibfnamefont {N.}~\bibnamefont {Gal}}\ and\ \bibinfo {author} {\bibfnamefont {D.}~\bibnamefont {Weihs}},\ }\bibfield  {title} {\bibinfo {title} {Experimental evidence of strong anomalous diffusion in living cells},\ }\href@noop {} {\bibfield  {journal} {\bibinfo  {journal} {Phys. Rev. E}\ }\textbf {\bibinfo {volume} {81}},\ \bibinfo {pages} {020903} (\bibinfo {year} {2010})}\BibitemShut {NoStop}%
\bibitem [{\citenamefont {Pini}\ \emph {et~al.}(2024)\citenamefont {Pini}, \citenamefont {Mazzamuto}, \citenamefont {Riboli}, \citenamefont {Wiersma},\ and\ \citenamefont {Pattelli}}]{pini2023breakdown}%
  \BibitemOpen
  \bibfield  {author} {\bibinfo {author} {\bibfnamefont {E.}~\bibnamefont {Pini}}, \bibinfo {author} {\bibfnamefont {G.}~\bibnamefont {Mazzamuto}}, \bibinfo {author} {\bibfnamefont {F.}~\bibnamefont {Riboli}}, \bibinfo {author} {\bibfnamefont {D.~S.}\ \bibnamefont {Wiersma}},\ and\ \bibinfo {author} {\bibfnamefont {L.}~\bibnamefont {Pattelli}},\ }\bibfield  {title} {\bibinfo {title} {Non-self-similar light transport in scattering media},\ }\href@noop {} {\bibfield  {journal} {\bibinfo  {journal} {Phys. Rev. Res.}\ }\textbf {\bibinfo {volume} {6}},\ \bibinfo {pages} {L032026} (\bibinfo {year} {2024})}\BibitemShut {NoStop}%
\bibitem [{\citenamefont {Vilk}\ \emph {et~al.}(2022{\natexlab{c}})\citenamefont {Vilk}, \citenamefont {Orchan}, \citenamefont {Charter}, \citenamefont {Ganot}, \citenamefont {Toledo}, \citenamefont {Nathan},\ and\ \citenamefont {Assaf}}]{vilk2022ergodicity}%
  \BibitemOpen
  \bibfield  {author} {\bibinfo {author} {\bibfnamefont {O.}~\bibnamefont {Vilk}}, \bibinfo {author} {\bibfnamefont {Y.}~\bibnamefont {Orchan}}, \bibinfo {author} {\bibfnamefont {M.}~\bibnamefont {Charter}}, \bibinfo {author} {\bibfnamefont {N.}~\bibnamefont {Ganot}}, \bibinfo {author} {\bibfnamefont {S.}~\bibnamefont {Toledo}}, \bibinfo {author} {\bibfnamefont {R.}~\bibnamefont {Nathan}},\ and\ \bibinfo {author} {\bibfnamefont {M.}~\bibnamefont {Assaf}},\ }\bibfield  {title} {\bibinfo {title} {Ergodicity breaking in area-restricted search of avian predators},\ }\href@noop {} {\bibfield  {journal} {\bibinfo  {journal} {Phys. Rev. X}\ }\textbf {\bibinfo {volume} {12}},\ \bibinfo {pages} {031005} (\bibinfo {year} {2022}{\natexlab{c}})}\BibitemShut {NoStop}%
\bibitem [{\citenamefont {Breed}\ \emph {et~al.}(2017)\citenamefont {Breed}, \citenamefont {Golson},\ and\ \citenamefont {Tinker}}]{breed2017predicting}%
  \BibitemOpen
  \bibfield  {author} {\bibinfo {author} {\bibfnamefont {G.~A.}\ \bibnamefont {Breed}}, \bibinfo {author} {\bibfnamefont {E.~A.}\ \bibnamefont {Golson}},\ and\ \bibinfo {author} {\bibfnamefont {M.~T.}\ \bibnamefont {Tinker}},\ }\bibfield  {title} {\bibinfo {title} {Predicting animal home-range structure and transitions using a multistate ornstein-uhlenbeck biased random walk},\ }\href@noop {} {\bibfield  {journal} {\bibinfo  {journal} {Ecology}\ }\textbf {\bibinfo {volume} {98}},\ \bibinfo {pages} {32} (\bibinfo {year} {2017})}\BibitemShut {NoStop}%
\bibitem [{\citenamefont {Jonsen}\ \emph {et~al.}(2005)\citenamefont {Jonsen}, \citenamefont {Flemming},\ and\ \citenamefont {Myers}}]{jonsen2005robust}%
  \BibitemOpen
  \bibfield  {author} {\bibinfo {author} {\bibfnamefont {I.~D.}\ \bibnamefont {Jonsen}}, \bibinfo {author} {\bibfnamefont {J.~M.}\ \bibnamefont {Flemming}},\ and\ \bibinfo {author} {\bibfnamefont {R.~A.}\ \bibnamefont {Myers}},\ }\bibfield  {title} {\bibinfo {title} {Robust state--space modeling of animal movement data},\ }\href@noop {} {\bibfield  {journal} {\bibinfo  {journal} {Ecology}\ }\textbf {\bibinfo {volume} {86}},\ \bibinfo {pages} {2874} (\bibinfo {year} {2005})}\BibitemShut {NoStop}%
\bibitem [{\citenamefont {Benhamou}(2014)}]{benhamou2014scales}%
  \BibitemOpen
  \bibfield  {author} {\bibinfo {author} {\bibfnamefont {S.}~\bibnamefont {Benhamou}},\ }\bibfield  {title} {\bibinfo {title} {Of scales and stationarity in animal movements},\ }\href@noop {} {\bibfield  {journal} {\bibinfo  {journal} {Ecol. Lett.}\ }\textbf {\bibinfo {volume} {17}},\ \bibinfo {pages} {261} (\bibinfo {year} {2014})}\BibitemShut {NoStop}%
\bibitem [{\citenamefont {Wunderle~Jr}(1991)}]{wunderle1991age}%
  \BibitemOpen
  \bibfield  {author} {\bibinfo {author} {\bibfnamefont {J.}~\bibnamefont {Wunderle~Jr}},\ }\bibfield  {title} {\bibinfo {title} {Age-specific foraging proficiency in birds.},\ }\href@noop {} {\bibfield  {journal} {\bibinfo  {journal} {Current ornithology}\ }\textbf {\bibinfo {volume} {8}},\ \bibinfo {pages} {273} (\bibinfo {year} {1991})}\BibitemShut {NoStop}%
\bibitem [{\citenamefont {Weiser}\ \emph {et~al.}(2016)\citenamefont {Weiser}, \citenamefont {Orchan}, \citenamefont {Nathan}, \citenamefont {Charter}, \citenamefont {Weiss},\ and\ \citenamefont {Toledo}}]{weiser2016characterizing}%
  \BibitemOpen
  \bibfield  {author} {\bibinfo {author} {\bibfnamefont {A.~W.}\ \bibnamefont {Weiser}}, \bibinfo {author} {\bibfnamefont {Y.}~\bibnamefont {Orchan}}, \bibinfo {author} {\bibfnamefont {R.}~\bibnamefont {Nathan}}, \bibinfo {author} {\bibfnamefont {M.}~\bibnamefont {Charter}}, \bibinfo {author} {\bibfnamefont {A.~J.}\ \bibnamefont {Weiss}},\ and\ \bibinfo {author} {\bibfnamefont {S.}~\bibnamefont {Toledo}},\ }\bibfield  {title} {\bibinfo {title} {Characterizing the accuracy of a self-synchronized reverse-gps wildlife localization system},\ }in\ \href@noop {} {\emph {\bibinfo {booktitle} {2016 15th ACM/IEEE International Conference on Information Processing in Sensor Networks (IPSN)}}}\ (\bibinfo {organization} {IEEE},\ \bibinfo {year} {2016})\ pp.\ \bibinfo {pages} {1--12}\BibitemShut {NoStop}%
\bibitem [{\citenamefont {Toledo}\ \emph {et~al.}(2016)\citenamefont {Toledo}, \citenamefont {Kishon}, \citenamefont {Orchan}, \citenamefont {Shohat},\ and\ \citenamefont {Nathan}}]{toledo2016lessons}%
  \BibitemOpen
  \bibfield  {author} {\bibinfo {author} {\bibfnamefont {S.}~\bibnamefont {Toledo}}, \bibinfo {author} {\bibfnamefont {O.}~\bibnamefont {Kishon}}, \bibinfo {author} {\bibfnamefont {Y.}~\bibnamefont {Orchan}}, \bibinfo {author} {\bibfnamefont {A.}~\bibnamefont {Shohat}},\ and\ \bibinfo {author} {\bibfnamefont {R.}~\bibnamefont {Nathan}},\ }\bibfield  {title} {\bibinfo {title} {Lessons and experiences from the design, implementation, and deployment of a wildlife tracking system},\ }in\ \href@noop {} {\emph {\bibinfo {booktitle} {2016 IEEE International Conference on Software Science, Technology and Engineering (SWSTE)}}}\ (\bibinfo {organization} {IEEE},\ \bibinfo {year} {2016})\ pp.\ \bibinfo {pages} {51--60}\BibitemShut {NoStop}%
\bibitem [{\citenamefont {Charter}\ and\ \citenamefont {Rozman}(2022)}]{charter2022importance}%
  \BibitemOpen
  \bibfield  {author} {\bibinfo {author} {\bibfnamefont {M.}~\bibnamefont {Charter}}\ and\ \bibinfo {author} {\bibfnamefont {G.}~\bibnamefont {Rozman}},\ }\bibfield  {title} {\bibinfo {title} {The importance of nest box placement for barn owls (\textit{Tyto alba})},\ }\href@noop {} {\bibfield  {journal} {\bibinfo  {journal} {Animals}\ }\textbf {\bibinfo {volume} {12}},\ \bibinfo {pages} {2815} (\bibinfo {year} {2022})}\BibitemShut {NoStop}%
\bibitem [{\citenamefont {Taylor}(2004)}]{taylor2004barn}%
  \BibitemOpen
  \bibfield  {author} {\bibinfo {author} {\bibfnamefont {I.}~\bibnamefont {Taylor}},\ }\href@noop {} {\emph {\bibinfo {title} {Barn owls: predator-prey relationships and conservation}}}\ (\bibinfo  {publisher} {Cambridge University Press},\ \bibinfo {year} {2004})\BibitemShut {NoStop}%
\bibitem [{\citenamefont {Cain}\ \emph {et~al.}(2023)\citenamefont {Cain}, \citenamefont {Solomon}, \citenamefont {Leshem}, \citenamefont {Toledo}, \citenamefont {Arnon}, \citenamefont {Roulin},\ and\ \citenamefont {Spiegel}}]{cain2023movement}%
  \BibitemOpen
  \bibfield  {author} {\bibinfo {author} {\bibfnamefont {S.}~\bibnamefont {Cain}}, \bibinfo {author} {\bibfnamefont {T.}~\bibnamefont {Solomon}}, \bibinfo {author} {\bibfnamefont {Y.}~\bibnamefont {Leshem}}, \bibinfo {author} {\bibfnamefont {S.}~\bibnamefont {Toledo}}, \bibinfo {author} {\bibfnamefont {E.}~\bibnamefont {Arnon}}, \bibinfo {author} {\bibfnamefont {A.}~\bibnamefont {Roulin}},\ and\ \bibinfo {author} {\bibfnamefont {O.}~\bibnamefont {Spiegel}},\ }\bibfield  {title} {\bibinfo {title} {Movement predictability of individual barn owls facilitates estimation of home range size and survival},\ }\href@noop {} {\bibfield  {journal} {\bibinfo  {journal} {Mov. Ecol.}\ }\textbf {\bibinfo {volume} {11}},\ \bibinfo {pages} {10} (\bibinfo {year} {2023})}\BibitemShut {NoStop}%
\bibitem [{\citenamefont {Rozman}\ \emph {et~al.}(2021)\citenamefont {Rozman}, \citenamefont {Izhaki}, \citenamefont {Roulin},\ and\ \citenamefont {Charter}}]{rozman2021movement}%
  \BibitemOpen
  \bibfield  {author} {\bibinfo {author} {\bibfnamefont {G.}~\bibnamefont {Rozman}}, \bibinfo {author} {\bibfnamefont {I.}~\bibnamefont {Izhaki}}, \bibinfo {author} {\bibfnamefont {A.}~\bibnamefont {Roulin}},\ and\ \bibinfo {author} {\bibfnamefont {M.}~\bibnamefont {Charter}},\ }\bibfield  {title} {\bibinfo {title} {Movement ecology, breeding, diet, and roosting behavior of barn owls (\textit{Tyto alba}) in a transboundary conflict region},\ }\href@noop {} {\bibfield  {journal} {\bibinfo  {journal} {Regional Environmental Change}\ }\textbf {\bibinfo {volume} {21}},\ \bibinfo {pages} {1} (\bibinfo {year} {2021})}\BibitemShut {NoStop}%
\bibitem [{\citenamefont {Barraquand}\ and\ \citenamefont {Benhamou}(2008)}]{barraquand2008animal}%
  \BibitemOpen
  \bibfield  {author} {\bibinfo {author} {\bibfnamefont {F.}~\bibnamefont {Barraquand}}\ and\ \bibinfo {author} {\bibfnamefont {S.}~\bibnamefont {Benhamou}},\ }\bibfield  {title} {\bibinfo {title} {Animal movements in heterogeneous landscapes: identifying profitable places and homogeneous movement bouts},\ }\href@noop {} {\bibfield  {journal} {\bibinfo  {journal} {Ecology}\ }\textbf {\bibinfo {volume} {89}},\ \bibinfo {pages} {3336} (\bibinfo {year} {2008})}\BibitemShut {NoStop}%
\bibitem [{\citenamefont {Lavielle}(2005)}]{lavielle2005using}%
  \BibitemOpen
  \bibfield  {author} {\bibinfo {author} {\bibfnamefont {M.}~\bibnamefont {Lavielle}},\ }\bibfield  {title} {\bibinfo {title} {Using penalized contrasts for the change-point problem},\ }\href@noop {} {\bibfield  {journal} {\bibinfo  {journal} {Signal Processing}\ }\textbf {\bibinfo {volume} {85}},\ \bibinfo {pages} {1501} (\bibinfo {year} {2005})}\BibitemShut {NoStop}%
\bibitem [{Note1()}]{Note1}%
  \BibitemOpen
  \bibinfo {note} {Study of SAND for individual animals requires collecting additional data per individual over longer periods and employing time averages instead of ensemble averages, and is not discussed in this manuscript.}\BibitemShut {Stop}%
\bibitem [{SIr()}]{SIref}%
  \BibitemOpen
  \href@noop {} {\bibinfo  {journal} {See Supplementary Information at [URL] for additional figures supporting the discussion in the main text}\ }\BibitemShut {NoStop}%
\bibitem [{\citenamefont {Rebenshtok}\ \emph {et~al.}(2014)\citenamefont {Rebenshtok}, \citenamefont {Denisov}, \citenamefont {H{\"a}nggi},\ and\ \citenamefont {Barkai}}]{rebenshtok2014non}%
  \BibitemOpen
\bibfield  {journal} {  }\bibfield  {author} {\bibinfo {author} {\bibfnamefont {A.}~\bibnamefont {Rebenshtok}}, \bibinfo {author} {\bibfnamefont {S.}~\bibnamefont {Denisov}}, \bibinfo {author} {\bibfnamefont {P.}~\bibnamefont {H{\"a}nggi}},\ and\ \bibinfo {author} {\bibfnamefont {E.}~\bibnamefont {Barkai}},\ }\bibfield  {title} {\bibinfo {title} {Non-normalizable densities in strong anomalous diffusion: beyond the central limit theorem},\ }\href@noop {} {\bibfield  {journal} {\bibinfo  {journal} {Phys. Rev. Lett.}\ }\textbf {\bibinfo {volume} {112}},\ \bibinfo {pages} {110601} (\bibinfo {year} {2014})}\BibitemShut {NoStop}%
\bibitem [{\citenamefont {Burioni}\ \emph {et~al.}(2013)\citenamefont {Burioni}, \citenamefont {Gradenigo}, \citenamefont {Sarracino}, \citenamefont {Vezzani},\ and\ \citenamefont {Vulpiani}}]{burioni2013rare}%
  \BibitemOpen
  \bibfield  {author} {\bibinfo {author} {\bibfnamefont {R.}~\bibnamefont {Burioni}}, \bibinfo {author} {\bibfnamefont {G.}~\bibnamefont {Gradenigo}}, \bibinfo {author} {\bibfnamefont {A.}~\bibnamefont {Sarracino}}, \bibinfo {author} {\bibfnamefont {A.}~\bibnamefont {Vezzani}},\ and\ \bibinfo {author} {\bibfnamefont {A.}~\bibnamefont {Vulpiani}},\ }\bibfield  {title} {\bibinfo {title} {Rare events and scaling properties in field-induced anomalous dynamics},\ }\href@noop {} {\bibfield  {journal} {\bibinfo  {journal} {J Stat. Mech.: Theor. Exp.}\ }\textbf {\bibinfo {volume} {2013}},\ \bibinfo {pages} {P09022} (\bibinfo {year} {2013})}\BibitemShut {NoStop}%
\bibitem [{\citenamefont {Blackwell}(1997)}]{blackwell1997random}%
  \BibitemOpen
  \bibfield  {author} {\bibinfo {author} {\bibfnamefont {P.}~\bibnamefont {Blackwell}},\ }\bibfield  {title} {\bibinfo {title} {Random diffusion models for animal movement},\ }\href@noop {} {\bibfield  {journal} {\bibinfo  {journal} {Ecological Modelling}\ }\textbf {\bibinfo {volume} {100}},\ \bibinfo {pages} {87} (\bibinfo {year} {1997})}\BibitemShut {NoStop}%
\bibitem [{\citenamefont {Hanks}\ \emph {et~al.}(2011)\citenamefont {Hanks}, \citenamefont {Hooten}, \citenamefont {Johnson},\ and\ \citenamefont {Sterling}}]{hanks2011velocity}%
  \BibitemOpen
  \bibfield  {author} {\bibinfo {author} {\bibfnamefont {E.~M.}\ \bibnamefont {Hanks}}, \bibinfo {author} {\bibfnamefont {M.~B.}\ \bibnamefont {Hooten}}, \bibinfo {author} {\bibfnamefont {D.~S.}\ \bibnamefont {Johnson}},\ and\ \bibinfo {author} {\bibfnamefont {J.~T.}\ \bibnamefont {Sterling}},\ }\bibfield  {title} {\bibinfo {title} {Velocity-based movement modeling for individual and population level inference},\ }\href@noop {} {\bibfield  {journal} {\bibinfo  {journal} {PLoS One}\ }\textbf {\bibinfo {volume} {6}},\ \bibinfo {pages} {e22795} (\bibinfo {year} {2011})}\BibitemShut {NoStop}%
\bibitem [{\citenamefont {Eisaguirre}\ \emph {et~al.}(2021)\citenamefont {Eisaguirre}, \citenamefont {Booms}, \citenamefont {Barger}, \citenamefont {Goddard},\ and\ \citenamefont {Breed}}]{eisaguirre2021multistate}%
  \BibitemOpen
  \bibfield  {author} {\bibinfo {author} {\bibfnamefont {J.~M.}\ \bibnamefont {Eisaguirre}}, \bibinfo {author} {\bibfnamefont {T.~L.}\ \bibnamefont {Booms}}, \bibinfo {author} {\bibfnamefont {C.~P.}\ \bibnamefont {Barger}}, \bibinfo {author} {\bibfnamefont {S.~D.}\ \bibnamefont {Goddard}},\ and\ \bibinfo {author} {\bibfnamefont {G.~A.}\ \bibnamefont {Breed}},\ }\bibfield  {title} {\bibinfo {title} {Multistate ornstein--uhlenbeck approach for practical estimation of movement and resource selection around central places},\ }\href@noop {} {\bibfield  {journal} {\bibinfo  {journal} {Methods in Ecology and Evolution}\ }\textbf {\bibinfo {volume} {12}},\ \bibinfo {pages} {507} (\bibinfo {year} {2021})}\BibitemShut {NoStop}%
\bibitem [{\citenamefont {McClintock}\ \emph {et~al.}(2012)\citenamefont {McClintock}, \citenamefont {King}, \citenamefont {Thomas}, \citenamefont {Matthiopoulos}, \citenamefont {McConnell},\ and\ \citenamefont {Morales}}]{mcclintock2012general}%
  \BibitemOpen
  \bibfield  {author} {\bibinfo {author} {\bibfnamefont {B.~T.}\ \bibnamefont {McClintock}}, \bibinfo {author} {\bibfnamefont {R.}~\bibnamefont {King}}, \bibinfo {author} {\bibfnamefont {L.}~\bibnamefont {Thomas}}, \bibinfo {author} {\bibfnamefont {J.}~\bibnamefont {Matthiopoulos}}, \bibinfo {author} {\bibfnamefont {B.~J.}\ \bibnamefont {McConnell}},\ and\ \bibinfo {author} {\bibfnamefont {J.~M.}\ \bibnamefont {Morales}},\ }\bibfield  {title} {\bibinfo {title} {A general discrete-time modeling framework for animal movement using multistate random walks},\ }\href@noop {} {\bibfield  {journal} {\bibinfo  {journal} {Ecological Monographs}\ }\textbf {\bibinfo {volume} {82}},\ \bibinfo {pages} {335} (\bibinfo {year} {2012})}\BibitemShut {NoStop}%
\bibitem [{\citenamefont {Dray}\ \emph {et~al.}(2010)\citenamefont {Dray}, \citenamefont {Royer-Carenzi},\ and\ \citenamefont {Calenge}}]{dray2010exploratory}%
  \BibitemOpen
  \bibfield  {author} {\bibinfo {author} {\bibfnamefont {S.}~\bibnamefont {Dray}}, \bibinfo {author} {\bibfnamefont {M.}~\bibnamefont {Royer-Carenzi}},\ and\ \bibinfo {author} {\bibfnamefont {C.}~\bibnamefont {Calenge}},\ }\bibfield  {title} {\bibinfo {title} {The exploratory analysis of autocorrelation in animal-movement studies},\ }\href@noop {} {\bibfield  {journal} {\bibinfo  {journal} {Ecol. Res.}\ }\textbf {\bibinfo {volume} {25}},\ \bibinfo {pages} {673} (\bibinfo {year} {2010})}\BibitemShut {NoStop}%
\bibitem [{\citenamefont {Cristol}\ \emph {et~al.}(2017)\citenamefont {Cristol}, \citenamefont {Akst}, \citenamefont {Curatola}, \citenamefont {Dunlavey}, \citenamefont {Fisk},\ and\ \citenamefont {Moody}}]{cristol2017age}%
  \BibitemOpen
  \bibfield  {author} {\bibinfo {author} {\bibfnamefont {D.~A.}\ \bibnamefont {Cristol}}, \bibinfo {author} {\bibfnamefont {J.~G.}\ \bibnamefont {Akst}}, \bibinfo {author} {\bibfnamefont {M.~K.}\ \bibnamefont {Curatola}}, \bibinfo {author} {\bibfnamefont {E.~G.}\ \bibnamefont {Dunlavey}}, \bibinfo {author} {\bibfnamefont {K.~A.}\ \bibnamefont {Fisk}},\ and\ \bibinfo {author} {\bibfnamefont {K.~E.}\ \bibnamefont {Moody}},\ }\bibfield  {title} {\bibinfo {title} {Age-related differences in foraging ability among clam-dropping herring gulls (larus argentatus)},\ }\href@noop {} {\bibfield  {journal} {\bibinfo  {journal} {The Wilson Journal of Ornithology}\ }\textbf {\bibinfo {volume} {129}},\ \bibinfo {pages} {301} (\bibinfo {year} {2017})}\BibitemShut {NoStop}%
\bibitem [{\citenamefont {Martins}\ \emph {et~al.}(2024)\citenamefont {Martins}, \citenamefont {Soriano-Redondo}, \citenamefont {Franco},\ and\ \citenamefont {Catry}}]{martins2024age}%
  \BibitemOpen
  \bibfield  {author} {\bibinfo {author} {\bibfnamefont {B.~H.}\ \bibnamefont {Martins}}, \bibinfo {author} {\bibfnamefont {A.}~\bibnamefont {Soriano-Redondo}}, \bibinfo {author} {\bibfnamefont {A.~M.}\ \bibnamefont {Franco}},\ and\ \bibinfo {author} {\bibfnamefont {I.}~\bibnamefont {Catry}},\ }\bibfield  {title} {\bibinfo {title} {Age mediates access to landfill food resources and foraging proficiency in a long-lived bird species},\ }\href@noop {} {\bibfield  {journal} {\bibinfo  {journal} {Animal Behaviour}\ }\textbf {\bibinfo {volume} {207}},\ \bibinfo {pages} {23} (\bibinfo {year} {2024})}\BibitemShut {NoStop}%
\bibitem [{\citenamefont {Grecian}\ \emph {et~al.}(2018)\citenamefont {Grecian}, \citenamefont {Lane}, \citenamefont {Michelot}, \citenamefont {Wade},\ and\ \citenamefont {Hamer}}]{grecian2018understanding}%
  \BibitemOpen
  \bibfield  {author} {\bibinfo {author} {\bibfnamefont {W.~J.}\ \bibnamefont {Grecian}}, \bibinfo {author} {\bibfnamefont {J.~V.}\ \bibnamefont {Lane}}, \bibinfo {author} {\bibfnamefont {T.}~\bibnamefont {Michelot}}, \bibinfo {author} {\bibfnamefont {H.~M.}\ \bibnamefont {Wade}},\ and\ \bibinfo {author} {\bibfnamefont {K.~C.}\ \bibnamefont {Hamer}},\ }\bibfield  {title} {\bibinfo {title} {Understanding the ontogeny of foraging behaviour: insights from combining marine predator bio-logging with satellite-derived oceanography in hidden markov models},\ }\href@noop {} {\bibfield  {journal} {\bibinfo  {journal} {Journal of the Royal Society Interface}\ }\textbf {\bibinfo {volume} {15}},\ \bibinfo {pages} {20180084} (\bibinfo {year} {2018})}\BibitemShut {NoStop}%
\bibitem [{\citenamefont {Franks}\ and\ \citenamefont {Thorogood}(2018)}]{franks2018older}%
  \BibitemOpen
  \bibfield  {author} {\bibinfo {author} {\bibfnamefont {V.~R.}\ \bibnamefont {Franks}}\ and\ \bibinfo {author} {\bibfnamefont {R.}~\bibnamefont {Thorogood}},\ }\bibfield  {title} {\bibinfo {title} {Older and wiser? age differences in foraging and learning by an endangered passerine},\ }\href@noop {} {\bibfield  {journal} {\bibinfo  {journal} {Behavioural Processes}\ }\textbf {\bibinfo {volume} {148}},\ \bibinfo {pages} {1} (\bibinfo {year} {2018})}\BibitemShut {NoStop}%
\bibitem [{\citenamefont {Belthoff}\ \emph {et~al.}(1993)\citenamefont {Belthoff}, \citenamefont {Sparks},\ and\ \citenamefont {Ritchison}}]{belthoff1993home}%
  \BibitemOpen
  \bibfield  {author} {\bibinfo {author} {\bibfnamefont {J.~R.}\ \bibnamefont {Belthoff}}, \bibinfo {author} {\bibfnamefont {E.~J.}\ \bibnamefont {Sparks}},\ and\ \bibinfo {author} {\bibfnamefont {G.}~\bibnamefont {Ritchison}},\ }\bibfield  {title} {\bibinfo {title} {Home ranges of adult and juvenile eastern screech-owls: Size, seasonal},\ }\href@noop {} {\bibfield  {journal} {\bibinfo  {journal} {J. Raptor Res}\ }\textbf {\bibinfo {volume} {27}},\ \bibinfo {pages} {8} (\bibinfo {year} {1993})}\BibitemShut {NoStop}%
\bibitem [{\citenamefont {Kr{\"u}ger}\ \emph {et~al.}(2014)\citenamefont {Kr{\"u}ger}, \citenamefont {Reid},\ and\ \citenamefont {Amar}}]{kruger2014differential}%
  \BibitemOpen
  \bibfield  {author} {\bibinfo {author} {\bibfnamefont {S.}~\bibnamefont {Kr{\"u}ger}}, \bibinfo {author} {\bibfnamefont {T.}~\bibnamefont {Reid}},\ and\ \bibinfo {author} {\bibfnamefont {A.}~\bibnamefont {Amar}},\ }\bibfield  {title} {\bibinfo {title} {Differential range use between age classes of southern african bearded vultures gypaetus barbatus},\ }\href@noop {} {\bibfield  {journal} {\bibinfo  {journal} {PLoS One}\ }\textbf {\bibinfo {volume} {9}},\ \bibinfo {pages} {e114920} (\bibinfo {year} {2014})}\BibitemShut {NoStop}%
\bibitem [{\citenamefont {Paiva}\ \emph {et~al.}(2010)\citenamefont {Paiva}, \citenamefont {Geraldes}, \citenamefont {Ram{\'\i}rez}, \citenamefont {Garthe},\ and\ \citenamefont {Ramos}}]{paiva2010area}%
  \BibitemOpen
  \bibfield  {author} {\bibinfo {author} {\bibfnamefont {V.~H.}\ \bibnamefont {Paiva}}, \bibinfo {author} {\bibfnamefont {P.}~\bibnamefont {Geraldes}}, \bibinfo {author} {\bibfnamefont {I.}~\bibnamefont {Ram{\'\i}rez}}, \bibinfo {author} {\bibfnamefont {S.}~\bibnamefont {Garthe}},\ and\ \bibinfo {author} {\bibfnamefont {J.~A.}\ \bibnamefont {Ramos}},\ }\bibfield  {title} {\bibinfo {title} {How area restricted search of a pelagic seabird changes while performing a dual foraging strategy},\ }\href@noop {} {\bibfield  {journal} {\bibinfo  {journal} {Oikos}\ }\textbf {\bibinfo {volume} {119}},\ \bibinfo {pages} {1423} (\bibinfo {year} {2010})}\BibitemShut {NoStop}%
\bibitem [{\citenamefont {Dorfman}\ \emph {et~al.}(2022)\citenamefont {Dorfman}, \citenamefont {Hills},\ and\ \citenamefont {Scharf}}]{dorfman2022guide}%
  \BibitemOpen
  \bibfield  {author} {\bibinfo {author} {\bibfnamefont {A.}~\bibnamefont {Dorfman}}, \bibinfo {author} {\bibfnamefont {T.~T.}\ \bibnamefont {Hills}},\ and\ \bibinfo {author} {\bibfnamefont {I.}~\bibnamefont {Scharf}},\ }\bibfield  {title} {\bibinfo {title} {A guide to area-restricted search: a foundational foraging behaviour},\ }\href@noop {} {\bibfield  {journal} {\bibinfo  {journal} {Biological Reviews}\ }\textbf {\bibinfo {volume} {97}},\ \bibinfo {pages} {2076} (\bibinfo {year} {2022})}\BibitemShut {NoStop}%
\bibitem [{\citenamefont {Sims}\ \emph {et~al.}(2019)\citenamefont {Sims}, \citenamefont {Humphries}, \citenamefont {Hu}, \citenamefont {Medan},\ and\ \citenamefont {Berni}}]{sims2019optimal}%
  \BibitemOpen
  \bibfield  {author} {\bibinfo {author} {\bibfnamefont {D.~W.}\ \bibnamefont {Sims}}, \bibinfo {author} {\bibfnamefont {N.~E.}\ \bibnamefont {Humphries}}, \bibinfo {author} {\bibfnamefont {N.}~\bibnamefont {Hu}}, \bibinfo {author} {\bibfnamefont {V.}~\bibnamefont {Medan}},\ and\ \bibinfo {author} {\bibfnamefont {J.}~\bibnamefont {Berni}},\ }\bibfield  {title} {\bibinfo {title} {Optimal searching behaviour generated intrinsically by the central pattern generator for locomotion},\ }\href@noop {} {\bibfield  {journal} {\bibinfo  {journal} {Elife}\ }\textbf {\bibinfo {volume} {8}},\ \bibinfo {pages} {e50316} (\bibinfo {year} {2019})}\BibitemShut {NoStop}%
\bibitem [{\citenamefont {Campeau}\ \emph {et~al.}(2022)\citenamefont {Campeau}, \citenamefont {Simons},\ and\ \citenamefont {Stevens}}]{campeau2022evolutionary}%
  \BibitemOpen
  \bibfield  {author} {\bibinfo {author} {\bibfnamefont {W.}~\bibnamefont {Campeau}}, \bibinfo {author} {\bibfnamefont {A.~M.}\ \bibnamefont {Simons}},\ and\ \bibinfo {author} {\bibfnamefont {B.}~\bibnamefont {Stevens}},\ }\bibfield  {title} {\bibinfo {title} {The evolutionary maintenance of l{\'e}vy flight foraging},\ }\href@noop {} {\bibfield  {journal} {\bibinfo  {journal} {PLoS computational biology}\ }\textbf {\bibinfo {volume} {18}},\ \bibinfo {pages} {e1009490} (\bibinfo {year} {2022})}\BibitemShut {NoStop}%
\bibitem [{\citenamefont {Alessandretti}\ \emph {et~al.}(2020)\citenamefont {Alessandretti}, \citenamefont {Aslak},\ and\ \citenamefont {Lehmann}}]{alessandretti2020scales}%
  \BibitemOpen
  \bibfield  {author} {\bibinfo {author} {\bibfnamefont {L.}~\bibnamefont {Alessandretti}}, \bibinfo {author} {\bibfnamefont {U.}~\bibnamefont {Aslak}},\ and\ \bibinfo {author} {\bibfnamefont {S.}~\bibnamefont {Lehmann}},\ }\bibfield  {title} {\bibinfo {title} {The scales of human mobility},\ }\href@noop {} {\bibfield  {journal} {\bibinfo  {journal} {Nature}\ }\textbf {\bibinfo {volume} {587}},\ \bibinfo {pages} {402} (\bibinfo {year} {2020})}\BibitemShut {NoStop}%
\bibitem [{\citenamefont {Campeau}\ \emph {et~al.}(2024)\citenamefont {Campeau}, \citenamefont {Simons},\ and\ \citenamefont {Stevens}}]{campeau2024intermittent}%
  \BibitemOpen
  \bibfield  {author} {\bibinfo {author} {\bibfnamefont {W.}~\bibnamefont {Campeau}}, \bibinfo {author} {\bibfnamefont {A.~M.}\ \bibnamefont {Simons}},\ and\ \bibinfo {author} {\bibfnamefont {B.}~\bibnamefont {Stevens}},\ }\bibfield  {title} {\bibinfo {title} {Intermittent search, not strict l{\'e}vy flight, evolves under relaxed foraging distribution constraints},\ }\href@noop {} {\bibfield  {journal} {\bibinfo  {journal} {The American Naturalist}\ }\textbf {\bibinfo {volume} {203}},\ \bibinfo {pages} {513} (\bibinfo {year} {2024})}\BibitemShut {NoStop}%
\bibitem [{\citenamefont {Viswanathan}\ \emph {et~al.}(1996)\citenamefont {Viswanathan}, \citenamefont {Afanasyev}, \citenamefont {Buldyrev}, \citenamefont {Murphy}, \citenamefont {Prince},\ and\ \citenamefont {Stanley}}]{viswanathan1996levy}%
  \BibitemOpen
  \bibfield  {author} {\bibinfo {author} {\bibfnamefont {G.~M.}\ \bibnamefont {Viswanathan}}, \bibinfo {author} {\bibfnamefont {V.}~\bibnamefont {Afanasyev}}, \bibinfo {author} {\bibfnamefont {S.}~\bibnamefont {Buldyrev}}, \bibinfo {author} {\bibfnamefont {E.}~\bibnamefont {Murphy}}, \bibinfo {author} {\bibfnamefont {P.}~\bibnamefont {Prince}},\ and\ \bibinfo {author} {\bibfnamefont {H.~E.}\ \bibnamefont {Stanley}},\ }\bibfield  {title} {\bibinfo {title} {L{\'e}vy flight search patterns of wandering albatrosses},\ }\href@noop {} {\bibfield  {journal} {\bibinfo  {journal} {Nature}\ }\textbf {\bibinfo {volume} {381}},\ \bibinfo {pages} {413} (\bibinfo {year} {1996})}\BibitemShut {NoStop}%
\bibitem [{\citenamefont {Pacheco-Cobos}\ \emph {et~al.}(2019)\citenamefont {Pacheco-Cobos}, \citenamefont {Winterhalder}, \citenamefont {Cuatianquiz-Lima}, \citenamefont {Rosetti}, \citenamefont {Hudson},\ and\ \citenamefont {Ross}}]{pacheco2019nahua}%
  \BibitemOpen
  \bibfield  {author} {\bibinfo {author} {\bibfnamefont {L.}~\bibnamefont {Pacheco-Cobos}}, \bibinfo {author} {\bibfnamefont {B.}~\bibnamefont {Winterhalder}}, \bibinfo {author} {\bibfnamefont {C.}~\bibnamefont {Cuatianquiz-Lima}}, \bibinfo {author} {\bibfnamefont {M.~F.}\ \bibnamefont {Rosetti}}, \bibinfo {author} {\bibfnamefont {R.}~\bibnamefont {Hudson}},\ and\ \bibinfo {author} {\bibfnamefont {C.~T.}\ \bibnamefont {Ross}},\ }\bibfield  {title} {\bibinfo {title} {Nahua mushroom gatherers use area-restricted search strategies that conform to marginal value theorem predictions},\ }\href@noop {} {\bibfield  {journal} {\bibinfo  {journal} {Proceedings of the National Academy of Sciences}\ }\textbf {\bibinfo {volume} {116}},\ \bibinfo {pages} {10339} (\bibinfo {year} {2019})}\BibitemShut {NoStop}%
\bibitem [{\citenamefont {Song}\ \emph {et~al.}(2010)\citenamefont {Song}, \citenamefont {Koren}, \citenamefont {Wang},\ and\ \citenamefont {Barab{\'a}si}}]{song2010modelling}%
  \BibitemOpen
  \bibfield  {author} {\bibinfo {author} {\bibfnamefont {C.}~\bibnamefont {Song}}, \bibinfo {author} {\bibfnamefont {T.}~\bibnamefont {Koren}}, \bibinfo {author} {\bibfnamefont {P.}~\bibnamefont {Wang}},\ and\ \bibinfo {author} {\bibfnamefont {A.-L.}\ \bibnamefont {Barab{\'a}si}},\ }\bibfield  {title} {\bibinfo {title} {Modelling the scaling properties of human mobility},\ }\href@noop {} {\bibfield  {journal} {\bibinfo  {journal} {Nature Physics}\ }\textbf {\bibinfo {volume} {6}},\ \bibinfo {pages} {818} (\bibinfo {year} {2010})}\BibitemShut {NoStop}%
\bibitem [{\citenamefont {Brockmann}\ \emph {et~al.}(2006)\citenamefont {Brockmann}, \citenamefont {Hufnagel},\ and\ \citenamefont {Geisel}}]{brockmann2006scaling}%
  \BibitemOpen
  \bibfield  {author} {\bibinfo {author} {\bibfnamefont {D.}~\bibnamefont {Brockmann}}, \bibinfo {author} {\bibfnamefont {L.}~\bibnamefont {Hufnagel}},\ and\ \bibinfo {author} {\bibfnamefont {T.}~\bibnamefont {Geisel}},\ }\bibfield  {title} {\bibinfo {title} {The scaling laws of human travel},\ }\href@noop {} {\bibfield  {journal} {\bibinfo  {journal} {Nature}\ }\textbf {\bibinfo {volume} {439}},\ \bibinfo {pages} {462} (\bibinfo {year} {2006})}\BibitemShut {NoStop}%
\bibitem [{\citenamefont {Li}\ \emph {et~al.}(2008)\citenamefont {Li}, \citenamefont {N{\o}rrelykke},\ and\ \citenamefont {Cox}}]{li2008persistent}%
  \BibitemOpen
  \bibfield  {author} {\bibinfo {author} {\bibfnamefont {L.}~\bibnamefont {Li}}, \bibinfo {author} {\bibfnamefont {S.~F.}\ \bibnamefont {N{\o}rrelykke}},\ and\ \bibinfo {author} {\bibfnamefont {E.~C.}\ \bibnamefont {Cox}},\ }\bibfield  {title} {\bibinfo {title} {Persistent cell motion in the absence of external signals: a search strategy for eukaryotic cells},\ }\href@noop {} {\bibfield  {journal} {\bibinfo  {journal} {PLoS one}\ }\textbf {\bibinfo {volume} {3}},\ \bibinfo {pages} {e2093} (\bibinfo {year} {2008})}\BibitemShut {NoStop}%
\bibitem [{\citenamefont {Matth{\"a}us}\ \emph {et~al.}(2011)\citenamefont {Matth{\"a}us}, \citenamefont {Mommer}, \citenamefont {Curk},\ and\ \citenamefont {Dobnikar}}]{matthaus2011origin}%
  \BibitemOpen
  \bibfield  {author} {\bibinfo {author} {\bibfnamefont {F.}~\bibnamefont {Matth{\"a}us}}, \bibinfo {author} {\bibfnamefont {M.~S.}\ \bibnamefont {Mommer}}, \bibinfo {author} {\bibfnamefont {T.}~\bibnamefont {Curk}},\ and\ \bibinfo {author} {\bibfnamefont {J.}~\bibnamefont {Dobnikar}},\ }\bibfield  {title} {\bibinfo {title} {On the origin and characteristics of noise-induced l{\'e}vy walks of e. coli},\ }\href@noop {} {\bibfield  {journal} {\bibinfo  {journal} {PloS one}\ }\textbf {\bibinfo {volume} {6}},\ \bibinfo {pages} {e18623} (\bibinfo {year} {2011})}\BibitemShut {NoStop}%
\bibitem [{\citenamefont {Watkins}\ and\ \citenamefont {Rose}(2013)}]{watkins2013evaluating}%
  \BibitemOpen
  \bibfield  {author} {\bibinfo {author} {\bibfnamefont {K.~S.}\ \bibnamefont {Watkins}}\ and\ \bibinfo {author} {\bibfnamefont {K.~A.}\ \bibnamefont {Rose}},\ }\bibfield  {title} {\bibinfo {title} {Evaluating the performance of individual-based animal movement models in novel environments},\ }\href@noop {} {\bibfield  {journal} {\bibinfo  {journal} {Ecological Modelling}\ }\textbf {\bibinfo {volume} {250}},\ \bibinfo {pages} {214} (\bibinfo {year} {2013})}\BibitemShut {NoStop}%
\bibitem [{\citenamefont {Martens}\ \emph {et~al.}(2012)\citenamefont {Martens}, \citenamefont {Angelani}, \citenamefont {Di~Leonardo},\ and\ \citenamefont {Bocquet}}]{martens2012probability}%
  \BibitemOpen
  \bibfield  {author} {\bibinfo {author} {\bibfnamefont {K.}~\bibnamefont {Martens}}, \bibinfo {author} {\bibfnamefont {L.}~\bibnamefont {Angelani}}, \bibinfo {author} {\bibfnamefont {R.}~\bibnamefont {Di~Leonardo}},\ and\ \bibinfo {author} {\bibfnamefont {L.}~\bibnamefont {Bocquet}},\ }\bibfield  {title} {\bibinfo {title} {Probability distributions for the run-and-tumble bacterial dynamics: An analogy to the lorentz model},\ }\href@noop {} {\bibfield  {journal} {\bibinfo  {journal} {The European Physical Journal E}\ }\textbf {\bibinfo {volume} {35}},\ \bibinfo {pages} {84} (\bibinfo {year} {2012})}\BibitemShut {NoStop}%
\bibitem [{\citenamefont {Cates}(2012)}]{cates2012diffusive}%
  \BibitemOpen
  \bibfield  {author} {\bibinfo {author} {\bibfnamefont {M.~E.}\ \bibnamefont {Cates}},\ }\bibfield  {title} {\bibinfo {title} {Diffusive transport without detailed balance in motile bacteria: does microbiology need statistical physics?},\ }\href@noop {} {\bibfield  {journal} {\bibinfo  {journal} {Reports on Progress in Physics}\ }\textbf {\bibinfo {volume} {75}},\ \bibinfo {pages} {042601} (\bibinfo {year} {2012})}\BibitemShut {NoStop}%
\bibitem [{\citenamefont {Postlethwaite}\ \emph {et~al.}(2013)\citenamefont {Postlethwaite}, \citenamefont {Brown},\ and\ \citenamefont {Dennis}}]{postlethwaite2013new}%
  \BibitemOpen
  \bibfield  {author} {\bibinfo {author} {\bibfnamefont {C.~M.}\ \bibnamefont {Postlethwaite}}, \bibinfo {author} {\bibfnamefont {P.}~\bibnamefont {Brown}},\ and\ \bibinfo {author} {\bibfnamefont {T.~E.}\ \bibnamefont {Dennis}},\ }\bibfield  {title} {\bibinfo {title} {A new multi-scale measure for analysing animal movement data},\ }\href@noop {} {\bibfield  {journal} {\bibinfo  {journal} {Journal of theoretical biology}\ }\textbf {\bibinfo {volume} {317}},\ \bibinfo {pages} {175} (\bibinfo {year} {2013})}\BibitemShut {NoStop}%
\bibitem [{\citenamefont {Fleming}\ \emph {et~al.}(2014)\citenamefont {Fleming}, \citenamefont {Calabrese}, \citenamefont {Mueller}, \citenamefont {Olson}, \citenamefont {Leimgruber},\ and\ \citenamefont {Fagan}}]{fleming2014fine}%
  \BibitemOpen
  \bibfield  {author} {\bibinfo {author} {\bibfnamefont {C.~H.}\ \bibnamefont {Fleming}}, \bibinfo {author} {\bibfnamefont {J.~M.}\ \bibnamefont {Calabrese}}, \bibinfo {author} {\bibfnamefont {T.}~\bibnamefont {Mueller}}, \bibinfo {author} {\bibfnamefont {K.~A.}\ \bibnamefont {Olson}}, \bibinfo {author} {\bibfnamefont {P.}~\bibnamefont {Leimgruber}},\ and\ \bibinfo {author} {\bibfnamefont {W.~F.}\ \bibnamefont {Fagan}},\ }\bibfield  {title} {\bibinfo {title} {From fine-scale foraging to home ranges: a semivariance approach to identifying movement modes across spatiotemporal scales},\ }\href@noop {} {\bibfield  {journal} {\bibinfo  {journal} {The American Naturalist}\ }\textbf {\bibinfo {volume} {183}},\ \bibinfo {pages} {E154} (\bibinfo {year} {2014})}\BibitemShut {NoStop}%
\bibitem [{\citenamefont {Glennie}\ \emph {et~al.}(2023)\citenamefont {Glennie}, \citenamefont {Adam}, \citenamefont {Leos-Barajas}, \citenamefont {Michelot}, \citenamefont {Photopoulou},\ and\ \citenamefont {McClintock}}]{glennie2023hidden}%
  \BibitemOpen
  \bibfield  {author} {\bibinfo {author} {\bibfnamefont {R.}~\bibnamefont {Glennie}}, \bibinfo {author} {\bibfnamefont {T.}~\bibnamefont {Adam}}, \bibinfo {author} {\bibfnamefont {V.}~\bibnamefont {Leos-Barajas}}, \bibinfo {author} {\bibfnamefont {T.}~\bibnamefont {Michelot}}, \bibinfo {author} {\bibfnamefont {T.}~\bibnamefont {Photopoulou}},\ and\ \bibinfo {author} {\bibfnamefont {B.~T.}\ \bibnamefont {McClintock}},\ }\bibfield  {title} {\bibinfo {title} {Hidden markov models: Pitfalls and opportunities in ecology},\ }\href@noop {} {\bibfield  {journal} {\bibinfo  {journal} {Methods in Ecology and Evolution}\ }\textbf {\bibinfo {volume} {14}},\ \bibinfo {pages} {43} (\bibinfo {year} {2023})}\BibitemShut {NoStop}%
\bibitem [{\citenamefont {Langrock}\ \emph {et~al.}(2012)\citenamefont {Langrock}, \citenamefont {King}, \citenamefont {Matthiopoulos}, \citenamefont {Thomas}, \citenamefont {Fortin},\ and\ \citenamefont {Morales}}]{langrock2012flexible}%
  \BibitemOpen
  \bibfield  {author} {\bibinfo {author} {\bibfnamefont {R.}~\bibnamefont {Langrock}}, \bibinfo {author} {\bibfnamefont {R.}~\bibnamefont {King}}, \bibinfo {author} {\bibfnamefont {J.}~\bibnamefont {Matthiopoulos}}, \bibinfo {author} {\bibfnamefont {L.}~\bibnamefont {Thomas}}, \bibinfo {author} {\bibfnamefont {D.}~\bibnamefont {Fortin}},\ and\ \bibinfo {author} {\bibfnamefont {J.~M.}\ \bibnamefont {Morales}},\ }\bibfield  {title} {\bibinfo {title} {Flexible and practical modeling of animal telemetry data: hidden markov models and extensions},\ }\href@noop {} {\bibfield  {journal} {\bibinfo  {journal} {Ecology}\ }\textbf {\bibinfo {volume} {93}},\ \bibinfo {pages} {2336} (\bibinfo {year} {2012})}\BibitemShut {NoStop}%
\bibitem [{\citenamefont {Patterson}\ \emph {et~al.}(2009)\citenamefont {Patterson}, \citenamefont {Basson}, \citenamefont {Bravington},\ and\ \citenamefont {Gunn}}]{patterson2009classifying}%
  \BibitemOpen
  \bibfield  {author} {\bibinfo {author} {\bibfnamefont {T.~A.}\ \bibnamefont {Patterson}}, \bibinfo {author} {\bibfnamefont {M.}~\bibnamefont {Basson}}, \bibinfo {author} {\bibfnamefont {M.~V.}\ \bibnamefont {Bravington}},\ and\ \bibinfo {author} {\bibfnamefont {J.~S.}\ \bibnamefont {Gunn}},\ }\bibfield  {title} {\bibinfo {title} {Classifying movement behaviour in relation to environmental conditions using hidden markov models},\ }\href@noop {} {\bibfield  {journal} {\bibinfo  {journal} {Journal of Animal Ecology}\ }\textbf {\bibinfo {volume} {78}},\ \bibinfo {pages} {1113} (\bibinfo {year} {2009})}\BibitemShut {NoStop}%
\bibitem [{\citenamefont {Fauchald}\ and\ \citenamefont {Tveraa}(2003)}]{fauchald2003using}%
  \BibitemOpen
  \bibfield  {author} {\bibinfo {author} {\bibfnamefont {P.}~\bibnamefont {Fauchald}}\ and\ \bibinfo {author} {\bibfnamefont {T.}~\bibnamefont {Tveraa}},\ }\bibfield  {title} {\bibinfo {title} {Using first-passage time in the analysis of area-restricted search and habitat selection},\ }\href@noop {} {\bibfield  {journal} {\bibinfo  {journal} {Ecology}\ }\textbf {\bibinfo {volume} {84}},\ \bibinfo {pages} {282} (\bibinfo {year} {2003})}\BibitemShut {NoStop}%
\bibitem [{\citenamefont {Johnson}\ \emph {et~al.}(2008)\citenamefont {Johnson}, \citenamefont {London}, \citenamefont {Lea},\ and\ \citenamefont {Durban}}]{johnson2008continuous}%
  \BibitemOpen
  \bibfield  {author} {\bibinfo {author} {\bibfnamefont {D.~S.}\ \bibnamefont {Johnson}}, \bibinfo {author} {\bibfnamefont {J.~M.}\ \bibnamefont {London}}, \bibinfo {author} {\bibfnamefont {M.-A.}\ \bibnamefont {Lea}},\ and\ \bibinfo {author} {\bibfnamefont {J.~W.}\ \bibnamefont {Durban}},\ }\bibfield  {title} {\bibinfo {title} {Continuous-time correlated random walk model for animal telemetry data},\ }\href@noop {} {\bibfield  {journal} {\bibinfo  {journal} {Ecology}\ }\textbf {\bibinfo {volume} {89}},\ \bibinfo {pages} {1208} (\bibinfo {year} {2008})}\BibitemShut {NoStop}%
\bibitem [{\citenamefont {Patterson}\ \emph {et~al.}(2008)\citenamefont {Patterson}, \citenamefont {Thomas}, \citenamefont {Wilcox}, \citenamefont {Ovaskainen},\ and\ \citenamefont {Matthiopoulos}}]{Pa08}%
  \BibitemOpen
  \bibfield  {author} {\bibinfo {author} {\bibfnamefont {T.}~\bibnamefont {Patterson}}, \bibinfo {author} {\bibfnamefont {L.}~\bibnamefont {Thomas}}, \bibinfo {author} {\bibfnamefont {C.}~\bibnamefont {Wilcox}}, \bibinfo {author} {\bibfnamefont {O.}~\bibnamefont {Ovaskainen}},\ and\ \bibinfo {author} {\bibfnamefont {J.}~\bibnamefont {Matthiopoulos}},\ }\bibfield  {title} {\bibinfo {title} {State-space models of individual animal movement},\ }\href@noop {} {\bibfield  {journal} {\bibinfo  {journal} {Trends Ecol. Evol.}\ }\textbf {\bibinfo {volume} {23}},\ \bibinfo {pages} {87} (\bibinfo {year} {2008})}\BibitemShut {NoStop}%
\bibitem [{\citenamefont {Morales}\ \emph {et~al.}(2004)\citenamefont {Morales}, \citenamefont {Haydon}, \citenamefont {Frair}, \citenamefont {Holsinger},\ and\ \citenamefont {Fryxell}}]{Mo04}%
  \BibitemOpen
  \bibfield  {author} {\bibinfo {author} {\bibfnamefont {J.~M.}\ \bibnamefont {Morales}}, \bibinfo {author} {\bibfnamefont {D.~T.}\ \bibnamefont {Haydon}}, \bibinfo {author} {\bibfnamefont {J.}~\bibnamefont {Frair}}, \bibinfo {author} {\bibfnamefont {K.~E.}\ \bibnamefont {Holsinger}},\ and\ \bibinfo {author} {\bibfnamefont {J.~M.}\ \bibnamefont {Fryxell}},\ }\bibfield  {title} {\bibinfo {title} {Extracting more out of relocation data: building movement models as mixtures of random walks},\ }\href@noop {} {\bibfield  {journal} {\bibinfo  {journal} {Ecology}\ }\textbf {\bibinfo {volume} {85}},\ \bibinfo {pages} {2436} (\bibinfo {year} {2004})}\BibitemShut {NoStop}%
\bibitem [{\citenamefont {Worton}(1989)}]{worton1989kernel}%
  \BibitemOpen
  \bibfield  {author} {\bibinfo {author} {\bibfnamefont {B.~J.}\ \bibnamefont {Worton}},\ }\bibfield  {title} {\bibinfo {title} {Kernel methods for estimating the utilization distribution in home-range studies},\ }\href@noop {} {\bibfield  {journal} {\bibinfo  {journal} {Ecology}\ }\textbf {\bibinfo {volume} {70}},\ \bibinfo {pages} {164} (\bibinfo {year} {1989})}\BibitemShut {NoStop}%
\bibitem [{\citenamefont {Bekoff}\ and\ \citenamefont {Mech}(1984)}]{bekoff1984simulation}%
  \BibitemOpen
  \bibfield  {author} {\bibinfo {author} {\bibfnamefont {M.}~\bibnamefont {Bekoff}}\ and\ \bibinfo {author} {\bibfnamefont {L.~D.}\ \bibnamefont {Mech}},\ }\bibfield  {title} {\bibinfo {title} {Simulation analyses of space use: home range estimates, variability, and sample size},\ }\href@noop {} {\bibfield  {journal} {\bibinfo  {journal} {Behavior Research Methods, Instruments, \& Computers}\ }\textbf {\bibinfo {volume} {16}},\ \bibinfo {pages} {32} (\bibinfo {year} {1984})}\BibitemShut {NoStop}%
\bibitem [{\citenamefont {Fieberg}\ and\ \citenamefont {B{\"o}rger}(2012)}]{fieberg2012could}%
  \BibitemOpen
  \bibfield  {author} {\bibinfo {author} {\bibfnamefont {J.}~\bibnamefont {Fieberg}}\ and\ \bibinfo {author} {\bibfnamefont {L.}~\bibnamefont {B{\"o}rger}},\ }\bibfield  {title} {\bibinfo {title} {Could you please phrase “home range” as a question?},\ }\href@noop {} {\bibfield  {journal} {\bibinfo  {journal} {Journal of mammalogy}\ }\textbf {\bibinfo {volume} {93}},\ \bibinfo {pages} {890} (\bibinfo {year} {2012})}\BibitemShut {NoStop}%
\bibitem [{\citenamefont {Seaman}\ and\ \citenamefont {Powell}(1996)}]{seaman1996evaluation}%
  \BibitemOpen
  \bibfield  {author} {\bibinfo {author} {\bibfnamefont {D.~E.}\ \bibnamefont {Seaman}}\ and\ \bibinfo {author} {\bibfnamefont {R.~A.}\ \bibnamefont {Powell}},\ }\bibfield  {title} {\bibinfo {title} {An evaluation of the accuracy of kernel density estimators for home range analysis},\ }\href@noop {} {\bibfield  {journal} {\bibinfo  {journal} {Ecology}\ }\textbf {\bibinfo {volume} {77}},\ \bibinfo {pages} {2075} (\bibinfo {year} {1996})}\BibitemShut {NoStop}%
\bibitem [{\citenamefont {Fleming}\ \emph {et~al.}(2015)\citenamefont {Fleming}, \citenamefont {Fagan}, \citenamefont {Mueller}, \citenamefont {Olson}, \citenamefont {Leimgruber},\ and\ \citenamefont {Calabrese}}]{fleming2015rigorous}%
  \BibitemOpen
  \bibfield  {author} {\bibinfo {author} {\bibfnamefont {C.~H.}\ \bibnamefont {Fleming}}, \bibinfo {author} {\bibfnamefont {W.~F.}\ \bibnamefont {Fagan}}, \bibinfo {author} {\bibfnamefont {T.}~\bibnamefont {Mueller}}, \bibinfo {author} {\bibfnamefont {K.~A.}\ \bibnamefont {Olson}}, \bibinfo {author} {\bibfnamefont {P.}~\bibnamefont {Leimgruber}},\ and\ \bibinfo {author} {\bibfnamefont {J.~M.}\ \bibnamefont {Calabrese}},\ }\bibfield  {title} {\bibinfo {title} {Rigorous home range estimation with movement data: a new autocorrelated kernel density estimator},\ }\href@noop {} {\bibfield  {journal} {\bibinfo  {journal} {Ecology}\ }\textbf {\bibinfo {volume} {96}},\ \bibinfo {pages} {1182} (\bibinfo {year} {2015})}\BibitemShut {NoStop}%
\end{thebibliography}%


\end{document}